# Topological Quantum Matter to Topological Phase Conversion: Fundamentals, Materials, Physical Systems for Phase Conversions, and Device Applications

Md Mobarak Hossain Polash,[1,2] Shahram Yalameha,[3] Haihan Zhou,[4] Kaveh Ahadi,[1] Zahra Nourbakhsh,[3] and Daryoosh Vashaee[1,2,*]

[1]Department of Materials Science and Engineering, NC State University, Raleigh, NC 27606, USA
[2]Department of Electrical and Computer Engineering, NC State University, Raleigh, NC 27606, USA
[3]Faculty of Physics, University of Isfahan, 81746-73441, Isfahan, Iran
[4]Department of Physics, NC State University, Raleigh, NC 27606, USA

**Outlines**

---

[*] Corresponding author: dvashae@ncsu.edu

**Abstract**
The spin-orbit coupling field, an atomic magnetic field inside a Kramers' system, or discrete symmetries can create a topological torus in the Brillouin Zone and provide protected edge or surface states, which can contain relativistic fermions, namely, Dirac and Weyl Fermions. The topology-protected helical edge or surface states and the bulk electronic energy band define different quantum or topological phases of matters, offering an excellent prospect for some unique device applications. Device applications of the quantum materials rely primarily on understanding the topological properties, their mutual conversion processes under different external stimuli, and the physical system for achieving the phase conversion. There have been tremendous efforts in finding new topological materials with exotic topological phases. However, the application of the topological properties in devices is still limited due to the slow progress in developing the physical structures for controlling the topological phase conversions. Such control systems often require extreme tuning conditions or the fabrication of complex multi-layered topological structures. This review article highlights the details of the topological phases, their conversion processes, along with their potential physical systems, and the prospective application fields. A general overview of the critical factors for topological phases and the materials properties are further discussed to provide the necessary background for the following sections.

**Key Index:** Topological Phases, Quantum Materials, Condensed Matter, Topological Phase Conversion, Topological Device Applications

## 1. Introduction

### 1.1. Overview

Symmetry, a magnificent and ubiquitous property of nature, provides a profound organizational principle to describe the stable phases of the matter based on the phenomenological Ginzburg-Landau theory on broken symmetry.[1-4] Till the 1980s, the broken symmetry theory was a complete success to explain all the known phases in condensed matters. The phases of condensed matter were explained by characterizing the phases with an internal local ordering parameter, while the phase transitions are defined by the minimization of the Landau free energy associated with the ground state of the Hamiltonian of the corresponding phase under certain external changes like temperature, pressure, and magnetic field.[1,2] Integer Quantum Hall Effects (IQHE) in 1980[5,6] and fractional quantum Hall effects (FQHE) in 1982[7,8] were the first experimental observations where quantum phases were appeared without breaking any local symmetry.[1-9] In QHE experiments, 2D electron gas (2DEG) systems showed quantized plateaus in the Hall conductance with the protected conducting edge states while the bulk states remain insulating.[1,2] This unique observation leads towards a new phase in condensed matter physics, known as topological phases, which are defined by the breaking of time-reversal symmetry (TRS), a long-range order, or global symmetry of the matter.[1-2] The global symmetry of the matter is constructed by the quantum superposition or long-range entanglement of the random particles of the system.[1,2] Later, more topological phases were discovered with interesting topology protected properties through the evolution of the ordinary Hall effect by considering spin and quantum effects. Like QHE helped to demonstrate the quantum Hall insulators, the generalization of quantum spin Hall effect to three dimensions helped to predict and demonstrate several topological phases, i.e., 2D topological insulator (in 2005),[10,11] 3D topological insulator (in 2007),[12,13,14] Dirac semimetal (in 2014),[15,17,18] and Weyl semimetal (in 2015).[19,20] These topological phases can be arising either from the extrinsic origin where a disorder-mediated spin-dependent scattering of the charge carriers plays the role or from the intrinsic origin, which is the spin-dependent band structure of the matter. Generally, quantum Hall or Chern insulator phases are caused by the external origin and defined by the Thouless-Kohmoto-Nightingale-Njis (TKKN) number or Chern number, a topological invariance property that is insensitive to the geometry and interaction of electrons.[21-24] On the other hand, topological insulator and semimetal phases are represented by the topological invariant, $Z_2$ index which is calculated from the Berry connection and Berry curvature of the surface of the reciprocal space, the Fourier space of crystals.[21-26] Berry connection and Berry curvature are calculated from the Berry phase of the ground state of the quantum Hamiltonian of a system.

Quantum phase conversions in condensed matter are typically defined by the breaking or conservation of the time-reversal symmetry (TRS). In Quantum Hall insulators, TRS is generally broken by an external magnetic field,[21-24] while, in the topological insulator and semimetal phases, TRS is either protected or broken by different external stimuli like E-field or B-field along with the protection or breaking of other symmetries.[23,24] According to Kramers' theorem, TRS protection can create a band degeneracy into the system,[23,24] a vital phenomenon in topological phases. In bulk topological phases, the external magnetic field is replaced by the spin-orbit coupling (SOC) field, another important aspect of the topological phases. Here, SOC plays the role of introducing the band inversion into a system by mixing the *s* and *p* orbitals (or mixing other orbitals; see 6.2) of the band structure, although it does not break the TRS.[23-26,27] The Dirac equation for the relativistic quantum mechanical effect combines both special relativity and quantum mechanics where space and time derivatives in the equation of motion are under the limits of relativity and the probabilistic interpretation of the wave function.[25,26] Dirac equation is the fundamental equation to define the different relativistic nature of electrons, known as Dirac fermions, Weyl fermions, and Majorana fermions. These relativistic fermions can have near zero-mass and high mobility. The spin-polarized relativistic fermions also can exhibit some unique behaviors like topology protected conducting state, no backscattering, magnetic monopole, fractional charge carrier, and special Fermi surface.[21-26] These protected properties cannot be destroyed by the external perturbations, surface defects, and impurities, as long as the topology of the Brillouin zone is protected.[9,21-26] Many interesting transport properties and applications were reported associated with these massless relativistic fermions like chiral anomaly transport, non-local transport, Shubnikov-de Haas (SdH) oscillation, negative magnetoresistance, and Fermi-arc transport.[21-30] Here, both TRS and SOC are the critical leads in the search for new topological materials, while tuning of TRS and SOC plays a vital role in describing the topological phase conversion. Topological phases and their conversion techniques open the door for making extraordinary devices like near zero dissipationless transistors,[31-34] spin-filter transistors,[35-39] super-lens,[40-43] thermoelectric devices,[44-47] and optoelectronic devices.[48-51] For successful device operation, it is essential to have a physical system with a controlled tuning mechanism to realize the topological phase conversions. Despite the extensive ongoing research on the fundamentals of the topological state of matter and their phase conversions, fewer works addressed the physical device structures to realize the phase conversion.

Multi-disciplinary teamwork in topological research is a current gap in the scientific community. The review is written with that in mind to provide a comprehensive article with easy-to-understand illustrations that can help to close the gap among different disciplines and elaborate on the prospect of topological phenomena in real-life applications. The review will cover the prospects of the topological matter from the fundamental physics, materials aspect, and device applications.

We start the article with a concise but complete discussion on topological physics in a less mathematical tone. We will elaborate on the essential topological terms and then discuss the material systems which have already shown the topological characters. A comprehensive list of the previously discovered topological materials and their properties is presented. This can provide the necessary information for the material science community in one place to focus on both the critical material aspects and their further development to meet the application needs. We will then extend the article for device applications, which is less covered in previous reviews. We will elaborate on the conversion/tuning mechanism to obtain and control the topological phases, the existing challenges, the requirements to observe, amplify, and control the topological properties from the perspective of device applications.

### 1.2. Outline of the Review

This review aims to provide a complete and fundamental picture of the topological materials from fundamental theories to device applications, which will be helpful to a wider scientific community, including condensed matter, materials science, and electrical engineering. The topological/quantum phase of matter is a rapidly growing field where the existing theories like topological band theories are still developing. New theories like non-abelian band topology[52,53,54,55] and topological quantum chemistry[56] are expanding by multidisciplinary research and studies. It is nearly impossible to include all aspects of the topological field into a single review article with limited space; hence, we will keep our focus on the fundamental aspects of the predominant topological phases. To fulfill the objectives of this article, we discuss the central aspects of the topological/quantum materials: underlying physics, different quantum/topological states of matter, the conversion techniques, and device applications. The state-of-the-art quantum/topological phases of matter are reviewed in section 2, along with their fundamental properties. Section 3 highlights the key parameters to define the concept of topological invariants with topological spaces, homeomorphisms, and topological invariants. Section 4 defines the fundamentals of Berry phases, including the parallel transport, Berry connection, and Berry curvature. Based on the fundamentals of topology described in earlier sections, we will then introduce the fundamentals of symmetries in topological materials and the topological classifications based on the corresponding symmetry classes. Section 6 describes the key parameters like SOC, band inversion, and topological invariant to observe the topological phases. Section 7 summarizes different Hall characteristics of associated topological phases, which is one of the important indicators of topological phases. In section 8, we discuss the topology protected properties of different topological phases. Section 9 summarizes the existing topological materials according to their topological phases, dimensions, and symmetry classes. Finally, section 10 reviews the existing techniques for topological phase conversions, including the fundamentals, state-of-the-art examples, and potential device applications.

2. **Quantum/Topological Phases of Matter**

Quantum or topological phases are defined by the breaking of the global symmetries instead of the local symmetries, which are common in conventional phases of matter like solid, liquid, gas, ferromagnetic order, antiferromagnetic order, etc.[1-9] Among the common quantum phases of matter, superconductors,[57,58] quantum Hall phases,[59,60] and superfluid[61,62] have received considerable attention among the scientific community. The global symmetry in these quantum materials is insensitive to the geometry and the external perturbations, which cannot change the topology of the system.[9,21-26] This topology is defined by the Berry phase of the gapped ground state of the quantum Hamiltonian of the system.[1-9] Topological order is a global pattern where every particle (or spin) is moving around each other in a very organized way by following some rules, a) all moving particles/spins will create a local pattern to lower the energy of the local Hamiltonian, b) all moving particles/spins will form a global pattern by following the local pattern which corresponds to the topological order, and c) the global pattern of the collective particles/spins is a pattern of quantum fluctuation of random particles/spins that defines by the long-range entanglements.[1-9,63] According to the last condition, it can be stated that topological order is the notion that describes the long-range entangled states (long range entanglements = topological order). It is worth noting that the short-range entangled states are trivial in the sense that they all belong to one phase. However, in the presence of symmetry, even short-range entangled states are nontrivial and can belong to different phases (see Sec. 7 for more details). Those phases are said to contain symmetry-protected topological order. Hence, quantum orders that contain long-range entanglement patterns are called topological orders (such as fractional quantum Hall effect), and quantum orders that have short-range entanglement patterns (such as topological insulator, Dirac/Weyl, and nodal line semimetal, etc.), in which symmetry plays an important role, are symmetry-protected topological orders.[1-9,63] The first observed topology protected quantum phase is the quantum Hall insulator. Other observed quantum phases are topological insulators, and topological semimetals, superconductors, super-fluids, etc.[57-62,64] Topological ordering offers some new unique properties. For example, it can create new kinds of quasiparticles (known as anyons) from the collective excitations of particles/spins in the topologically ordered ground states. These new topological quasiparticles can carry fractional charges by obeying fractional statistics such as non-Abelian statistics.[1-9] Some topological orders can also support topologically protected gapless boundary excitations.[1-9]

Among the topological phases, quantum Hall insulator (QHI) phases are explained by disorder-controlled spin-scattering with electron-electron interaction in the presence of an external magnetic field. The external field creates different landau levels in the 2D electron gas (2DEG) QHI system.[65-67] Two kinds of QHI phases are observed: integer quantum Hall insulator (IQHI) and fractional quantum Hall insulator (FQH). QHI states are the gapped ground states of 2DEG under an adiabatically changing strong magnetic field.[2,24,25] QHIs are considered as quantum liquids where strong quantum fluctuations of electrons prevent the formation of the crystal due to their light mass.[2] Quantum Hall liquid is incompressible with well-defined density.[2,24,25] In QHIs, current density creates a transverse Hall electric field with a quantized Hall coefficient, which is precisely quantized as an integer or rational number (also known as filling factor) in the unit of quantum resistance ($h/e^2$).[2,24,25,68] Topological order in QHIs helps to create metallic chiral edge states from the skipping orbits via which relativistic fermions can flow without experiencing any backscattering.[1-12] Different QHIs states have the same local symmetry and hence cannot be defined by symmetry breaking. Instead, the presence of an external magnetic field introduces the breaking of TRS. Integer filling factor in IQHI states can be explained by Landau levels in the presence of a strong magnetic field and disorder of the system, while filling factor in FQHI states are described by the Laughlin levels arising from the electron-electron interaction.[1-12] FQHI states form new quasiparticles due to the formation of composite fermions that can carry fractional charges.[1-12,69]

One of the significant challenges of QHIs in device applications is the need for a strong magnetic field. Later, it was found that spin-orbit coupling (SOC) can replace the need for an external magnetic field, which helps to observe the other remarkable topological order phases (or symmetry-protected topological order phases): topological insulators and topological semimetals.[1-12] Topological phases get more attraction due to their prospective for different device applications. Topological insulator and semimetal phases are defined with the Bloch levels rather than the Landau levels along with some fascinating physical characteristics,[70] which are discussed in the following sections. *Figure 1* illustrates different quantum or topological families along with their spin-polarized edge or surface states, spin-momentum locking behavior, and band structures.

In topological insulators, the topology-protected helical and chiral surface conducting states exist along with the bulk insulating counterparts; however, the surface conduction has high mobility due to the absence of backscattering.[9,11] Here, it is essential to note that the helicity is defined by the direction of the spin with respect to the direction of the motion. However, the chirality is set by the mirror-difference between the spins. Helicity and chirality can influence parity symmetry (PS), time-reversal symmetry (TRS), and inversion symmetry, as shown in the legend of *Figure 1*. Both TRS and inversion symmetry along with rotational symmetry, can create the Dirac node into a system having the band inversion.[23-26] The Dirac nodes are typically formed at the points on the rotation axis of a crystal, where electrons act as relativistic fermions and show chiral and helical nature considering their spin texture to momentum.[23-

[26] Based on the bulk bandgap, topological materials can be categorized into topological insulators and topological semimetals. By breaking either time-reversal or inversion symmetry, degeneracy at Dirac nodes is lifted, which introduces a new topological phase known as Weyl semimetals (WSM).[23-26,71] Dirac semimetals (DSMs) form a special spin-momentum locked texture in the Fermi surface while Weyl semimetals form Fermi arcs between the chiral Weyl fermions.[9,26,72] To realize and utilize such unique quantum or topological properties of matters in physical applications, it is essential to have a device that can allow controlling the topological nature of the system. One important application area of topological materials is in electronics, where topological properties can make devices like near-dissipationless low-power transistors, spin-filter transistors, quantum gates for quantum computing,[73-75] etc. Topological transistors with a reliable and controllable switching mechanism can make the building-block of a new generation of electronic systems.

| Phases | System | Sym. | Edge/Surface | Band | FS | DOS |
|---|---|---|---|---|---|---|
| Integer Quantum Hall Insulator (1980) | Heterostructure with 2DEG and Disorder | TRS ✗ | Chiral Edge State / 2D bulk State | E, $E_F$, k | • | Extended State, Disorder State |
| Fractional Quantum Hall Insulator (1982) | Clean Heterostructure With 2DEG | TRS ✗ | Chiral Edge State / 2D bulk State | E, $E_F$, k | • | $E_F$ |
| Quantum Spin Hall Insulator (2005) | Clean Heterostructure with band inversion | TRS ✓ | Chiral Helical Edge State / 2D bulk State | E, $E_F$, k | • | $E_F$ |
| 2D Dirac Semimetal (2005) | 2D Materials (Graphene) With Intrinsic SOC | DCS | | E, $E_F$, k | Fermi Surface | $E_F$ |
| 3D Topological Insulator (2007) | 3D Insulator Materials | TRS ✓ | | E, $E_F$, k | ○ | $E_F$ |
| Topological Magnetic Insulator (2010) | Topological Insulators with magnetic doping | TRS ✗ | | E, $E_F$, k | | $E_F$ |
| Topological Kondo Insulator (2010) | Strong Correlated System with Heavy Fermions | TRS ✓ | | E, $E_F$, Kondo gap, df hybrid, k | | Kondo State |
| Topological Superconducting Insulator (2010) | Topological Insulators with Superconducting doping | TRS ✓ | Superconducting pair | E, $E_F$, k | ○ | Majorana Bound State |
| Topological Crystalline Insulator (2010) | Bulk Insulators | DCS | | E, $E_F$, k | • ○ | $E_F$ |
| Dirac Semimetals (2012) | Bulk Semimetals | II+TRS + DCS | | E, $E_F$, k | | $E_F$ |

2D ↕ 3D

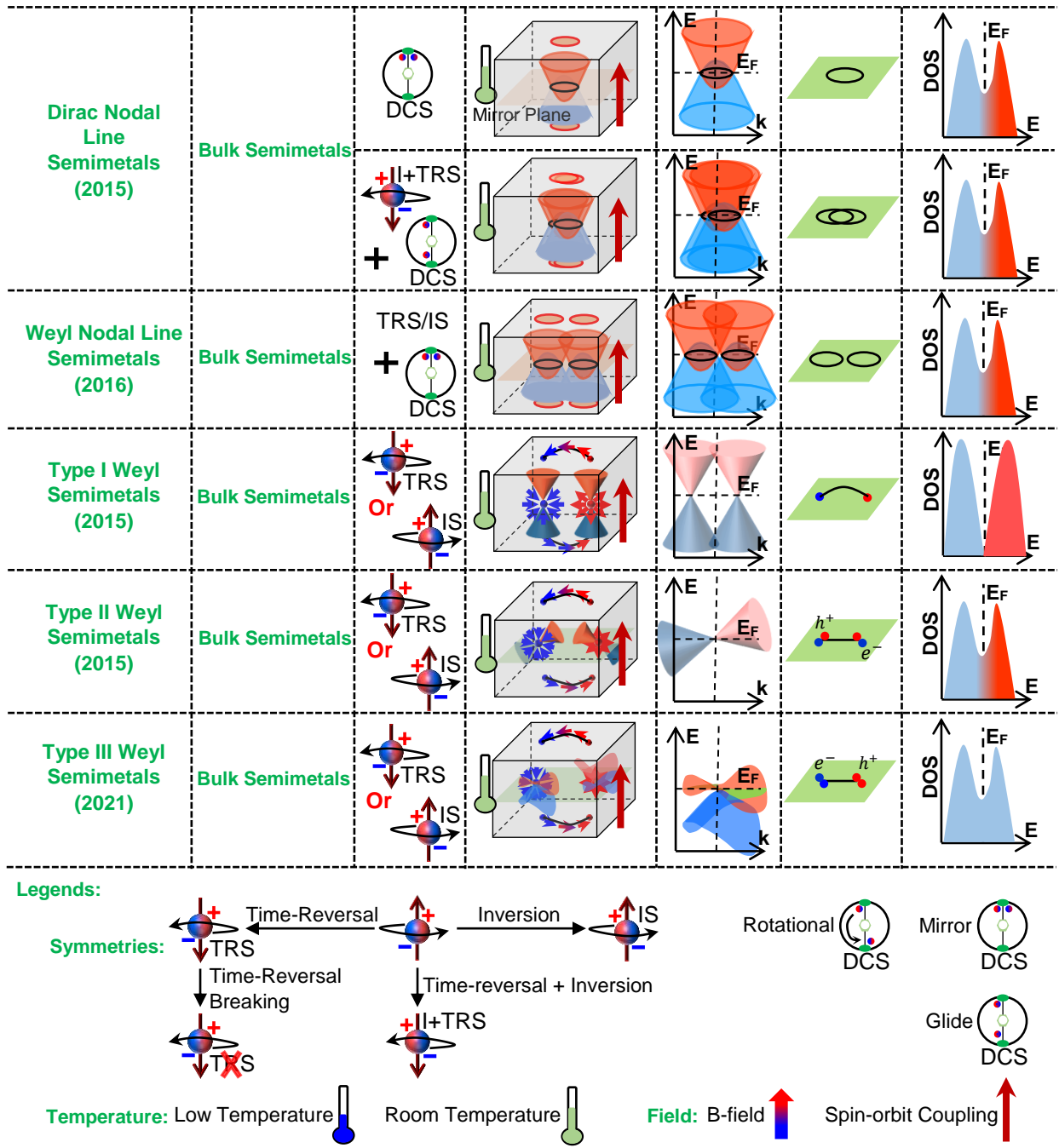

Figure 1: Tabular illustration of different 2D and 3D quantum/topological families along with the corresponding material systems, symmetries that help to create the phase, topological edge or surface states in real and momentum space, band structures, Fermi surfaces (FS), and density of states (DOS). All topological phases are also categorized according to topological insulators (black text) and semimetals (green text). In Legends, TRS, IS, I+TRS, and DCS represents time-reversal, inversion, inversion + time-reversal, and discrete crystal symmetry, respectively.

In some crystalline systems, nontrivial topological phases (known as topological crystalline insulators) can be originated from the discrete type of space group symmetry instead of TRS.[76] Discrete crystal symmetry (DCS) includes rotational, mirror, glide, and rotoinversion symmetries. Topological crystalline insulators (TCIs) can exhibit several unique features like nonlinear surface band crossing, topological phases without SOC, and topological superconductivity protected by crystal symmetry.[76] Among different crystal symmetries, four-fold or six-fold rotational symmetry and mirror symmetry show the possibilities of providing TCIs theoretically, which were later experimentally observed by spin-resolved angle-resolved photoemission spectroscopy.[76] Topological phases can also appear in Kondo insulator materials with strong electron correlations having heavy fermions (*f*-electrons) behavior, which means Kondo insulators behave like a correlated metal at high temperatures and like an insulator at low temperatures due to the hybridization between localized *f*-bands and dispersive *d*-bands.[77,78] Topological Kondo insulators appear at low temperatures where topological surface states exist with the bulk Kondo gap and show anomalous residual conductivity.[79-81]

Magnetic topological insulators are another type of quantum phases obtained in magnetically doped TIs where a magnetic gap is opened at the Dirac point of the surface states, and magnetically-driven spin texture is observed near the gap edge.[9] Due to the TRS breaking by magnetic doping, magnetic TIs show some novel effects like the quantum anomalous Hall effect, axion electrodynamics, and topological magneto-electrical effect.[9,82,83] Magnetic energy gap opened at Dirac point is challenging to observe experimentally due to spatial energy-momentum fluctuation near the Dirac point[84] and surface chemical modifications.[85] Similar to the 3D topological insulator phases, magnetic TIs have a quantized Berry phase that is defined by the Hedgehog-like spin texture of the surface states.[86] Superconducting TIs show the superconducting gap in bulk with topological surface states at the boundaries having Majorana fermions, which are their antiparticles.[87,88,89,90] Superconducting gap is an energy gap that opens when electrons at the Fermi level attract each other, forming Cooper pairs and condense. Superconducting TIs demonstrate several novel quantum effects like a time-reversal invariant topological superconductor, supersymmetry physics, and fault-tolerant quantum computation.[91] Superconducting TIs can be realized in two ways, (i) bulk doping into a TI material and (ii) interfacing a TI with a superconductor to get a superconducting proximity effect.[9] Helical-Cooper pairing, the superconducting Bose condensation of spin-momentum locked Dirac electron gas, is the key feature to demonstrate superconducting TIs, which was shown by spin and momentum-resolved photoemission spectroscopy.[92] An odd number of Helical-Cooper pairing channels can guarantee the presence of Majorana bound states in superconducting TIs.[93,94] All the shown topological phases in *Figure 1* can be transitioned to some other phases by using some external stimuli. However, before driving into more details on topological phase conversions, some fundamental factors of topological materials need to be discussed, which will also be essential for understanding the concepts of topological phase conversion. Therefore, the following sections are dedicated to discussing those critical factors in the quantum or topological phases of matter.

3. **The concept of topological invariant**

Topological invariance is an important parameter to define quantum phases of matters which does not change within the phase. This quantity determines the equivalence between two quantum states. A continuous deformation is possible between two topologically invariant quantum states of matter without changing the phase. The topological invariant is a mathematical quantity defined with respect to the dimension of the matter. For any dimensional physical system, topological invariance can be calculated from the Berry connection and Berry curvature, which has a relation with Berry phase hence with the Brillouin zone. The surface of the Brillouin zone represents the topology or the conserved symmetry of the quantum phase of a condensed matter. In two dimensional topological systems (such as Chern insulator), $Z$ or the Chern number is the topological invariance, while $Z_2$ represents the same in the three-dimensional topological systems (such as QSHI and 3D topological insulator). To understand the meaning of topological invariant, one must define the topological spaces, homeomorphism, and the equivalence relation. Furthermore, the mathematical definition of topology and topological invariant will give us more generalized notions of the topological invariant and the topological phase in condensed matter physics.

### 3.1. Topological Spaces

Topology is used to study abstract spaces. It introduces the unchanged mathematical structures under continuous deformation. It is often called *rubber sheet geometry* and focuses on the unchanged features when the parameters of the system change continuously.[95,96] In mathematics, a topological space $(X; \tau)$ is a set $X$ equipped with a collection $\tau$ of subsets of $X$ satisfying the following:[95]

1. The empty set $\emptyset$ and the space $X$ are sets in the topology
2. The union of any collection of sets in $\tau$ is contained in $\tau$
3. The intersection of any finitely many sets in $\tau$ is also contained in $\tau$

Hence, a topological space is a pair $(X; \tau)$ where $X$ is a set, and $\tau$ is a set of subsets of $X$ satisfying certain axioms. $\tau$ is called a topology. For example, consider the following set consisting of 3 points; $X = \{a, b, c\}$ and determine if the set $\tau = \{\emptyset, X, \{a\}, \{b\}\}$ satisfies the requirements for a topology. Definitely, this is not a topology because the union of the two sets $\{a\}$ and $\{b\}$ is the set $\{a, b\}$, which is not in the set $\tau$.

According to this definition, we can find some examples in condensed matter with ordered states, like crystals, liquid crystals, magnets, etc. These phases have a symmetry described by a group $S$, which is a subgroup of the expected full symmetry $G$. For example, if a similar Hamiltonian is considered for two-electron gas and crystals systems, the Hamiltonian of $N$-interacting particles is expected to be invariant under rotational and full translational symmetry, $G$, but a crystal system has only a discrete set of space group symmetries. This phenomenon of ordering is known as symmetry breaking. The crystal system is described by parameters that reflect this subgroup structure. The allowed values of the order parameter constitute the topological space of the ordered system, and this space is called the order parameter space. Another famous example where the topological space can be constituted by the order parameter is the quantum Hall effect, and the order parameter can be the filling factor ($\nu$), which will be explained further in the next sections.

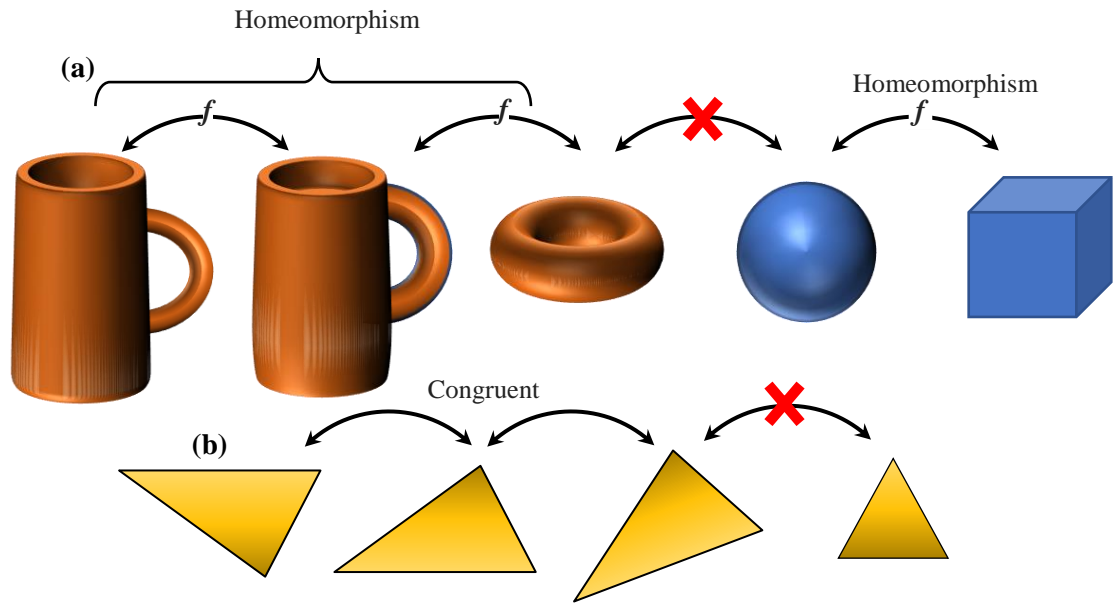

Figure 2: (a) The homeomorphism between a coffee cup and a doughnut. In contrast, a doughnut is not a homeomorph of a sphere, while a sphere and a cube are. (b) The concept of congruent among triangles in Euclidean Space.

### 3.2. Homeomorphisms

Homeomorphism is the notion of equality in topology where there are an equivalence relation and one-to-one correspondence between the points in two topological spaces. For example, a topologist cannot distinguish between a coffee cup and a doughnut (*Figure 2*(a)). This is because one object can be obtained by stretching and bending continuously from the other. The formal definition of homeomorphism in mathematics is as follows: a homeomorphism between two topological spaces X and Y is a function, $f: X \rightarrow Y$, and therefore, it is a continuous bijection and has a continuous inverse function $f^{-1}$. In other definitions, two topological spaces *X* and *Y* are said to be homeomorphic if there is continuous map $f: X \rightarrow Y$ and $g: Y \rightarrow X$ such that $fog: X \rightarrow X$ and $gof: Y \rightarrow Y$. Moreover, the maps *f* and *g* are homeomorphisms and are inverses of each other so that we can write $f^{-1}$ and $g^{-1}$ in place of *g* and *f*, respectively. This definition indicates that the homeomorphism forms an *equivalence relation* between all class of topological spaces, i.e., i) *Reflexivity*: *X* is homeomorphic to *X*, ii) *Symmetry*: If *X* is homeomorphic to *Y*, then *Y* is homeomorphic to *X*, and iii) *Transitivity*: If *X* is homeomorphic to *Y*, and *Y* is homeomorphic to *Z*, then *X* is homeomorphic to *Z*.

In Euclidean space, often equivalence is defined as congruent. For example, $A_1$ and $A_2$ sets are congruent if $A_2$ can be obtained by translating and rotating $A_1$. For congruent, we can define the equivalence relation and, therefore, the equivalence classes (for example, the equivalence class of triangles shown in *Figure 2*(b)). In the Euclidean space (or topological space), we can form an equivalence class using the equivalence relation and map, which can be separated by special property contained in these equivalence classes. This property is called geometric topological invariance.

### 3.3. Topological Invariants

A topological invariant is a property of topological spaces which is preserved under homeomorphisms. Each homeomorphism has an equivalence class. Therefore, in the equivalence class of triangles of Euclidean space, angles are geometrical invariants. Also, the hole of the doughnut and a coffee cup is considered to be a topological invariant. Therefore, instead of obtaining equivalence classes for homeomorphisms, we can get these topological invariants, which indicate whether the two topological spaces are equivalent or not. Before topological invariants are expressed in condensed-matter physics, it is helpful to get a feel for the topological invariants of one-dimensional objects in $\boldsymbol{R}^2$. The simplest topological invariant is the number and type of vertices in an object. In this case, a vertex is considered as a point where multiple curves intersect or join together. The number of intersect curves determines the vertex type. In fact, the number of *n*-vertices ($n \geq 3$) in the object is the topological invariants that we have identified. We can say that the number of 3-vertices or 4-vertices is a topological invariant because homeomorphisms preserve connectedness (*Figure 3*). Therefore, the connected set around a vertex must map to another connected set, and the set of *n* disjoint, connected pieces must map to another set of *n* disjoint connected pieces. In the study of topological materials, in each spatial dimension, symmetries are expressed, which by using these symmetries, the topological space can be classified by topological invariants. It should be noted, although the classification method is as in the previous example, the detection and calculation of invariants are more complex. In the following sections, we discuss in more detail the calculation of the topological invariance for different topological phases.

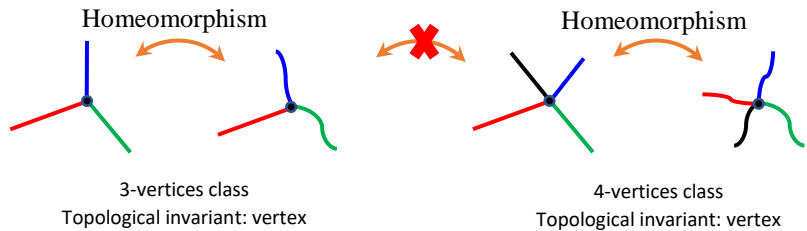

Figure 3: 3-vertex and 4-vertex homeomorphism examples. The number of 3-vertices or 4-vertices is a topological invariant because homeomorphisms preserve connectedness. Thus, the connected set around a vertex must map to another connected set, and the set of *n* disjoint, connected pieces must map to another set of *n* disjoint connected pieces.

4. **The concept of the Berry phase**

In 1983, "Berry" showed that a quantum system adiabatically transported around a closed contour *C* in the space of parameters acquires a non-integrable phase (Berry phase) besides the familiar dynamical phase depending only on the geometry of the contour *C*.[97] This phase provides deep insight on the geometric structure of the quantum mechanical systems and gives rise to various observable effects.[98-100] Berry phase has an important role in the concepts of the topological phases. Therefore, to describe the Berry phase, we first start with a simple example of the concept of parallel transport and then generalize it to quantum mechanics.

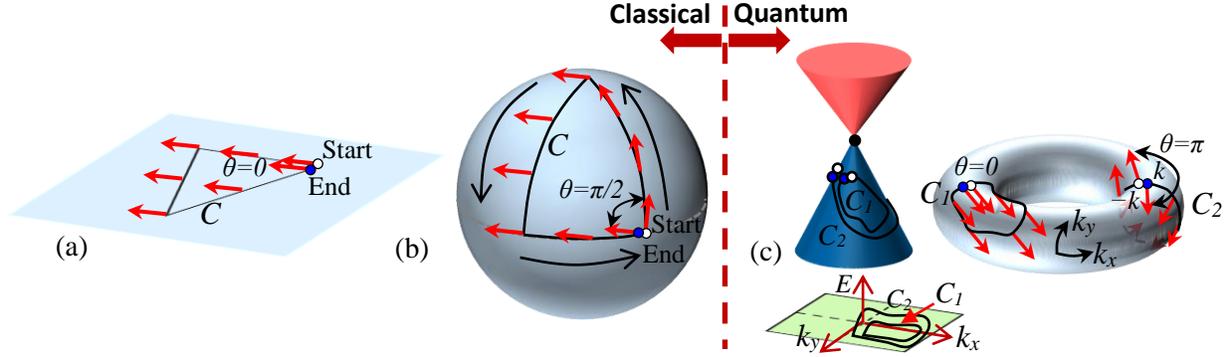

Figure 4: Classical illustration of parallel transport on (a) a plane and (b) a sphere and the anholonomy angles. Parallel transport of a vector refers to circulating a vector quantity along the contour *C* without any local change. (c) Quantum illustration of Berry phase from a parallel transport of electron in k-space. If the contour covers both +k and -k space, the electron will experience a π phase change after circulating the contour.

### 4.1. Parallel transport

The importance of the Berry phase comes from the fact that it unfolds the intimate geometrical structure of the underlying quantum mechanics. It is, therefore, appropriate to begin by introducing the basic concept of parallel transport in a purely geometrical context. Suppose we have a surface that can be a plane, a sphere, or a cone. Consider a tangent vector to this surface (*Figure 4*), we wish to transport the vector on the surface without rotating it around the axis normal to the surface, and the arrow is transported around a closed circuit *C*. There may be two different situations in this process: i) if the surface is flat, then the arrow always remains parallel to its original orientation, and the original orientation of the arrow isn't changed after the completion of the circuit *C* (see *Figure 4*(a)). ii) if the surface is curved, the arrow being constrained to lie in the local tangent plane cannot remain parallel to its original orientation. Therefore, after completion of the circuit *C*, it is seen to have been rotated by an angle $\theta(C)$. The angle of difference $\theta(C)$ is called the *anholonomy* angle. This concept is a simple expression of parallel transport in geometry. As shown in *Figure 4*(b), the arrow can be represented by a tangent unit vector $v_1$ and transported along a circuit $C = \{r(t): 0 \to T\}$ on the surface. *r* is a vector that after time *T* sweeps through closed path *C*. By defining $n(r)$ as the unit vector normal to the surface at point *r* and a second tangent unit vector $v_2 \equiv n \times v_1$, which is also parallel transported on the surface along *C*, we form three-unit vectors ($n$; $v_1$; $v_2$). As a rule of parallel transport, when $v_1$ and $v_2$ are moved, they have to rotate with an angular velocity ω. The equation of motion of $v_1$ and $v_2$ (not for the flat surface) is $\dot{e}_i = \omega \times e_i$ ($i = 1;2$) (where the over-dot indicates the time derivative). One can easily find that in order to fulfill the requirements of parallel transport, $v_1$ and $v_2$ remain tangent unit vectors and never rotate around *n* (i.e. $e_i \cdot n = 0$ and $\omega \cdot n = 0$). Therefore, the angular velocity is given by, $\omega = n \times \dot{n}$. Therefore, the parallel transport law is introduced as follows:[97,99]

$$\dot{\mathbf{e}}_i = (\mathbf{n} \times \dot{\mathbf{n}}) \times \mathbf{e}_i = -(\mathbf{e}_i \cdot \dot{\mathbf{n}})\mathbf{n} \qquad (1)$$

Eq. (1) can be expressed in a more appropriate forum for generalization to the case of quantum mechanics by defining a complex unit vector:

$$\psi = \tfrac{1}{\sqrt{2}}(\mathbf{v}_1 + i\mathbf{v}_2); \quad \psi^* \cdot \psi = 1 \qquad (2)$$

Which can be rewritten as the law of parallel transport by taking the complex conjugate:

$$\text{Im}(\psi^* \cdot \dot{\psi}) = 0 \text{ or } i\psi^* \cdot \dot{\psi} = 0 \tag{3}$$

The real part of $\psi^* \cdot \dot{\psi}$ is always zero because $v_1.v_1$ and $v_2.v_2$ are time independent. The $v_1$ and $v_2$ vectors are not dependent on the location vector $r(t)$; also, the closed contour $C$ on which they move is determined by the location vector $r(t)$. Therefore, we need vectors that depend on the location vector $r(t)$, i.e., a fixed local orthonormal frame $(n(r); t_1(r); t_2(r))$ on the surface; where $n$ is a normal unit vector perpendicular to the surface. We have an infinity of possible choices for $t_1(r)$, which corresponds to gauge freedom. Therefore, when we have selected $t_1(r)$, then $t_2(r)$ is, of course, uniquely determined. As before, define a complex vector as:

$$\mathbf{u}(\mathbf{r}) \equiv \frac{1}{\sqrt{2}}(\mathbf{t_1}(\mathbf{r}) + i\mathbf{t_2}(\mathbf{r}));$$
$$\mathbf{u}^*(\mathbf{r}) \cdot \mathbf{u}(\mathbf{r}) = 1. \tag{4}$$

Now, $\Theta(t)$ is the angle by which $t_1$ and $t_2$ must be rotated to coincide with $v_1$ and $v_2$. The relation between $(t_1; t_2)$ and $(v_1; v_2)$ can be written by the following phase factor:

$$\phi(t) = \mathbf{u}(\mathbf{r}(t))\exp[-i\Theta(t)] ; \tag{5}$$

Now using the parallel transport law (Eq. (3)) and Eq. (4):

$$0 = \text{Im}(\psi^* \cdot \dot{\psi}) = \text{Im}(\exp[-i\Theta(t)]\{-i\dot{\Theta}\mathbf{u}^* \cdot \mathbf{u} + \mathbf{u}^* \cdot \dot{\mathbf{u}}\}) \Rightarrow \dot{\Theta} = \text{Im}(\mathbf{u}^* \cdot \dot{\mathbf{u}}) \tag{6}$$

So that

$$\Theta(C) = \text{Im} \oint_C \mathbf{u}^* \cdot d\mathbf{u} \tag{7}$$

A coordinate system $(X_1; X_2)$ is chosen on the surface, and the vector field $A(r)$ (called a connection) on $\Sigma$ is defined as:

$$A_i(\mathbf{X}) \equiv \text{Im}\left[u_j^*(\mathbf{X})\frac{\partial u_j(\mathbf{X})}{\partial X}\right] \tag{8}$$

Given this relation, the first form of *anholonomy* angle $\theta(C)$ can be written as:[97]

$$\Theta(C) = \oint_C A(\mathbf{X}) \cdot d\mathbf{X} \tag{9}$$

Using the Stokes theorem, another relation is obtained:

$$\Theta(C) = \iint_S B(\mathbf{X})dX_1 dX_2 \tag{10}$$

where $B(\mathbf{X})$ is called the Berry curvature. Similar to the parallel transport in geometry, the phase of a quantum state may not return to its original value after a cyclic evolution in parameter space. In the next section, we introduce and discuss the basic concept of the Berry phase.

### 4.2. Berry phase, Berry connection, and Berry curvature

Consider a quantum mechanical system described by a Hamiltonian $H(R)$. This Hamiltonian is controlled by a set of external parameters such as magnetic, electric fields, SOC, etc., and $R = (R_1, R_2, \ldots )$ is described with some collective vectors in parameter space. For each $\mathbf{R}$, the $H(\mathbf{R})$ has eigenvalues $E_n(\mathbf{R})$ and eigenstates $|n(\mathbf{R})\rangle$ satisfying the independent-Schrodinger equation,

$$H(\mathbf{R})|n(\mathbf{R})\rangle = E_n(\mathbf{R})|n(\mathbf{R})\rangle \tag{11}$$

In fact, the eigenstate $|n(\mathbf{R}(t))\rangle$ of the $H(\mathbf{R}(\text{t}))$ varies along with the slow variation of $\mathbf{R}(\text{t})$ in the parameter space.

It should also be noted that we assume a gap existing between the eigenstates and all other states; therefore, the adiabatic flow of $\mathbf{R}(\text{t})$ introduces weak perturbation to prevent the transition to other states. In more detail, let us perform an adiabatic closed-circuit $C \equiv \{\mathbf{R}(t)|t = 0 \to T\}$ in the parameter space. The adiabatic theorem[101] tells us that if the rate of variation of the external parameters is low enough, a system that is initially in the $n^{\text{th}}$ stationary state $|n\rangle$ of the Hamiltonian will remain continuously in the state $|n\rangle$. Therefore, we assume an adiabatic variation where there is no transition to any other state. Now the question arises, what will be *the phase* of the state after completion of the circuit *C*?

Note that the expectation value of any observable quantity $B = \langle\psi|B|\psi\rangle$ does not depend on the phase of $|\psi\rangle$. This has caused this question to be of no interest. Indeed, this lack of interest is certainly the main reason why the Berry phase has been completely overlooked for more than half a century of quantum mechanics. Barry had a clever look at this question and came up with a remarkable answer that has become widely used today in atomic and molecular physics, nuclear physics, classical optics, photonics, and condensed-matter physics[102-106]. Therefore, following Berry, taking $|\Psi(t = 0)\rangle \equiv |n(\mathbf{R}(t = 0))\rangle$ at *t*=0, we can write the state $|\psi(t)\rangle$ at a later time *t*:

$$|\Psi(t)\rangle \equiv e^{-\frac{i}{\hbar}\int_0^t dt'E_n(\mathbf{r}(t'))}|\Phi_n(t)\rangle \tag{12}$$

where $|\Phi_n(t)\rangle$ (in t=0) is an auxiliary wavefunction with a zero dynamical phase. By applying the time-dependent Schrödinger equation and projecting it on $\langle\Phi_n(t)|$, we will have:

$$\begin{aligned}
0 &= \langle\Psi(t)|\left(H(t) - i\hbar\frac{\partial}{\partial t}\right)|\Psi(t)\rangle \\
&= \langle\Phi_n(t)|e^{\frac{i}{\hbar}\int_0^t dt'E_n(\mathbf{r}(t'))}\left(H(t) - i\hbar\frac{\partial}{\partial t}\right)e^{-\frac{i}{\hbar}\int_0^t dt'E_n(\mathbf{r}(t'))}|\Phi_n(t)\rangle \\
&= i\langle\Phi_n(t)|\dot{\Phi}_n(t)\rangle
\end{aligned} \tag{13}$$

where we used the relation, $\langle\Psi(t)|H(t)|\Psi(t)\rangle = E_n(t)$ which follows from the adiabatic theorem. Comparison between Eqs. (6) and (13) shows that, $|\Phi_n(t)\rangle$ obeys a quantum mechanical analog to the parallel transport law. In analogy with the problem of parallel transport, there are three-unit vectors ($\mathbf{n}$; $\mathbf{v}_1$; $\mathbf{v}_2$) on a surface. We now express the parallel transported state $|\Phi_n(t)\rangle$ in terms of the fixed eigenstates $|n(\mathbf{R}(t))\rangle$; therefore,

$$|\Phi_n(t)\rangle = e^{i\gamma_n(t)}|n(\mathbf{R}(t))\rangle \tag{14}$$

where the phase $\gamma_n(t)$ plays the same role as the angle $-i\Theta(t)$ in the Eq. (6) for parallel transport on a surface. Hence, we can write the equation of motion of $\gamma_n(t)$,

$$\dot{\gamma}_n(t) = i\langle n(\mathbf{R}(t))|\dot{n}(\mathbf{R}(t))\rangle$$
$$= -\text{Im}\langle n(\mathbf{R}(t))|\tfrac{d}{dt}n(\mathbf{R}(t))\rangle \quad (15)$$

which is similar to Eq. (7). Now to answer the question at the beginning of this section:

$$C \equiv \{\mathbf{R}(t)|t = 0 \to T\}: \quad |\Psi(T)\rangle = e^{i[\delta_n+\gamma_n(C)]}|\Psi(0)\rangle \quad (16)$$

where $C$ is a closed path in the **R**-space, $\delta_n$ is the *dynamical* phase, and is defined as follows:

$$\delta_n \equiv -\tfrac{1}{\hbar}\int_0^T E_n(\mathbf{R}(t))\,dt \quad (17)$$

Finally, the *Berry* phase is defined as follows,

$$\gamma_n(C) \equiv -\text{Im}\left[\oint_C \langle n(\mathbf{R}(t))|\partial_\mathbf{R}|n(\mathbf{R}(t))\rangle \cdot d\mathbf{R}\right]$$
$$= i\oint_C \langle n(\mathbf{R}(t))|\partial_\mathbf{R}|n(\mathbf{R}(t))\rangle \cdot d\mathbf{R} \quad (18)$$

It is noteworthy that this result obtained for the quantum system is similar to the classical system (an illustration is shown in *Figure 4*). The Hannay angle[107] in classical mechanics is analogous to the Berry phase in the quantum system. The integrand $-\text{Im}[\langle n(\mathbf{R}(t))|\partial_\mathbf{R}|n(\mathbf{R}(t))\rangle]$ or $i[\langle n(\mathbf{R}(t))|\partial_\mathbf{R}|n(\mathbf{R}(t))\rangle]$ is often called the *Berry connection*, $\mathbf{A}(\mathbf{R})$:

$$\gamma_n(C) \equiv i\oint_C \mathbf{A}(\mathbf{R}) \cdot d\mathbf{R} \quad (19)$$

If the **R**-space is two-dimensional, one can use Stokes' theorem to transform the line integral to a surface integral,

$$\gamma_n(C) \equiv i\int_S \left\langle \tfrac{\partial}{\partial \mathbf{R}}n(\mathbf{R}(t))\middle|\tfrac{\partial}{\partial \mathbf{R}}n(\mathbf{R}(t))\right\rangle \cdot d^2\mathbf{R} = i\int_S \nabla \times \mathbf{A}(\mathbf{R}) \cdot d^2\mathbf{R} = i\int_S \mathbf{F}(\mathbf{R}) \cdot d^2\mathbf{R} \quad (20)$$

The integrand $\mathbf{F}(\mathbf{R}) \equiv \nabla_\mathbf{R} \times \mathbf{A}(\mathbf{R})$ is usually called the *Berry curvature*. For **R**-spaces with higher dimensions, it can be generalized, and such a transformation can still be done using the language of the differential form.[99]

Important features of the Berry curvature and Berry connection are their examination under gauge transformations. This feature plays a subtle role in topological invariant's calculations. Therefore, if we use a new choice for the phase of the reference state, i.e., $|n(\mathbf{R}(t))\rangle' = e^{-i\xi(\mathbf{R})}|n(\mathbf{R}(t))\rangle$, where $\xi(\mathbf{R})$ is a single-valued function, the Berry curvature and Berry connection change as follows:

$$A'(\mathbf{R}) \to A(\mathbf{R}) + \nabla\xi_n(\mathbf{R})$$
$$F'(\mathbf{R}) \to \nabla_\mathbf{R} \times (A(\mathbf{R}) + \nabla\xi_n(\mathbf{R})) = F(\mathbf{R}) \quad (21)$$

This is similar to the gauge transformation in electromagnetism, i.e., one can choose different gauges for the potentials ($A(r)$) without changing the field ($B(r)$). As can be seen in Eq. (21), connection $A(R)$ is not gauge invariant. However, the Berry curvature and thus Berry phase $\gamma_n(C)$ are gauge invariant. Finally, the analogy between geometric anholonomy (Anholonomy angle) and quantum anholonomy (Berry phase), and the similarity between the theory of Berry phase and electromagnetic is given in *Table 1*. In general, Berry phase can be defined as the global change of phase angle of a gapped ground state that appears under adiabatically varying external field without any local change, and Berry connection represent the vector potential on the Brillouin zone, while Berry curvature is analogous to the flux of a vector quantity.

Table 1. Anholonomies in geometry and quantum state. Also, the similarity between the theory of Berry phase and electromagnetism

|  | Parallel transport in | | Electromagnetism |
|---|---|---|---|
|  | **Geometry** | **Quantum** |  |
| **Fixed basis** | $\mathbf{e}_i$ | $|n(\mathbf{R})\rangle$ | - |
| **Moving basis** | $\mathbf{v}$ or $\psi$ | $|\Phi_n\rangle$ | - |
| **Parallel-transport condition** | $i\psi^* \cdot \dot\psi = 0$ | $i\langle\Phi_n|\dot\Phi_n\rangle = 0$ | - |
| **Anholonomy** | Anholonomy angle $\Theta(C)$ | Berry phase $\gamma_n(C)$ | Magnetic flux $\Phi(C)$ |
|  | Curvature $\mathbf{F}(\mathbf{R})$ | Berry curvature $\mathbf{F}(\mathbf{R})$ | Magnetic field $\mathbf{B}(\mathbf{r})$ |
|  | Connection $\mathbf{A}(\mathbf{R})$ | Berry connection $\mathbf{A}(\mathbf{R})$ | Vector potential $\mathbf{A}(\mathbf{r})$ |

## 5. Topological Phase Determinants

Topological phases of matter arise into a system with spin-orbit coupling, which causes a band inversion in the presence of symmetry protection/breaking. Topological invariances are another important determinant of topological phases that define the phase conversions under the adiabatic deformation of the system. Therefore, it is obvious that the key factors or determinants of the topological phases into quantum materials are spin-orbit coupling, band inversion, and symmetry protected/breaking topological invariance. Most of the topological or quantum phases were demonstrated by the quantum version of different Hall measurements. The previous section has covered the brief concepts on symmetry, and the following section is dedicated to discussing the different observed Hall effects in topological/quantum materials. This section is focusing on the discussions on the other key factors of topological phases, namely, spin-orbit coupling, band inversion, and topological invariance.

### 5.1. Spin-orbit coupling

Quantum Hall insulators have been the first quantum phases that require an external magnetic field to demonstrate the quantum phases. Under the external magnetic field, the time-reversal symmetry of the system is broken, and the backscattered-less chiral edge states emerge into the matter due to the cyclotron motion of electrons. The topological states of matter have been realized by replacing the external magnetic field with the spin-orbit coupling (SOC) field, which now can be protected under the time-reversal symmetry and provides Kramers degeneracy. Therefore, spin-orbit coupling is the main backbone to develop the topological materials, which create the Kramers degenerate edge states with band inversion and provides the existence of the relativistic fermions.

SOC is a quantum magnetic field inside an atom of a Kramers' system having spin-orbit interactions which induced a transverse magnetic force on an orbiting electron in the external electric field created by the charge of the nucleus.[25,26] This relativistic quantum mechanical effect arises from the relativistic momentum dependent motion of the charged particle into the electric field of the atom. This intrinsic magnetic field of the atom can be determined from the Thomas-term:[108]

$$H_{SOC} = -\frac{1}{2m_0^2 c^2} S.L \times (\nabla V) = -\frac{\hbar}{4m_0^2 c^2} \sigma.p \times (\nabla V) \quad (22)$$

Where $m_0$ is the electron mass, $c$ is the speed of light, $S$ and $L$ are spin and orbital angular momentum, respectively, $V$ is the Coulomb potential, $\sigma$ is Pauli's spin matrices, and $p$ is the momentum vector. In the atomic world, SOC provides band splitting and band inversion in the energy band, also known as *LS* coupling, as shown in *Figure 5*(a). The energy eigenvalues due to the spin-orbit interaction can be expressed from the first-order perturbation theory as,[109]

$$E_{SOC} \approx Z2\mu_B^2 \langle\frac{1}{r^3}\rangle \langle L.S\rangle \approx \frac{1}{2}a\hbar^2[J(J+1) - L(L+1) - S(S+1)] \quad (23)$$

Here, $Z$ is the atomic number, and $a$ is the spin-orbit coupling constant, which is proportional to $Z^4$. Another type of energy band splitting exists in a molecule, which is known as an electrostatic effect due to *e-e* repulsion and is proportional to $Z$. Therefore, the electrostatic effect is dominant in light atoms, while spin-orbit coupling dominates in the heavier atom systems. Moreover, the spin-orbit coupling provides $(2J+1)$ degeneracy for each *J* energy levels. SOC is a symmetry-independent interaction that can exist in all crystals. But there is a symmetry dependent SOC, which only exists in crystals without inversion symmetry. These SOCs are known as Dresselhaus interaction and Rashba interaction. *Figure 5*(b) illustrates the three types of SOCs that exist in the material systems along with their band structure and spin orientations.[110]

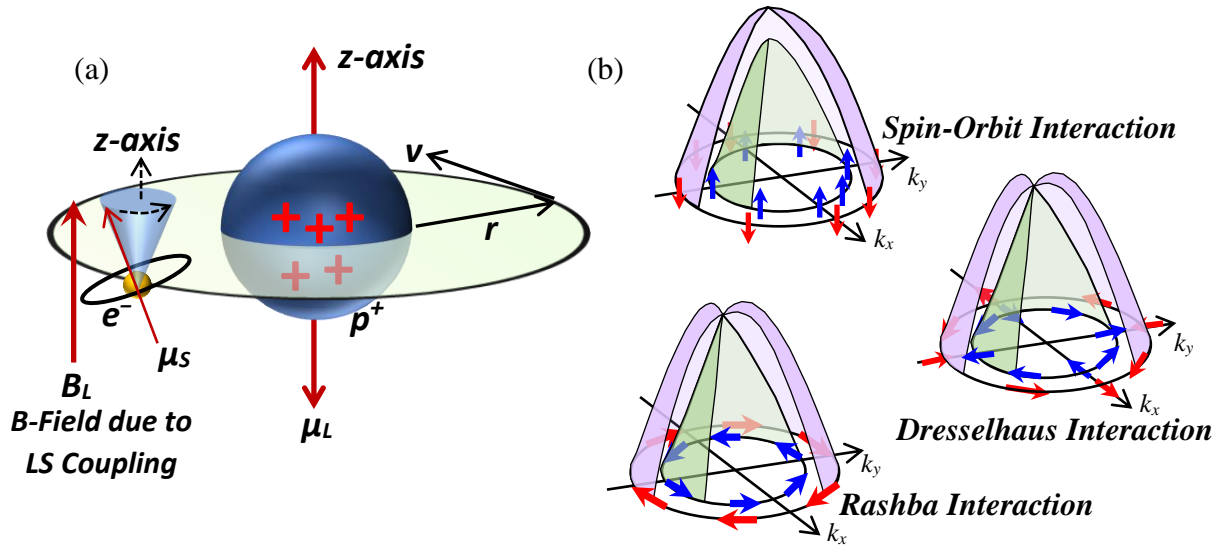

Figure 5: Illustration of (a) spin and orbital angular motions of an electron around nucleus giving spin-orbit or LS coupling effect, and (b) three types of spin-orbit relativistic interactions, namely, spin-orbit interaction, Dresselhaus interaction and Rashba interaction (adapted from 110).

## 5.2. Band Inversion

Band inversion is one of the key parameters in inversion symmetric topological insulators. Under band inversion, an electron-like band with positive effective mass is inverted to a hole-like band with negative effective mass. For a typical non-topological system, the highest valence band has odd parity, and the lowest conduction band has even parity. In narrow band-gap semiconductors with heavier atoms and strong spin-orbit coupling, odd parity bands are swapped with even parity bands, which eventually gives topological phases into the material. Band inversion is one of the essential means to find the topological phases into materials, though materials need to meet topological invariance condition to have the topological phases. Typically, the number of band inversions between opposite parity bands is odd in topological insulators. In PbSnTe$_2$, the number of band inversion is even and occurs at both Γ and three X points, which makes PbSnTe$_2$ topologically trivial.[23] On the other hand, topological crystalline insulators can have even band inversion as well as Dirac cone.[111,112] In topological materials, several types of band inversions can be observed: band inversion between $s$ and $p$ bands at Γ point,[113-115] band inversion between two $p_z$ orbitals with opposite parity at Γ point,[116,117] and $d$ and $p$ band inversion at Γ point.[118] Band inversion can also be obtained from lattice strain without SOC interaction.[119]

### 5.3. Topological Invariant

The topological invariant is a parameter of topological phases preserved under homeomorphism and does not change under adiabatic deformation. Typically, the topological invariant is defined by the integral of some geometric quantities like Gaussian curvature. In solids, the Brillouin zone represents the geometry, and the Berry phase provides curvature to get the topological invariant of the solids based on the Gauss-Bonnet theorem. The quantum mechanical curvature defined by Berry phase provides two other terms Berry connection, a gradient of Berry phase, and Berry curvature, a curl of Berry connection. The later term provides the curvature determining the topological invariant. The topological invariant for IQHI, the first observed quantum phase, was obtained from Berry curvature. The Chern number, $n$, is the invariant which is given by:[25,120]

$$n = \frac{1}{2\pi}\int_S F d^2k = \sum_m \frac{1}{2\pi}\int_S^{\phi_0} d\phi \int dk_x\, F(k_x, k_y^m(\phi)) \qquad (24)$$

Where $F$ is the Berry curvature, $k$ is the momentum space, $S$ is the closed surface in $k$-space, $\phi$ is the magnetic flux, and $\phi_0$ is the flux quantum. The Chern number is also known as Z invariant listed in *Table 2*. Berry phase defined topological invariant provides a nontrivial edge state or surface state at the boundary of an ordinary insulator or vacuum. In a strong SOC material system with time-reversal invariant, the topological invariant is given by the Chern parity or $Z_2$ invariant instead of the Chern number. Chern parity always refers to odd or even parity. In the 2D spin-orbit coupled system, $Z_2$ invariant is an integer value obtained from the integral over half of the Brillouin zone or effective Brillouin zone. The relation between $Z_2$ invariance and Berry phase is given as:[23]

$$Z_2 = \frac{1}{2\pi}\left[\oint_{edge\ of\ EBZ} A(k)dl - \oint_{EBZ} F(k)d\tau\right] mod(2) \qquad (25)$$

where, $A(k)$ and $F(k)$ are the Berry connection and Berry curvature, respectively, which can be obtained from the Berry phase. $Z_2$ invariant can also be defined by the total count of the number of Kramers' pairs that intersects the Fermi energy. In other words, the number of bands that cut the Fermi energy between two Kramers' points (consider as $N_{cut}$) can define the $Z_2$ index with the expression: $Z_2 = (N_{cut})mod(2)$; if $N_{cut}$ = even, $Z_2 = 0$ and if $N_{cut}$ = odd, $Z_2 = 1$. Calculation of $Z_2$ invariant from parameter $v$ in 2D topological insulators is given by,[23]

$$(-1)^\nu = \prod_{i=1}^{4} \delta_i \quad (26)$$

Formally, $\delta_i$ is defined as[12]

$$\delta_i = \frac{\sqrt{Det(w(\Gamma_i))}}{Pf(w(\Gamma_i))} = \pm 1 \quad i = n_1 n_2 \quad (i = n_1 n_2 n_3 \; for \; 3D) \quad (27)$$

with the unitary matrix $w_{mn} = \langle u_m(-k)|\Theta|u_n(k)\rangle$. Here, $u_n(k)$ are the eigenvectors of the occupied Bloch function, and $Pf(w)$ is the pfaffian over all 4 (8 in 3D) Kramers' points.[12,23] If the system has inversion symmetry, there exists a shortcut to obtaining $\delta_i$ in the Eq. (27).[23] At the Kramers points $\Gamma_i$, the occupied Bloch states $u_m(\Gamma_i)$ are also the eigenstates of the parity operator with the eigenvalue $\xi_m = \pm 1$. Therefore, $\delta_i$ is given by[23]

$$\delta_i = \prod_m^{occ} \xi_m(\Gamma_i) \quad (28)$$

Here, *occ* refers to the parities of pairs of occupied Kramers' doublets coming from the TRS at the TRIM points. Hence, the $Z_2$ invariant then simply follows from Eq. (26) with

$$(-1)^\nu = \prod_i^{4}\left(\prod_m^{occ} \xi_m(\Gamma_i)\right) = \left(\prod_m^{occ} \xi_m(\Gamma_{(0,0)})\right)\left(\prod_m^{occ} \xi_m(\Gamma_{(0,1)})\right)\left(\prod_m^{occ} \xi_m(\Gamma_{(1,0)})\right)\left(\prod_m^{occ} \xi_m(\Gamma_{(1,1)})\right) \quad (29)$$

Eq. (26) is limited to four terms related to the four TRIMs in 2D Kramers' system. The odd sum of the overall $Z_2$ sum represents the topological insulators. Similar to the 2D topological insulators, in 3D topological systems with inversion symmetry, $Z_2$ invariant is obtained from two expressions:[12]

$$(-1)^{\nu_0} = \prod_{i=1}^{8} \delta_i = \prod_i^{8}\left(\prod_m^{occ} \xi_m(\Gamma_i)\right) \quad (30)$$

$$(-1)^{\nu_{i=1,2,3}} = \prod_{n_{j\neq i}=0,1;n_i} \delta_{n_1 n_2 n_3} = \left(\prod_m^{occ} \xi_m(\Gamma_{(n_1 n_2 n_3)})\right)\left(\prod_m^{occ} \xi_m(\Gamma_{(n_1 n_2 n_3)})\right)\left(\prod_m^{occ} \xi_m(\Gamma_{(n_1 n_2 n_3)})\right)\left(\prod_m^{occ} \xi_m(\Gamma_{(n_1 n_2 n_3)})\right) \quad (31)$$

Where $n_i$ (i = 1,2,3) are parameters of reciprocal lattice vectors. Here, $\nu_0$ is independent of the choice of reciprocal vectors, while $\nu_i$'s can be defined from the reciprocal lattice. Based on $\nu_0$, 3D topological insulators (TIs) can be categorized into weak TI ($\nu_0 = 0$) and strong TI ($\nu_0 = 1$). Physically, the Chern number refers to the change in the electrical polarization of the quantum Hall insulators under an adiabatic change in magnetic flux, while $Z_2$ invariant refers to the change in the time-reversal polarization or local fermion parity of time-reversal invariant SOC material system under the adiabatic change in magnetic flux.

### 6. Hall effects and topological phases of matter

From the beginning of the discovery of the Hall effect, it has been associated with the breaking of time-reversal symmetry. In ordinary Hall effects, there are two currents existing in the presence of mutually perpendicular electric and magnetic fields: longitudinal and transverse current. The dissipationless transverse current directly breaks the time-reversal symmetry, while the longitudinal current breaks time-reversal symmetry based on the second law of thermodynamics by dissipating heat. Therefore, Hall measurements are considered as a vital fingerprint of the topological or quantum phases, which were initially observed in Hall measurement and determined from the Hall resistance. Based on different Hall observations, different topological phases are defined. This section is focused on different types of Hall families that correspond to different topological classes. All the Hall families are summarized in *Figure 6*, along with the inter-relations with other Hall effects. A generalized longitudinal and transverse Hall resistivity for all Hall families is illustrated in the figure to demonstrate the quantization of the Hall resistance, which is equivalent to the quantum resistance of $h/e^2$, where $e$ and $h$ are the elementary charge and Planck's constant, respectively.

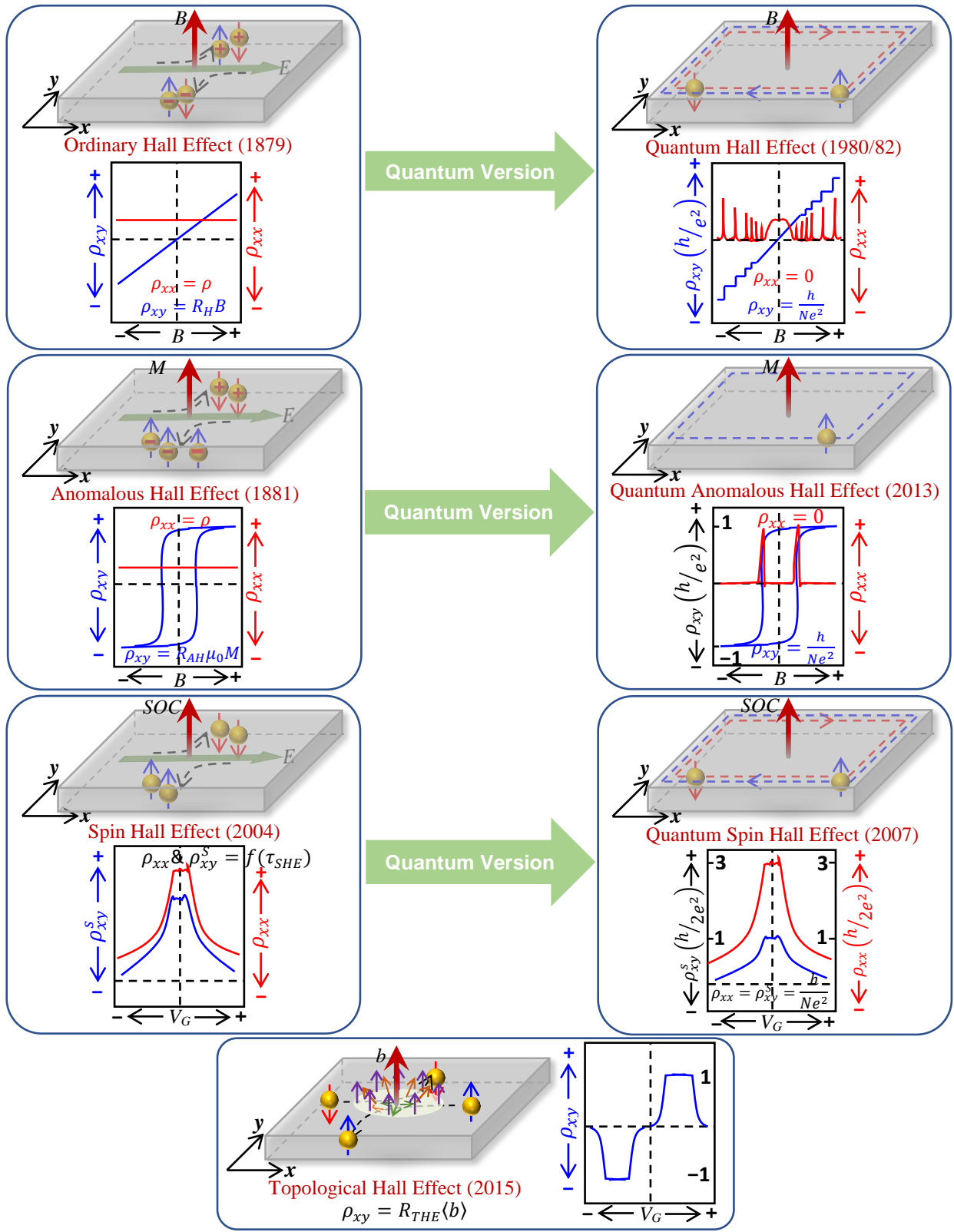

Figure 6: Family of different Hall effects along with their working principles, transverse and longitudinal resistivity, and mutual relations.

## 6.1. Time-reversal symmetry loss, chiral edge states, and quantum hall effect

A non-magnetic system in the absence of an external magnetic field is one example of a system with time-reversal symmetry. The two-dimensional electron system shown in *Figure 1* is subjected to an out-of-plane magnetic field, B, breaking the time-reversal symmetry. Electrons go through a circular trajectory in the classical description. Bulk electrons finish the full circle, but the edge electronic-states are not able to finish a full circle, hitting the edges repeatedly. In the quantum picture, however, the open orbital trajectory forms the chiral states at the edge of the sample field while the bulk stays insulating. The quasi-one-dimensional edge state motion of electrons is quantized and protected against backscattering (i. e., $\boldsymbol{p} \mapsto -\boldsymbol{p}$) since there is no counter-propagating state. In the quantum limit (e.g., strong magnetic field), the Hall conductivity quantizes and forms plateaus with integer units of quantum conductance $(2e^2/h)$, .[5]

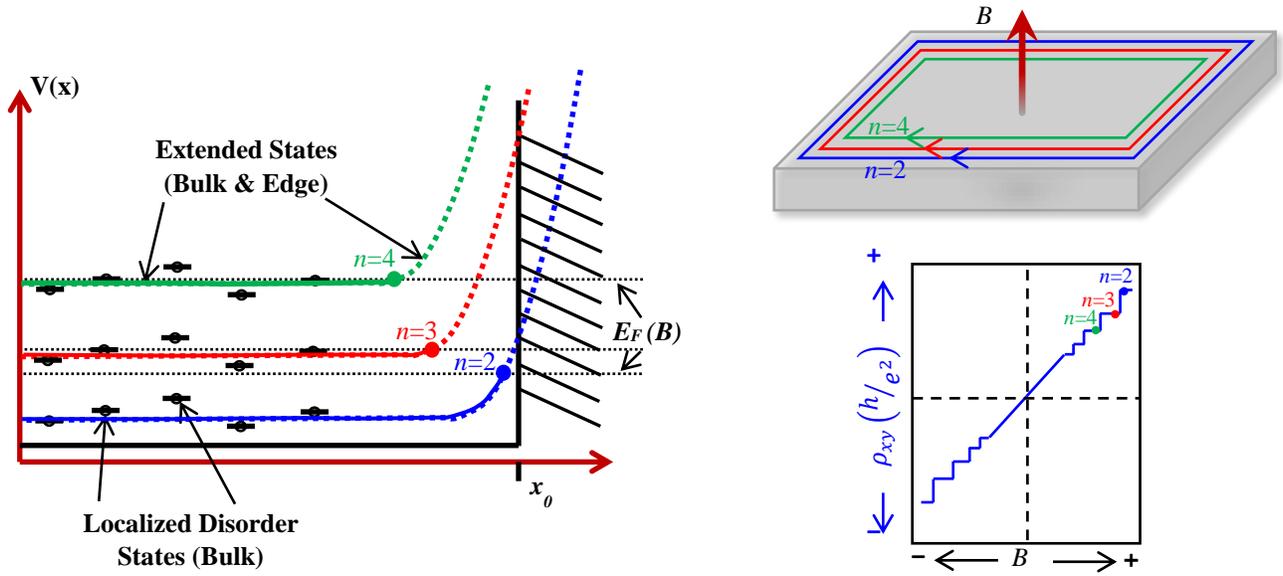

Figure 7: Illustration of the origin of quantization of Hall resistivity and plateaus in integer quantum Hall effect along with the landau band structure, shifting of Fermi level with magnetic field and the filling of the Landau levels.

$$\sigma_H = \frac{2e^2}{h} n \quad (32)$$

The dissipationless edge transport (i.e., without resistance) occurs due to the absence of the scattering events and counter-propagating states. The bulk mobile charge carriers could form a parasitic channel, obscuring quantum transport signatures of the dissipationless edge transport. Finally, the dissipationless edge transport is topologically protected against the continuous deformation of the band structure. The integer quantized *n* represents the filling factor of the $n^{th}$ landau levels, and the plateaus observed in the Hall resistance is associated with the band structure of the quantized landau edge states and the shifting of the Fermi energy as a function of a magnetic field as shown in *Figure 7*.[121]

## 6.2. Fractional quantum Hall effect: edge states with fractional charge

The fractional quantum Hall effect was observed just after two years of the observation of the integer quantum Hall effect. In a 2D electron gas system with fewer impurities and disorder, similar Hall resistivity plateaus were observed with a fractional filling factor of less than 2. A smaller disorder caused the disappearing of the higher filling factors. This exotic Hall feature was explained with strong *e-e* interaction and Landau theory. Considerable Coulomb interaction compare to the Landau energy ($E_{disorder} \ll E_{e-e} \leq n\hbar\omega_B$) can lift the degeneracy of the Landau level. Under magnetic flux quantum (quantized magnetic flux loop shown in *Figure 8*), now electrons try to minimize their Coulomb force by exchanging the positions and forming composite fermions, also known as *anyon*. These composite fermions are associated with a certain number of electron and flux quantum, which determine their filling factor and charge. Composite fermions of fractional quantum Hall insulators can carry a fractional charge. With the composite fermion theory, the fractional filling factors can be associated with integer filling factors for composite fermions, as shown in *Figure 8*. The relation between integer filling factor, *n,* and fractional filling for a composite fermion, $n_{CF}$, and its charge can be obtained as:[122]

$$n = \frac{n_{CF}}{2n_{CF} \pm 1}; \quad q_{CF} = \pm\frac{e}{2n_{CF} \pm 1} \qquad (33)$$

The formation of composite fermions is illustrated in *Figure 8*.

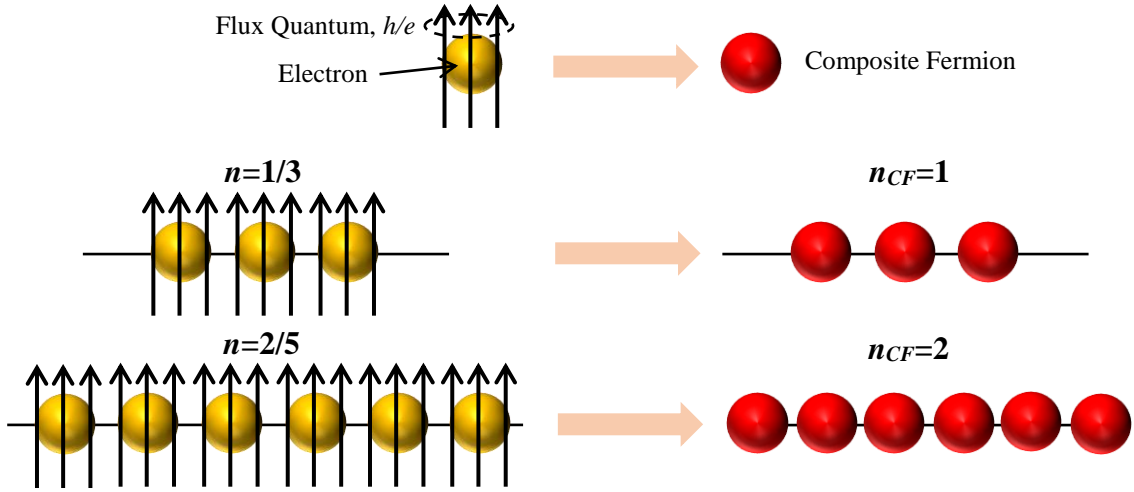

Figure 8: Composite Fermion theory for fractional quantum Hall effect: formation of composite fermion from electron and associated flux quantum, equivalent filling factor of composite fermion compare to fractional filling factors.

### 6.3. Quantum spin Hall effect: edge modes with time-reversal invariance

Two-dimensional topological insulators host edge modes in the absence of applied magnetic field without breaking time-reversal symmetry (*Figure 9*(a)). Here intrinsic spin-orbit coupling plays the role of an effective magnetic field, which is in the opposite direction for down- and up- spin species. In this configuration, the spin of electrons is locked to their momentum, which is also known as helicity.[10,11] These helical edge states form time-reversal conjugates, also referred to as Kramers' doublets. Time-reversal partner helical edge states were initially observed in HgTe quantum wells.[59,123,124]

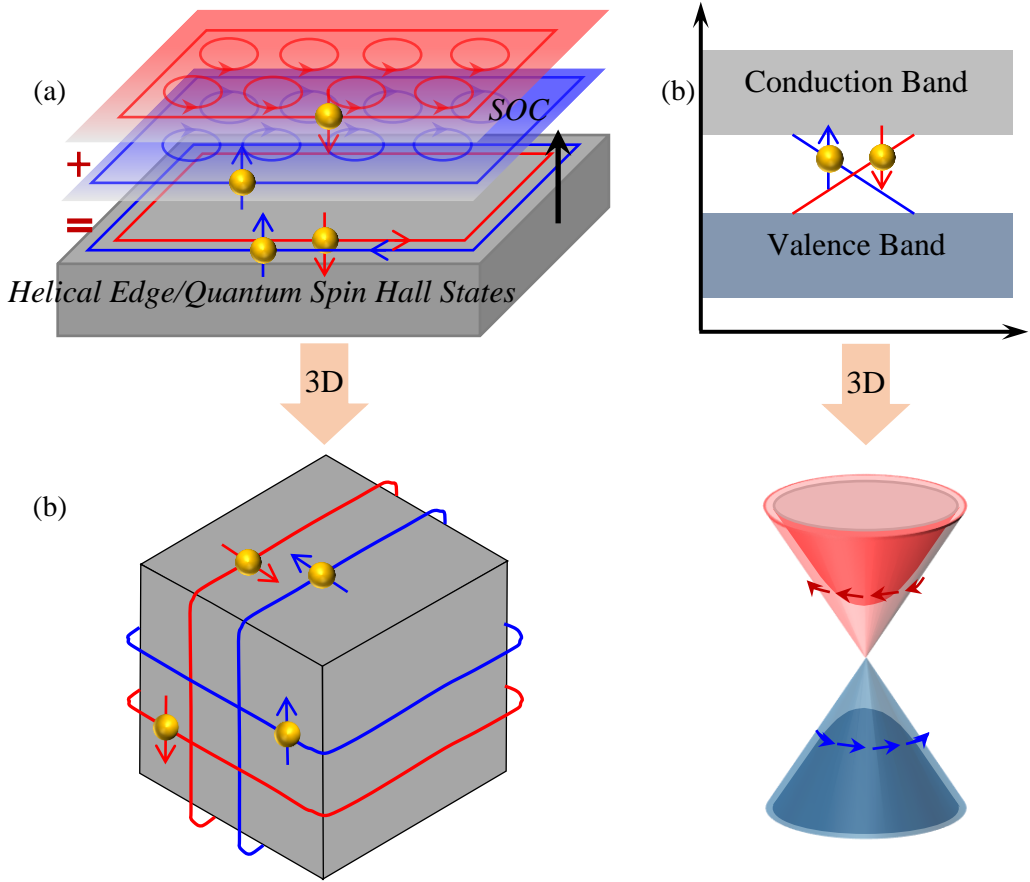

Figure 9: Illustration of Helical Edge or quantum spin Hall states in real and momentum space for (a) 2D topological insulators and (b) its 3D version known as 3D topological insulators.

Here, while the counter-propagating state does exist, the scattering between them is not possible, and thus time-reversal symmetry is preserved. Since the reversal of momentum and spin transforms each helical state to the other one, these states are related through the time-reversal operation. Furthermore, the lack of scattering between the two states can be proven:

$$\langle -|\hat{V}|+\rangle = \langle -|\mathbb{T}^\dagger V \mathbb{T}^{-1}|+\rangle = \langle -|\mathbb{T}^\dagger V \mathbb{T}|+\rangle = -\langle -|\mathbb{T}^\dagger \mathbb{T} V|+\rangle = \\ -\langle -|\hat{V}|+\rangle^* = -\langle -|\hat{V}^\dagger|+\rangle = -\langle -|\hat{V}|+\rangle \qquad (34)$$

This means that helical edge states are protected against any irregular scattering potential. Furthermore, the trajectory of helical edges may change near physical barriers, but the two states remain time-reversal conjugates. It is known that disorders can reduce the mean free path of the charge carriers. In the limit of the mean-free-path becoming comparable to the lattice constant, charge carriers localize, and an insulating ground state emerges, also referred to as Anderson localization.[125,126] Eq. (34) emphasizes that no matter the shape of the scattering potential, Anderson localization does not occur in the spin-momentum locked helical time-reversal partner states. Finally, higher-dimensional analogs of the two-dimensional topological insulators are possible. For example, in the case of a three-dimensional topological insulator, the spin and momentum of the electrons are locked in surface edge electronic states.[12,13] Three-dimensional topological insulator was initially observed, using angle-resolved photoemission spectroscopy in $Bi_2Se_3$.[127]

## 6.4. Topological invariants: Thouless-Kohmoto-Nishtingale-den Nijs relationship for Hall conductivity

Thouless, Kohmoto, Nishtingale, and den Nijs (TKKN) developed a relationship for the Hall conductivity, using the Kubo formula, in the limit of zero frequency and temperature, relating Berry curvature and Hall conductivity:[128]

$$\sigma_{xy} = \frac{e^2}{h} \int \frac{dk_x dk_y}{2\pi} (\nabla_k \times A_{ks})_z \qquad (35)$$

where the integration is over the whole Brillouin zone. The TKNN relationship establishes a connection between the band structure topology and the Hall conductivity. Two aspects of Eq. (35) need special attention.

- The Berry curvature ($\Omega(k) = \nabla_k \times A_{ks}$) is an odd function of the momentum ($\Omega(k) = -\Omega(-k)$), highlighting the disappearance of Hall conductivity with time-reversal symmetry. Hall conductivity, as a result, can only be nonzero for a system with broken time-reversal and nonzero Berry curvature.
- The Hall conductivity quantizes and forms plateaus with integer units of quantum conductance ($e^2/h$). Starting from the TKNN formula and applying a periodic boundary condition, one can connect the edges of the Brillouin zone. The Fermi surface, as a result, maps into the surface of a torus. The Brillouin zone might be divided into two arbitrary manifolds and the Hall conductivity, accordingly, only depends on the boundary of the two manifolds, as follows:

$$\sigma_{xy} = \frac{e^2}{h} \frac{1}{2\pi} \left[ \oint_C A_k^I \cdot d\mathbf{k} - \oint_C A_k^{II} \cdot d\mathbf{k} \right] \qquad (36)$$

where $C$ is the boundary of the two arbitrary manifolds. The phase difference of the Berry connections at the boundary is arbitrary and due to the gauge invariance. The quantization of the Hall conductance is related to the winding number of the phase, $n$.

$$\sigma_{xy} = \frac{e^2}{h} n, \quad n = \frac{1}{2\pi} \oint_C (A_k^I - A_k^{II}) \cdot d\mathbf{k} = \int \frac{dk_x dk_y}{2\pi} (\nabla_k \times A_{ks})_z \qquad (37)$$

Finally, $n$ is insensitive to the continuous deformation of the band structure and is determined based on the topology of the filled bands.

### 6.5. Anomalous and Topological Hall effects

#### 6.5.1. Anomalous Hall effect: Hall effect with band structure Berry curvature

Shortly after the initial discovery of the Hall effect, Edwin Hall observed an extraordinary transverse voltage in magnetic metals, including iron and nickel. The anomalous transverse voltage with magnetic field, qualitatively, follows the same trend as magnetization. An empirical relationship relates the magnetization to anomalous Hall voltage as follows:

$$H_e = H_0 + 4\pi M\alpha, \qquad R_H = R_0(H_0 + 4\pi M\alpha) \qquad (38)$$

where $M$ and $H$ are the magnetization and the magnetic field, respectively. Karplus and Luttinger revisited the problem in 1954[129] and observed a direct relationship between anomalous Hall conductivity and the spin-orbit coupling, which was in good agreement with experimental results.

$$R_{AHE} \sim \frac{B_{s.o}}{m^*} R_{xx}^2 \qquad (39)$$

The electrons acquire an *anomalous velocity* perpendicular to the electric field (second term in Eq. (40)). The anomalous velocity is finite over occupied bands and stays independent of the scattering rate and, as a result, longitudinal conductivity. It is, hence, referred to as the intrinsic anomalous Hall effect. Using the TKNN formula, one can describe the Berry curvature as an effective magnetic field, causing a Hall conductivity:

$$\frac{d\langle r \rangle}{dt} = \frac{\partial E}{\hbar \partial k} + \frac{e}{\hbar} E \times (\nabla_k \times A_{ks})_z \qquad (40)$$

The problem was revisited recently by Haldane as a "topological Fermi-liquid" property. Haldane took into account the accumulated Berry phase by an adiabatic motion of the quasiparticles only at the Fermi surface.[130] A complete theoretical picture, considering both the intrinsic and extrinsic anomalous Hall effects, was recently proposed by Nagaosa et al.[131-133] The intrinsic anomalous Hall conductivity is insensitive to scattering and is directly proportional to the Berry curvature. The anomalous Hall conductivity, using the TKNN formula, is also quantized with integer quantum conductance values ($e^2/h$):

$$\sigma_{xy}^{AHE-int} = -\epsilon_{ijl} \frac{e^2}{\hbar} \sum_n \int \frac{dk}{(2\pi)^d} f(\epsilon_n(k))(\nabla_k \times A_{ks}) \qquad (41)$$

where $\epsilon_{ijl}$ and $f(\epsilon_n(k))$ are the antisymmetric tensor and the Fermi-Dirac statistics with summation going through all the filled bands, *n*. The extrinsic anomalous Hall effect is typically due to the skew scattering or side jump. Skew scattering refers to an *effective* TRS breaking of charge carriers due to the asymmetric scattering of electrons originating from the spin-orbit coupling to impurity atoms. Side jump (SJ) is the time-integrated velocity deflection of electrons interacting with an impurity or disorder. The anomalous Hall voltage contains both extrinsic and intrinsic components.

$$\sigma_{xy}^{AHE} = \sigma_{xy}^{AHE-int} + \sigma_{xy}^{AHE-skew} + \sigma_{xy}^{AHE-SJ} \qquad (42)$$

Only intrinsic AHE contains information about the topology of the band structure. Three regimes and two crossovers have been established for constant spin-orbit coupling energy ($E_{SO} < E_F$):[131] (i) the clean regime, characterized by an insignificant scattering rate ($\hbar/\tau \leq u_{imp} E_{so} D$); (ii) the moderately dirty regime which has a modest scattering rate ($u_{imp} E_{so} D \leq \hbar/\tau \leq E_F$); (iii) the dirty regime ($E_F \leq \hbar/\tau$). Skew scattering dominates the AHE in the clean regime. Skew scattering is inversely proportional to the density of impurities. $u_{imp}$ and $D$ are the impurity potential and the density of states. Hall conductivity becomes insensitive to scattering (i.e., longitudinal conductivity, $\sigma_{xx}$) in moderately dirty regime due to the intrinsic Berry phase contribution.[134] A power-law scaling of the transverse conductivity with longitudinal conductivity is predicted in the dirty regime. *Figure 10* shows these three regimes versus the longitudinal conductivity and impurity density.[135-137]

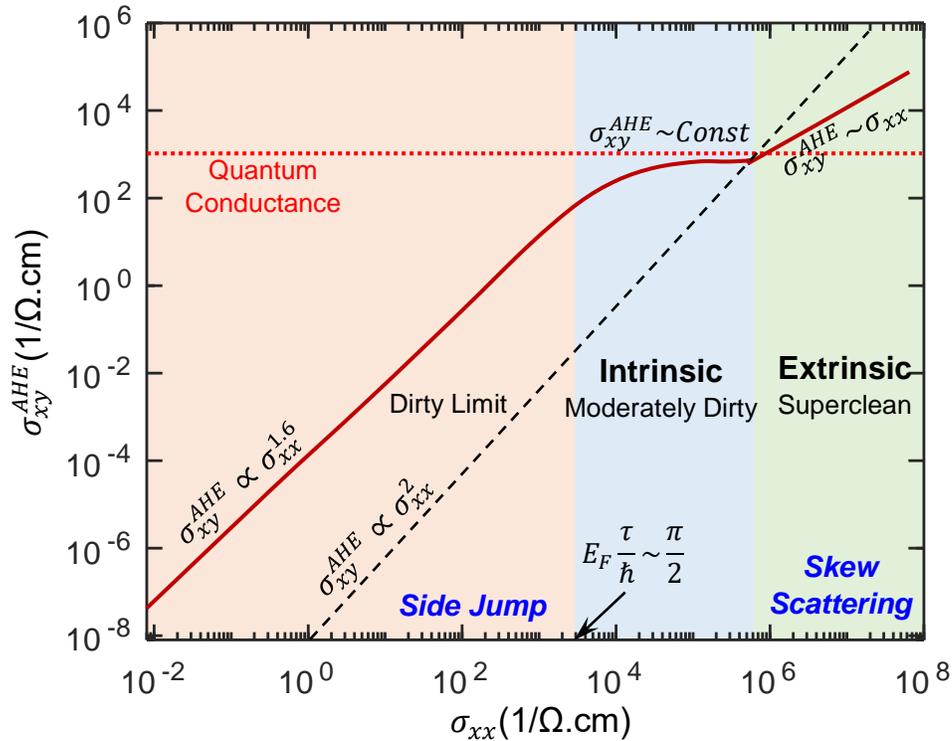

Figure 10: Anomalous Hall conductivity with respect to longitudinal conductivity. Hall conductivity becomes insensitive to scattering (longitudinal conductivity), highlighting intrinsic dissipationless topological transport governed by Berry curvature (adapted from 135-137).

### 6.5.2. Topological Hall effect: Hall effect with non-trivial spin texture

A combination of nonzero Berry curvature and broken are necessary for a finite Hall conductivity. Realspace spin texture, similar to the momentum space band structure, could be a source for large Berry curvature. The adiabatic motion of electrons in a smoothly varying magnetic field produces nonzero Berry curvature in the absence of spin-orbit coupling.[138] The spin of conduction electrons couple adiabatically and acquire the Berry phase. The effective Lorentz force, as a result, is independent of the spin-orbit coupling. The Berry phase is due to the topologically nontrivial spin texture and is proportional to the spin chirality of electrons (*Figure 11*).[139,140] This effect is commonly referred to as the *topological* Hall effect.

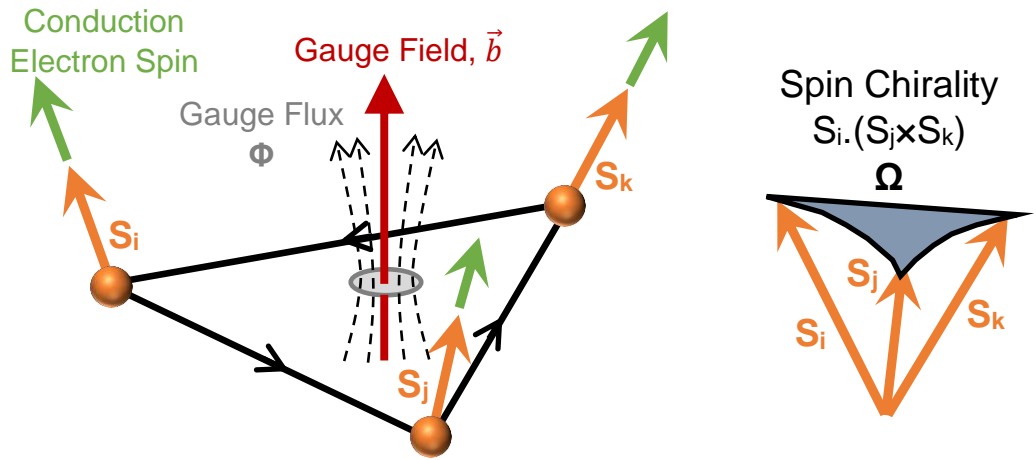

Figure 11: Conduction electrons (green arrow) moving in a smoothly varying internal magnetization due to the magnetic ions (brown arrows) couple adiabatically and accumulate Berry curvature, directly proportional to the magnitude of spin chirality. Reproduced with permission from [132].

The topological Hall effect (THE) reflects the topologically nontrivial spin textures, e.g., skyrmions.[141] The origin of THE is the accumulated Berry phase by the conduction electrons due to the motion in a non-coplanar spin texture.[142,143] The topological field generates an effective magnetic field, affecting the conduction electrons in spin-polarized subbands. The topological field is proportional to $(\frac{\partial \boldsymbol{n}}{\partial x} \times \frac{\partial \boldsymbol{n}}{\partial y})$ and is nonzero for non-coplanar spin textures.[141] Dzyaloshinskii-Moriya interaction causes helical structures in non-centrosymmetric crystals, e.g., MnSi, triggering skyrmions[142,144]. The magnetic frustration could also cause helical structures in centrosymmetric crystals, including $SrFe_{1-x}Co_xO_3$ and pyrochlores.[145,146] Thermal spin fluctuations are necessary for skyrmion formation in B20 non-centrosymmetric crystals. Observation of skyrmion crystals SkX at near 0 K is explained by magnetic frustration in B20 non-centrosymmetric crystals.[147] The inversion symmetry is lifted by interfaces of thin films, causing SkX formation.[148,149] The carrier density control of nontrivial spin textures in oxide films and interfaces gives rise to the possibility of topological devices within all-epitaxial heterostructures.[150-154]

## 7. Symmetries and Topological Classification

Topological states of quantum materials induced by the spin-orbit coupling (SOC) are defined by the symmetry-defined topological invariant, which is continuous under adiabatic deformations of a gapped system without closing its gap.[155] Generally, phases of matter can be distinguished by the spontaneous symmetry breaking of the local order parameters. Atoms and electrons can be organized in various ways to form different phases, such as gas, liquid, solid, magnet, superconductor, etc. It is well known that according to Landau's symmetry-breaking theory (LSBT),[156] different phases are corresponding to different symmetries in the organization of the materials. Symmetry is also an important parameter to define the topological states but with a different definition. In topological phases of condensed matter, the smooth deformations are not related to the tearing and cutting of material but to symmetry breaking such as time-reversal symmetry (TRS), particle-hole symmetry (PHS), and chiral symmetry. Like the continuous symmetry operators (operations that can be obtained by applying successively infinitesimal symmetry operations) in quantum mechanics, discrete symmetries like the parity (inversion symmetry/space inversion), lattice translation, and time-reversal symmetries are also played essential roles in condensed matter.[157]

Here, it is mention-worthy that both short-range and long-range quantum entanglement can characterize topological states. Quantum entanglement can be defined as the physical state where particles/spins of many-body systems with long-range interaction are connected in a way that the quantum state of individual particles cannot be defined separately.[158,159] The long-range and short-range entanglement are defined with respect to the local unitary transformation. Local unitary transformation defines whether two gapped quantum states are in the same phase (or adiabatically connected).[158,159] The gapped quantum systems without any symmetry can have two classes of quantum phases: long-range entangled and short-range entangled phases.[158,159] The long-range entangled states cannot be transformed into other long-range entangled systems via local unitary transformation. Therefore, they belong to different quantum phases, which are also called topologically ordered phases, such as fractional quantum Hall insulators, spin-liquids, etc.[158,159] On the other hand, the short-range entangled gapped quantum states without symmetry are trivial as they can be transformed into each other via local unitary transformation, and hence, they all belong to the same phase.[158,159] Interestingly, the gapped quantum states with symmetry, both long-range and short-range entangled states, become nontrivial and belong to different phases.[158,159] In gapped quantum systems with symmetry, the long-range entangled states are called symmetry enriched topological (SET) phases, and the short-range entangled states are categorized into two states: trivial symmetry-breaking short-range entangled states defined by the Landau symmetry-breaking theory and non-trivial symmetry protected short-range entangled states defined by the group cohomology theory.[158,159] The symmetry-protected short-range entangled systems include topological insulators and semimetals. The topology-protected long-range entangled system has a fractional charge, while the symmetry-protected short-range entangled system does not have a fractional charge.[155] Here, we mainly focused on the symmetry-protected topological phases, and therefore, different kinds of symmetries are discussed in the following subsections.

Without TRS, topological states can be deformed adiabatically into other topologically trivial states without closing the energy gap, emphasizing the characteristics of symmetry-protected topological states.[155] Besides nonspatial symmetries like TRS, several spatial space-group symmetries acting nonlocally in position space like rotation and reflection can induce nontrivial topological states. In quantum mechanics, symmetry transformation is defined by either linear and unitary or antilinear and antiunitary operators acting on fermionic operators in Hilbert space.[155] Both unitary and antiunitary symmetry operators can be nonspatial (on-site) or spatial and global (non-gauge), or local (gauge).[155] Different types of symmetries that define different topological states are briefly discussed in the following subsections.

### 7.1. Point-group symmetry

Point-group symmetries are unitary operators that keep the lattice unchanged after acting on the lattice. Point-group symmetry operators generally provide short-range ordered symmetry-protected topological states.[160] Point-group symmetries in a crystal form a subgroup in the large symmetry group.[160] Rotation, reflection, inversion, and rotoreflection operators are typical point-group symmetry operators. Point-group symmetry operators can make degeneracies in the energy bands. General rules for point-group symmetry operators (shown in *Figure 12*) are: $2\pi$ rotation in operators gives unity, square of $2\pi$ rotation on operators is also unity, and $2\pi$ rotation in spinor representation of the operators gives $-1$.[161,162]

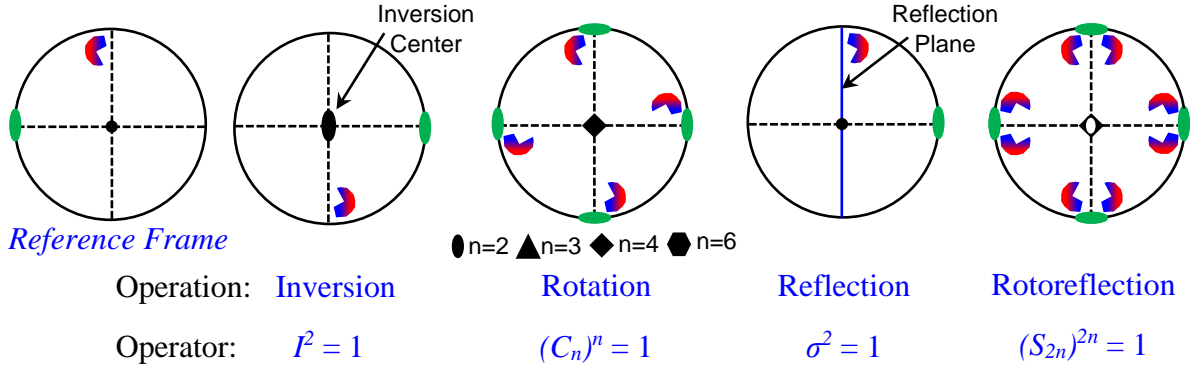

Figure 12: Illustration of different point group symmetry along with the generators of the operators.

Among the point-group symmetry operators, inversion operators play an important role in developing some key features for some topological states of matters. Inversion symmetry is a discrete symmetry (parity operation or space inversion) that changes a right-handed system into a left-handed system. Unitary inversion operator, $\mathbb{I}$, satisfies the conditions of $k \to -k$ and $\mathbb{I}^2 = 1$. In quantum mechanics, the expectation value of $x$, $\langle x \rangle$, can be calculated from the space-inverted state with the opposite sign,

$$\langle \beta | P^{-1} x P | \beta \rangle = -\langle \beta | x | \beta \rangle \qquad (43)$$

where $P^{-1} = P^\dagger = P$ ($P$ is now not only unitary but also Hermitian). Therefore, it can be deduced from the above equation that $x$ and $P$ must be anticommuting, i.e.,

$$P^{-1} x P = -x, \text{ or } xP = -Px \qquad (44)$$

Also, according to Eq. (44), we can say that $x$ is odd under parity. It is useful to look at the parity properties of energy eigenstates. In a space inversion invariant system, i.e., $[H, P]=0$, if we suppose $|n\rangle$ is a nondegenerate eigenstate of $H$ with eigenvalue $E_n$, $H|n\rangle = E_n|n\rangle$, then $|n\rangle$ is also a parity eigenstate. Therefore, $H|p\rangle = \pm 1|p\rangle$, where $\pm 1$ and $|p\rangle$ are eigenvalue and eigenstate of parity, respectively. Besides, Berry curvature within valence bands of the time-reversal invariant and space inversion system of spin 1/2 particles is odd and even, respectively:

$$TR: \quad F(k) = -F(-k), \quad IS: \quad F(k) = +F(-k) \qquad (45)$$

This feature can cause different kinds of non-trivial topological phases to emerge. Inversion symmetry operation on energy band provides:

$$E_{n,s}(-\vec{k}) = E_{n,s}(\vec{k}) \qquad (46)$$

Here, $E_{n,s}$ is the energy levels with non-spin quantum numbers, $n$, and spin quantum number, $s$. A special inversion symmetry operator also exists, which holds $\mathbb{I}^2 = -1$ and implies a system with an odd number of fermions. This type of inversion symmetry can exist in 1D/2D electrons with spin-orbit coupling under a particular crystal rotation symmetry along an axis perpendicular to the system.[163] Crystal lattice without inversion symmetry does not guarantee the spin degeneracy of the bands.

### 7.2. Time reversal symmetry and time-reversal invariant momenta

Time reversal symmetry (TRS) is an antiunitary involution fermionic operator[164] and simply indicates the invariance of a physical property with the time arrow ($t \mapsto -t$). An unchanged Hamiltonian ($\widehat{H}$) under this transformation suggests a time-reversal invariant system. Time reversal transformation leaves position ($\mathbf{r}$) unchanged but reverses the linear momentum ($\mathbf{p}$) and spin ($\mathbf{s}$). A time-reversal invariant system is, for example, an even function of the momentum and independent of spin (e.g., non-relativistic free particles).

$$TR: \quad \mathbf{r} \mapsto \mathbf{r}, \mathbf{p} \mapsto -\mathbf{p}, \text{and } \mathbf{s} \mapsto -\mathbf{s} \tag{47}$$

Considering $\mathbb{T}$ as a TRS operator, a system with an odd number of fermions satisfying $\mathbb{T}^2 = -1$ represents the system with Kramers degeneracy, which also means that two consecutive TRS operations on the spinor of the spin-1/2 system give a minus-sign without restoring the system into the initial configuration. In a spin-1/2 system, TRS operations can be written as:

$$\mathbb{T}\sigma_x \mathbb{T}^{-1} = -\sigma_x, \quad \mathbb{T}\sigma_y \mathbb{T}^{-1} = -\sigma_y, \quad \text{and} \quad \mathbb{T}\sigma_z \mathbb{T}^{-1} = -\sigma_z \tag{48}$$

Where σ are Pauli's spin matrices. TRS operation on crystal provides the following degeneracy condition:

$$E_{n,-s}(-\vec{k}) = E_{n,s}(\vec{k}) \tag{49}$$

In the presence of both TRS and IS operators, an antiunitary operator of $\mathbb{T} \otimes \mathbb{I}$ is obtained, which provides −1 when squared. Under both inversion and time-reversal symmetries operation on a crystal, we have:

$$E_{n,-s}(\vec{k}) = E_{n,s}(\vec{k}) \tag{50}$$

This condition guarantees the presence of spin degeneracy of the band, and the associated momenta points are known as time-reversal invariant momenta points (TRIMs), shown in *Figure 13*(d). For a 2D Brillouin zone, TRIMs are $(k_x, k_y) = (0,0), (0,\pi), (\pi, 0), (\pi, \pi)$ (*Figure 13*(a)). TRIM points of topological phases create the effective Brillouin zone (EBZ) from which the topological invariance, $Z_2$ can be determined. $Z_2$ invariance is determined from the Berry connection and Berry curvature under the boundary limit of the effective Brillouin zone. EBZ forms the torus, which represents the topology of the system. *Figure 13*(a)-(c) illustrates the TRIM points in the 2D and 3D Brillouin zone of the topological insulators along with the EBZ or the Fermi surface structures.

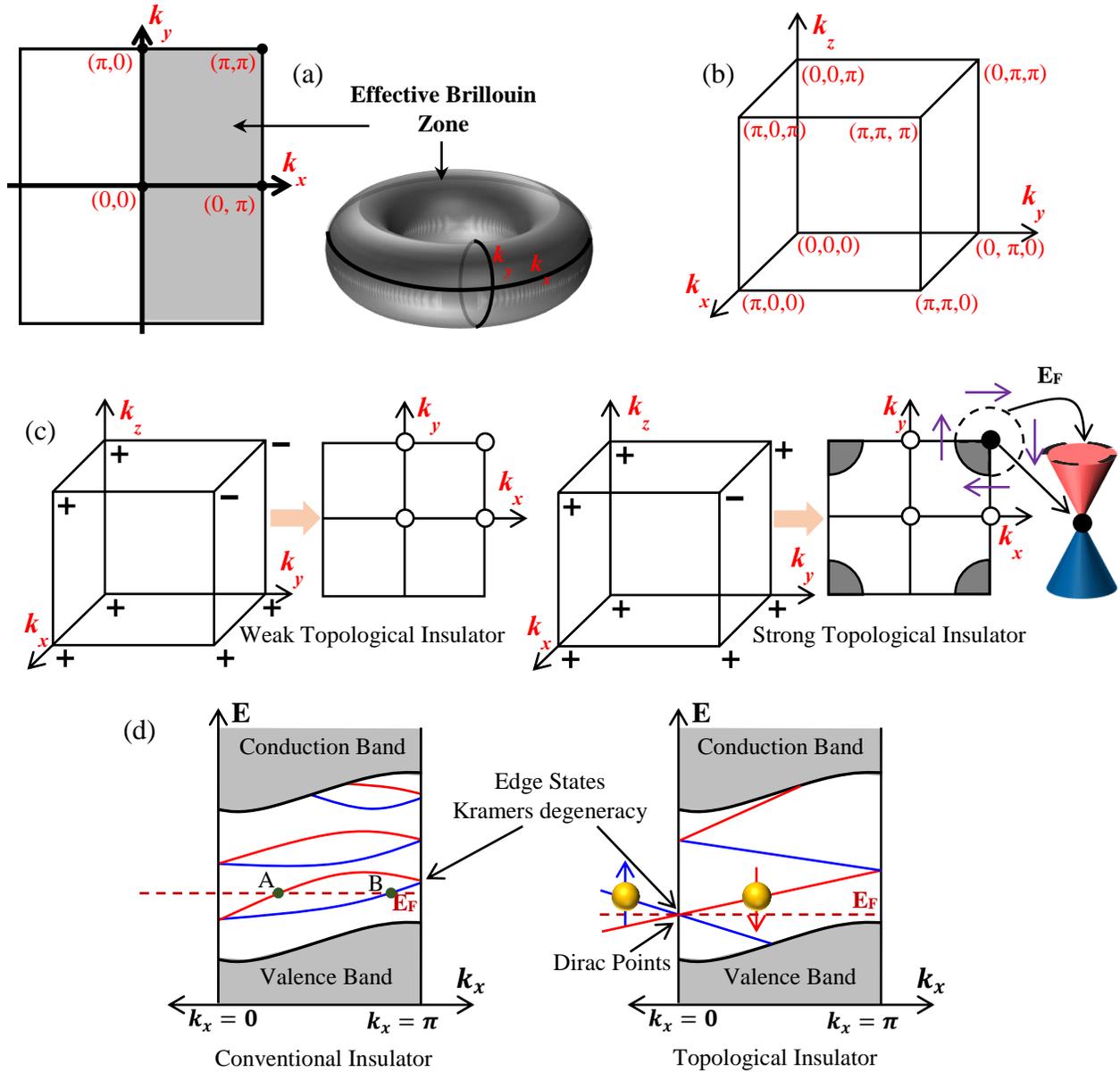

Figure 13: (a) Time reversal invariant momenta (TRIM) points in 2D topological insulator (TI) system along with the effective Brillouin Zone, (b) TRIM points in 3D TI system, (c) weak and strong topological insulator along with associated Fermi arcs enclosing TRIMs, and (d) band structures for conventional and topological insulator.

### 7.3. Kramers' Theorem

Each energy level is at least doubly degenerate for a time-reversal invariant system with half-integer spin, as stated by Kramers' theorem. In *Figure 13*(d), both Kramers partners (A and B in *Figure 13*(d) left) and Kramers points (Dirac point in *Figure 13*(d) right) are demonstrated, which have the same energy at different momenta positions, also known as Kramers' pair. Two degenerate states should be related to the time-reversal operation if time-reversal symmetry is intact. Accordingly, the Kramers points become their own time-reversal and provide the Dirac points (*Figure 13*(d) right). They are independent states and, therefore, they must be orthogonal ($\langle+|-\rangle = 0$). This can be proven as follows:

$$\langle+|-\rangle = \langle-|\mathbb{T}^\dagger\mathbb{T}^{-1}|+\rangle = -\langle-|\mathbb{T}^\dagger\mathbb{T}|+\rangle = -\langle-|+\rangle^* = -\langle+|-\rangle \quad (51)$$

In the band theory of electrons, TR operation relates momentum $k$ to $-k$, i.e., $-k = \mathbb{T}k\mathbb{T}^{-1}$. Therefore, in a system with time-reversal invariant, the Bloch Hamiltonian $H$ satisfies the following relation:

$$H(-k) = \mathbb{T}H(k)\mathbb{T}^{-1} \quad (52)$$

This equation shows that if $|n(k)\rangle$ is any Bloch-state of the $H(k)$, then $\mathbb{T}|n(k)\rangle$ is an eigenstate of the $H(-k)$ at $-k$, with the same energy. In topological states, counterpropagating time-reversal protected helical states are Kramers' pair, and the orthogonality relation between Kramers' pair supports the characteristics of no backscattering between two counterpropagating spin-filtered helical modes. TRS-protected Kramers' degeneracy can be lifted by breaking TRS via an external magnetic field. But TRS is protected against the spin-orbit coupling and crystal field. Note that away from the Kramers' points, the spin-orbit interaction can lift the degeneracy. Also, the specific TRIMs are dependent on the crystal structure. In general, TRIM points are directly related to high symmetry (HS) points in the Brillouin zone. In other words, the HS of the TRIMs is reflected by the electronic structure symmetry. Consider an electronic state with Energy $E(k)$, and Spin ↑. According to Kramers' degeneracy, it is guaranteed that there is a degenerate state with $E(-k)$ ↓. Therefore, if $k$ is a TRIM, $-k$ can also be reached by adding a reciprocal lattice vector to $k$, implying that there must not only be a degenerate state $E(-k)$ ↓ but also one $E(-k)$ ↑, i.e., the state is double-degenerate at a TRIM.

### 7.4. Other Symmetry Operators

Among the other symmetry operators, particle-hole symmetry (PHS) and chiral symmetry also play an important role in forming new classes of topological states. Particle-hole symmetry or charge-conjugation symmetry flips the sign of the charge after the antiunitary transformation action on fermions in a system conserving the particle numbers.[155] An antiunitary operator can be considered as a product of a unitary and a complex conjugate operator ($\mathbb{U}\mathcal{K}$, where $\mathbb{U}$ is the unitary and $\mathcal{K}$ is the complex conjugate operators). Two consecutive PHS operation gives a similar condition of TRS, another unitary operator. In general, PHS operators satisfy the following conditions:[155]

$$\mathbb{C}^{-1}\hat{Q}\mathbb{C} = -\hat{Q}, \mathbb{C}^2 = (\pm 1)^N \text{ when } \mathbb{U}_\mathbb{C}^*\mathbb{U}_\mathbb{C} = \pm 1 \quad (53)$$

Where $\mathbb{C}$ is the PHS or charge-conjugation operator, Q is a function of particle number operator, N, and $\mathbb{U}$ is the unitary matrix associated with the PHS operator. Like TRS, PHS can provide three characteristics for a system: i) system without PHS, ii) system with PHS, and the square of PHS is +1, and iii) system with PHS, and PHS squares to $-1$. Any superconductors with PHS always satisfy $\mathbb{C}^2 = -\mathbb{T}^2$. On the other hand, the Chiral symmetry operator is the combination of both TRS and PHS operators. Therefore, chiral symmetry is not an independent symmetry operator, which means a system with either TRS or PHS cannot have chiral symmetry, or a system with any two symmetries always has the third symmetry.[155] Under chiral symmetry, rotation of left-handed chiral fermions transforms into right-handed chiral fermions, which also implies that chiral symmetry holds for only massless Dirac fermions. The chiral symmetry operator, $\mathbb{S}$, holds the similar conditions of TRS and PHS, as the unitary matrix of the chiral symmetry is the product of unitary matrices of TRS and PHS.[155] Square of the chiral symmetry is always 1. Therefore, chiral symmetry holds the following condition:[155]

$$\mathbb{S}^{-1}\hat{Q}\mathbb{S} = -\hat{Q}, \mathbb{S}^2 = 1, \text{where } \mathbb{S} = U_\mathbb{S} \qquad (54)$$

In summary, TRS is an antiunitary symmetry operator that can commute with the system Hamiltonian, PHS is an antiunitary symmetry operator which anti-commutes with the Hamiltonian, and finally, chiral is the unitary symmetry operator that also anti-commutes with Hamiltonian. Both PHS and Chiral symmetry gives new classes of topological states, which are discussed in the following subsection.

### 7.5. Topological Classification

All the symmetry operators mentioned above can define different classes of topological states, which are characterized by respective topological invariants. Based on the presence of individual or combined global symmetries such as time-reversal, particle-hole, and chiral symmetry, ten classes of topological phases can be categorized for non-interacting fermionic systems and defined with their distinct topological invariance with respect to the dimensionality of the system. The topological classes are also called Altland and Zirnbauer (AZ) symmetry classes. All topological classes can be classified into three subgroups: standard or Wigner-Dyson (WD) symmetry class, a chiral class derived from the standard based on chiral symmetry, and Bogoliubov-de Gennes (BdG) classes for superconducting (SC) states (illustrated in *Figure 14*).[165,166] Standard and chiral classes can also be defined with three additional symmetry classes: orthogonal (Or.), unitary (Un.), and symplectic (or spin-orbit (SO)).[165] Under certain symmetries, some phases become trivial and can be deformed into other topological phases without closing the gap and breaking any symmetry. Ten topological classes have distinct dimensional characteristics for strong gapped (insulators and superconductors) and gapless (semimetals) topological phases.[167] Gapless topological classes are generally arranged with the modified dimensionality determined from the dimension of Brillouin zone and Fermi surfaces, while the dimension for gapped topological classes is determined from the space dimension. Among the gapped topological classes, only some classes are physically realized like 1D topological insulator known as Su-Schrieffer-Heeger (SSH) state (observed in polyacetylene), 2D topological insulators (known as quantum Hall insulators (QHI) and quantum spin Hall insulators (QSHI)), 3D topological insulators (TI), 2D and 3D topological superconducting insulators (TSC), and superfluid (SF). On the other hand, the observed gapless topological phases are nodal superconductors (NSC), Weyl semimetal (WSM), and Dirac semimetal (DSM). There are four types of topological invariances used to define the topological phases: empty entry, *Z*, *2Z*, and *Z₂*. Empty entry refers to the trivial system without a topological phase, *Z* refers to a topological phase with integer topological invariance specified by the number of chiral edge states and their chirality, *2Z* refers to even integer topological invariance systems with Kramers degeneracy, and finally, *Z₂* refers to a topological system with distinct two topological phases like time-reversal invariance 2D/3D topological insulators.[165] Among the ten topological classes, eight are involved with either of the antiunitary operators and are called real due to the complex conjugate operator associated with the antiunitary operators. The remaining two are called complex and are associated with the unitary chiral symmetry operators. The nomenclature for the topological classes comes from the mathematical classification of symmetric spaces given by E. Cartan.[168] By adding dimensionality and/or symmetry, a periodicity known as Bott periodicity can be observed for ten topological classes up to d=7 for strong gapped topological phases (shown in *Figure 14* and *Table 2*). Both in the figure and table, the elements in the gray-colored cell are the topological classes associated with chiral symmetry.

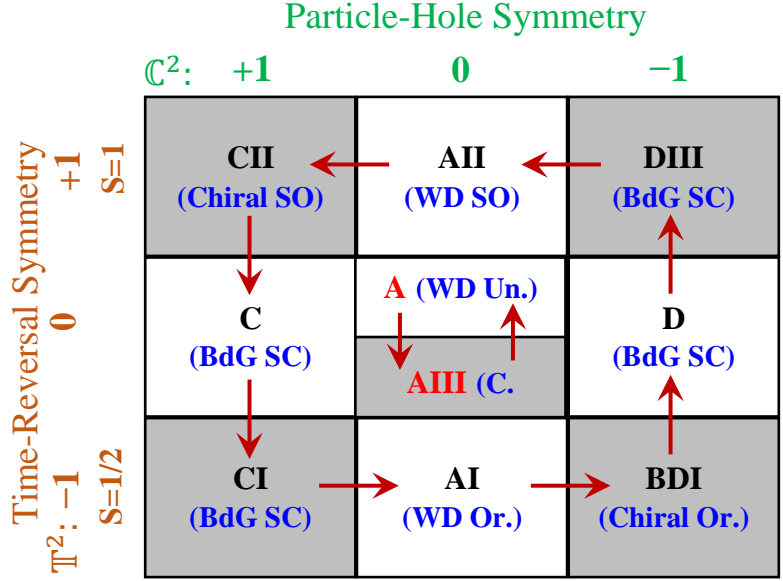

Figure 14: The Bott periodicity for ten gapped AZ topological classes based on time-reversal, particle-hole and chiral (gray entries) symmetries along with their sub-class symmetries. Red texts represent the complex classes and black texts represent real topological classes.

Table 2: Bott periodic table of ten topological classes for gapped strong topological phases (insulators and superconductors) based on global symmetries: time-reversal, particle-hole, and chiral. Table entries are arranged with respect to the corresponding space dimensions ($d_g$). Physically observed phases are written in parenthesis.

| Classes\$d_g$ | 0 | 1 | 2 | 3 | 4 | 5 | 6 | 7 | ($\mathbb{T}^2, \mathbb{C}^2, \mathbb{S}^2$) |
|---|---|---|---|---|---|---|---|---|---|
| **A** (WD) | $\mathbb{Z}$ | | $\mathbb{Z}$ (QHI) | | $\mathbb{Z}$ | | $\mathbb{Z}$ | | (0,0,0) |
| **AIII** | | $\mathbb{Z}$ (SSH) | | $\mathbb{Z}$ | | $\mathbb{Z}$ | | $\mathbb{Z}$ | (0,0,1) |
| **AI** (WD) | $\mathbb{Z}$ | | | | $2\mathbb{Z}$ | | $\mathbb{Z}_2$ | $\mathbb{Z}_2$ | (+1,0,0) |
| **BDI** | $\mathbb{Z}_2$ | $\mathbb{Z}$ | | | | $2\mathbb{Z}$ | | $\mathbb{Z}_2$ | (+1,+1,1) |
| **D** (BdG) | $\mathbb{Z}_2$ | $\mathbb{Z}_2$ (TSC) | $\mathbb{Z}$ (TSC) | | | | $2\mathbb{Z}$ | | (0,+1,0) (PHS triplet) |
| **DIII** (BdG) | | $\mathbb{Z}_2$ | $\mathbb{Z}_2$ (TSC) | $\mathbb{Z}$ (SF) | | | | $2\mathbb{Z}$ | (−1,+1,1) (PHS triplet) |
| **AII** (WD) | $2\mathbb{Z}$ | | $\mathbb{Z}_2$ (QSHI) | $\mathbb{Z}_2$ (TI) | $\mathbb{Z}$ | | | | (−1,0,0) |
| **CII** | | $2\mathbb{Z}$ | | $\mathbb{Z}_2$ (TI) | $\mathbb{Z}_2$ | $\mathbb{Z}$ | | | (−1,−1,1) |
| **C** (BdG) | | | $2\mathbb{Z}$ (TSC) | | $\mathbb{Z}_2$ | $\mathbb{Z}_2$ | $\mathbb{Z}$ | | (0,−1,0) (PHS singlet) |
| **CI** (BdG) | | | | $2\mathbb{Z}$ | | $\mathbb{Z}_2$ | $\mathbb{Z}_2$ | $\mathbb{Z}$ | (+1,−1,1) (PHS singlet) |

A similar type of tabular representation for ten AZ gapless topological classes with corresponding invariants is presented in the literature for the modified dimensionality and Fermi surface dimensionality.[155] Among the physically observed topological gapless (semimetal) phases, nodal line semimetals are class DIII, AIII, or CI, 2D Dirac graphene is class AI with rotational symmetry protection, Weyl semimetals are class A, and Dirac semimetals are class AII.[155] The inclusion of the inversion symmetry with the given global symmetries modifies the topological invariance of the ten topological classes.[163] The symmetry eigenvalues can also be used to map out insulators and semi-metals on a general level based on their demonstrated relation to K-theory.[169] Note that the ten topological classes can also be defined with respect to the defects and disorders by considering modified dimensionality, detail of which can be found in the literature.[155]

## 8. Topological Insulators and Semimetals in 2D and 3D

Topological insulators and semimetals, both quantum phases, are characterized by their distinct surface/edge states in momentum space defined by the topology of the bulk band structures and can be probed by the quantized Hall conductivity, also known as transverse magneto-conductivity. As mentioned in the earlier sections, topological phases are well defined by their corresponding topological invariants or adiabatic variants that do not change under adiabatic deformation. Topological phases are already observed in 2D and 3D materials. Quantum phases in 2D known as 2D topological insulators (TIs) like integer quantum Hall (IQH) insulators, fractional quantum Hall (FQH) insulators, and quantum spin Hall (QSH) insulators, are characterized with single-valued topological invariants, namely Chern number or Z invariant, while 3D topological phases like topological semimetals are defined by four topological invariants $(v_0;v_1,v_2,v_3)$ where $v_0$ represents strong topological invariant, and $v_i$'s ($i=1,2,3$) are weak topological invariants. Despite early observations of 2D topological states (IQH, FQH, and QSH), the physical realization of these phases remained challenging due to the requirement of semiconductor heterostructures with ultraclean interfaces at low temperature, and the Hall measurement is the only way to observe topology protected metallic states.[123] Due to the experimental barriers in 2D topological states and easier experimental realization of 3D topological phases along with distinct topology protected properties, 3D topological materials are widely studied in recent condensed matter research. Tremendous efforts on the topological material research enable the demonstration of different exotic characteristics of topological phases theoretically and experimentally. Therefore, the focus of this section is to present the different salient features of corresponding topological phases along with their prospect and roadblocks, which can provide the readers a complete overview of the drive of current research trends on topological materials.

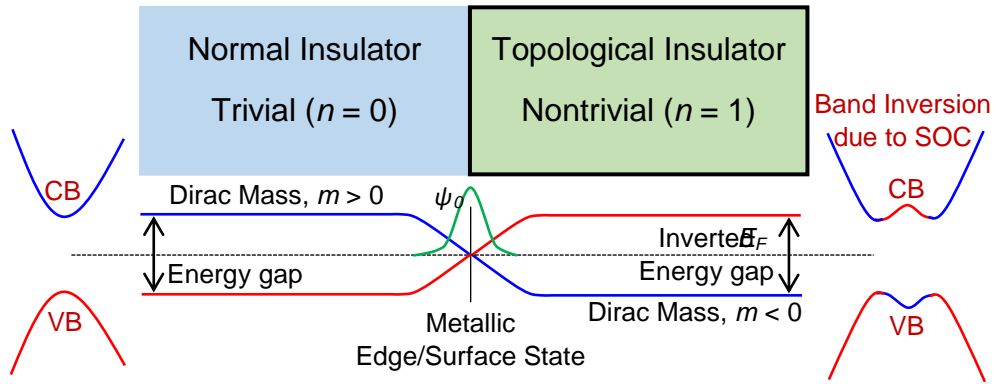

Figure 15: Illustration of topological insulating and metallic edge/surface states at the boundary of trivial topological insulator (normal insulator) and nontrivial topological insulator along with the inverted energy gap and closing of the gap at edge/surface state in energy-momentum space.

### 8.1. 2D and 3D topological Insulators

In general, topological insulator (TI) materials have an energy gap in bulk band structure opened by the spin-orbit coupling induced band inversion and a metallic state at the surface/interface where the energy gap closes at the boundary of the topological insulator (TI) and normal insulator (NI)/vacuum, as illustrated in *Figure 15*. In 2D material systems (mostly interface of heterostructures), 2D topological insulator phases like IQH, FQH, and QSH appear with spin-polarized helical edge states at low temperatures under the external magnetic field for IQH and FQH or in the presence of strong spin-orbit coupling for QSH. Helical edge states provide unique longitudinal or transverse quantized magneto-conductivity, which were observed in quantum Hall or quantum spin Hall measurement. Though spin polarization of these states was never observed experimentally, edge states in 2D topological phases have opposite spin-polarized and counter-propagating metallic states at *k* direction with an opposite sign which have characteristic 1D Dirac crossing. In IQH and FQH insulators, Dirac crossing happens within the cyclotron gap between two Landau energy levels in the magnetic Brillouin zone. QSH insulators can be imagined as a topological phase with two opposite spin-polarized integer quantum hall states under opposite magnetic fields and, therefore, are analogous to IQH and FQH insulators without the need of an external magnetic field (SOC plays the role of external field). In QSH insulators, Dirac points at Dirac crossing in 2D TIs generally appear at the TRIMs of the effective Brillouin zone. Dirac mass changes its sign from positive to negative after crossing the Dirac point from trivial to nontrivial topological phases.[24] Under crystalline symmetry preservation, Dirac points appear at high-symmetry *k* points, and theoretically, they can move freely along a line without breaking crystalline symmetry.[170] At reduced symmetry, energy dispersion relation becomes anisotropic at Dirac points.[170] Spin-orbit coupling (SOC) typically breaks crystal symmetry to open an energy gap at Dirac points, while Dirac points can only generate or annihilate in pairs of opposite chirality under no SOC.[170] The counter-propagating robust edge states do not allow backscattering and localization due to the symmetry protection, which provides dissipationless metallic states with high mobility and near-zero mass relativistic fermions. Under strong interactions, metallic states can have an electronic instability that can open an energy gap and break TRS. This symmetry defined characteristic provides unique electronic, spin, optical, and tunneling properties, which are challenging to observe in experiments due to their existence at low temperature and high magnetic field.[9] Three-dimensional generalization of 2D topological phases can be realized by 3D topological insulators, also known as $Z_2$ insulators.

3D topological insulators or $Z_2$ insulators, the 3D counterpart of 2D topological phases, can be categorized into weak topological insulators (WTIs) and strong topological insulators (STIs) based on the four $Z_2$ invariants. Among these adiabatic invariants, three invariants are related to the lattice translational symmetry and provide WTIs without robustness against disorder, while the fourth invariant gives STIs with robustness against nonmagnetic disorders. Like 2D TIs, 3D TIs (especially in STIs) have gapless 2D metallic topological surface states (TSS), which cannot be removed unless TRS is broken with bulk insulator bandgap under SOC. In TIs, TSSs contain massless Dirac fermions, which are helical, meaning the spin of the electron is perpendicular to its momentum. In 3D TIs, two-dimensional surface states become holographic, which cannot exist in the 2D system with TRS but can exist at the boundary of 3D materials. On surface states, the characteristic Fermi arc encloses an odd number of Dirac points, which provides a quantized π Berry phase to an electron after circulating the surface Fermi arc. This quantized Berry phase remains unchanged under continuous perturbations,[171] and the surface states show antilocalization with the disorder (even with a strong disorder in the absence of electron-electron interactions).[172] Under *e-e* interactions, localization can happen, and TRS can break.[173] STIs are considered as a genuine 3D topological state, which is fundamentally distinct from all 2D topological phases, which cannot be adiabatically connected to its 2D counterparts.[174] STIs have TRS protected odd number of gapless TSSs at all surfaces irrespective of the surface termination.[174] Moreover, experimental observation of 3D TIs make them unique from their 2D counterparts in several ways: a) TSSs of 3D TIs can be observed at the bare surface at room temperature without a magnetic field, b) electronic and spin properties of TSS along with topological invariants and other key features can be probed by spin and angle-resolved photoemission spectroscopy (spin-ARPES) and other experimental measurements, c) 3D TIs can be used to realize superconductivity or magnetism by doping or creating interfaces.[12,175] These experimental advantages have already helped to identify many prospective quantum materials with novel topological phases such as topological crystalline insulators (TCIs) where TSSs are protected by crystalline symmetry instead of TRS,[176,177] topological Kondo insulators (TKIs) where TSSs of strong electron correlation systems exist in bulk Kondo gap instead of bulk Bloch gap,[178,179] topological semimetals where TSSs exist with no bulk energy gap,[180,181] superconducting TIs,[182,183] and magnetic TIs.[184,185] Some of the later type topological phases are discussed in Section 2.

In materials having 3D STI phases with $v_0 = 1$, strong SOC acts as an internal magnetic field and causes band inversion at an odd number of high symmetry points in bulk Brillouin zone (BZ), which gives an odd number of surface states. The quantized Berry phase in STIs ensures the presence of spin-polarized surface states showing partner switching between a pair of TRIMs in BZ, which means the spin-down band is connected to and degenerate with different spin-up bands at different TRIMs.[173] Both surface and bulk states of 3D STIs can be controlled by doping without destroying topological nature like Na doping into $Bi_2Se_3$.[186] The exotic features of near massless, backscattering-less, high mobility, spin-polarized Dirac fermions in surface states make the TIs prospective for unique device applications, which will be discussed later in this manuscript.

The most straightforward approach to understanding the mechanism of WTI and STI is to stack several uncoupled 2D QSH states to form a 3D bulk. Therefore, a 3D TI can be obtained by stacking N copies of identical 2D QSH states along the out-of-plane direction (e.g., *z*-direction). In *k*-space, it means that the 2D Brillouin zone will gain periodicity in the $k_z$ direction and becomes a cube representing the 3D BZ. In the 3D BZ, the Kramers' points can be sorted by their origins from the 2D BZ before stacking. For instance, both the Γ and the Z points in the 3D BZ results from the Γ point in the 2D BZ (see Figure 16(a), (b), and (e)). For a WTI, we assume that the physical coupling between these 2D QSH slices is extremely weak. This physical coupling is a very weak van der Waals interaction in real materials such as in $Bi_{14}Rh_3I_9$ compound.[187] Before we turn on the coupling between these slices, each slice is a QSH that features two counter-propagating edge modes at its edge. To better understand, we fix the chemical potential at the crossing point of the edge modes and investigate the $k_x$-$k_z$ plane where all the QSH edges are stacked to form a surface (Figure 16 (b)). Therefore, all the 1D edge states from the QSH slices are stacked, and become a 2D surface state. As result, the surface state is a straight line that goes from Z-Γ-Z. Now if we turn on the inter-slice coupling (i.e., hybridization). This means that energy dispersion along the $k_z$ direction changes for some of the QSH slices, opening band gap (e.g., QSH slices between

-π < $k_z$ < or 0 < $k_z$ < -π). However, at and only at the Γ and the Z points, the edge state crossings are preserved because these two points are the Kramers' points. Consequently, the surface state at Fermi energy ($E_F$) of the $k_x$-$k_z$ plane becomes two dots (red color) at the Γ and the Z points (Figure 16 (d)). If the $E_F$ is shifted slightly away from the crossings, then the two Fermi dots will evolve into two circles that enclose the Γ and the Z points, respectively. It is noteworthy that at the top $k_x$-$k_y$ surface, no protected surface state is expected because both Γ and Z points project onto the $\overline{\Gamma}$ point (Γ point of the 2D BZ) of the top $k_x$-$k_y$ surface. Therefore, a WTI features protected 2D surface states only at certain crystalline surface terminations. From the topology point of view, the 3D WTI state is essentially equivalent to a number of independent 2D QSH slices. This is because there involves no band inversion by decomposing a WTI into uncoupled 2D QSH slices.

Unlike WTI, strong TI cannot be adiabatically reduced into a bunch of stacked QSH states. Moreover, a strong TI displays protected surface states at all surfaces, irrespective of the termination choice. To investigate this feature, we examine the strong coupling scenario (Figure 16 (e)). In this case, the inter-slice coupling is strong enough and affects the electronic structure in a way so that the band inversion at the Z point is removed. Therefore, in this case, there is only one band inversion at the Γ point throughout the 3D BZ. It is important to note that the 3D STI state is topologically distinct from both the 3D WTI and the NI (vacuum) states. This is because i) it is impossible to reduce the STI to uncoupled 2D QSH slices without going through a band inversion at the Z point; ii) it is impossible to change an STI to a normal band insulator without going through a band inversion at the Γ point. The possible existence of the protected surface states in the STI state is shown in Figure 16 (f). It is noteworthy that there is only one band inversion at the Γ point for the STI state, and that at any surface termination, the Γ point projects onto the surface 2D BZ center $\overline{\Gamma}$ point (such as Figure 13(c)). In general, an STI should always possess an odd number of band inversions (for a weak TI, the number of band inversions is even) at the Kramers' points in a 3D BZ, but the exact number of band inversions (e.g., 1, 3, 5, etc.) depends on the exact form of the inter-slice coupling as well as the properties of the 2D QSH state to start with. For example, one can imagine having a QSH state with a band inversion at the 2D BZ corner M point, not the Γ point.

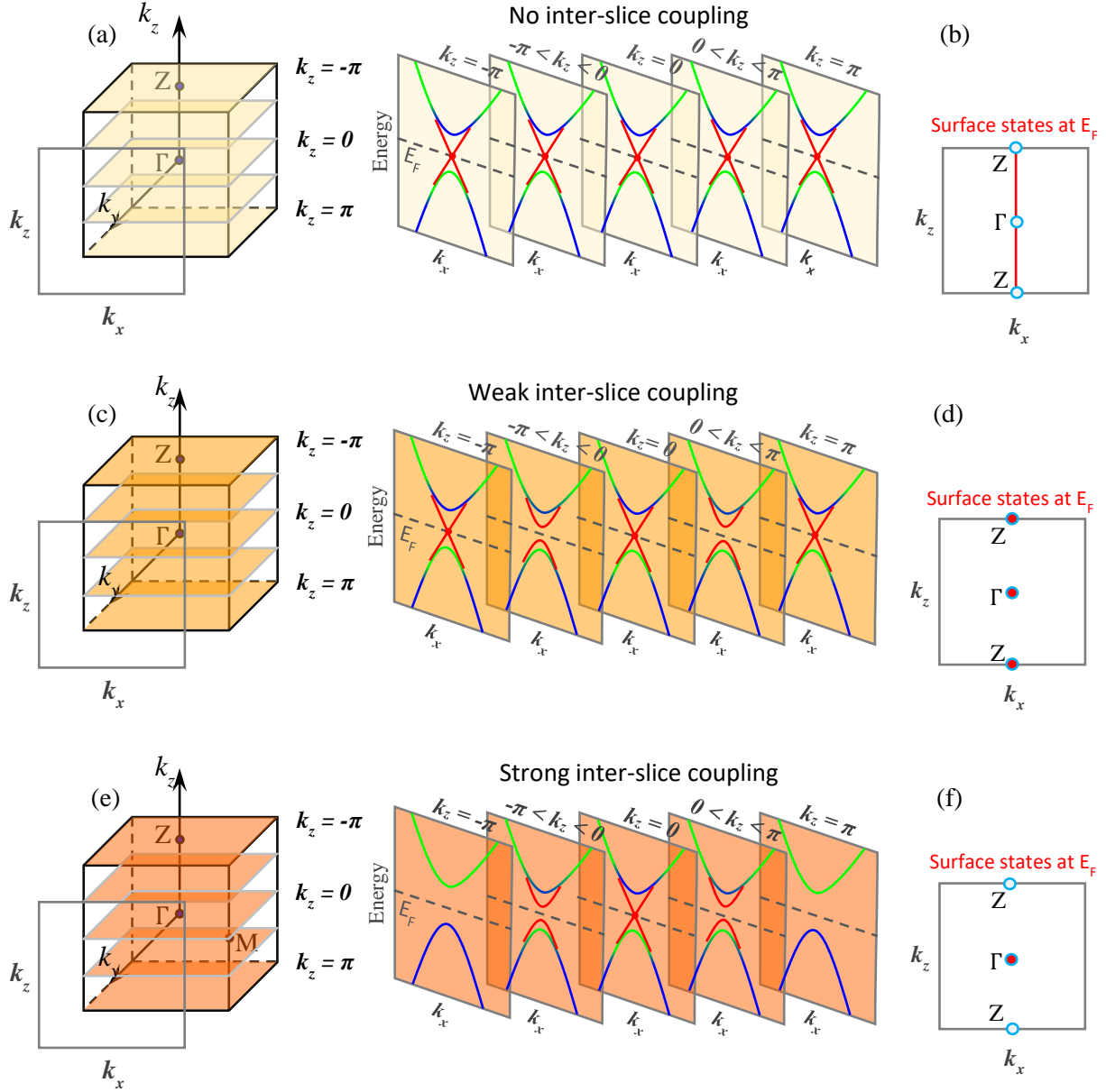

Figure 16: (a) Stacking N copies of such quantum spin Hall slices along the out-of-plane direction form a bulk material. In k-space, it means that the 2D BZ will gain periodicity in the $k_z$-direction and becomes a cube representing the 3D BZ. Five light-yellow slices are chosen for detailed studies. The edge projected electronic structure for the light-yellow slices shown in panel. Without any inter-slice coupling, these slices are identical and independent. (b) The surface states (nontrivial surface states) for the $k_x$-$k_z$ surface projection are a straight line (red line). (c) A bulk BZ of the weak TI; five yellow slices are chosen for detailed studies. Turning on the inter-slice coupling, the edge states can hybridize and open out gaps, except at the $\Gamma$ and the Z points due to the protection of the time-reversal symmetry. (d) Therefore, the surface state at energy Fermi is described by two dots at the $\Gamma$ and the Z points. (e) A bulk BZ of the strong TI. The orange slices are chosen for detailed studies. After turning on the inter-slice coupling, in the case of a strong TI, only the band inversion at the $\Gamma$ point remains whereas the band inversion at the Z point is removed. For $k$-space region near the $\Gamma$ point where the conduction and valence bands are inverted, the edge states can hybridize and open out gaps, except at the $\Gamma$ point itself due to the protection of the time-reversal symmetry. (f) Therefore, the surface state at energy Fermi is described by one dot (red point) at the $\Gamma$ point.

### 8.2. Dirac semimetals

Since 1928, it is well-known from the Dirac relativistic equation of electron ($H = cp.\alpha + (mc^2)\beta$, where, $c$ is the speed of light and $\alpha$, and $\beta$ are the coefficient matrices) that the energy dispersion relation becomes linear in the ultra-relativistic regime (kinetic energy dominates over mass, $cp \gg mc^2$).[170] In this regime with near-zero mass and proper chirality, Dirac fermion decouples into two Weyl fermions with certain chirality and helicity, as shown in *Figure 17*, which fix at their momentum positions.[170] Like TIs, Weyl points only can appear when two non-degenerate bands cross at a single point near the Fermi level. The first Dirac fermion in a solid-state band structure was observed in graphene,[188] also known as a 2D Dirac semimetal (DSM). Typically, topological surface states in Dirac semimetals exist without any bulk energy gap. TSSs of 2D Dirac semimetals are protected by discrete lattice symmetry, but the realization of Dirac semimetal in 3D requires extra symmetry protection, as $m\sigma_z$, a mass term in the electron Hamiltonian will open an energy gap at the presence of SOC (this mass term is absence in 2D Dirac semimetals). Generally, the 3D Dirac semimetal phase appears during normal insulator (NI) to topological insulator (TI) phase conversion or as an intrinsic phase at the presence of extra symmetry protection, mostly by rotational symmetry. 3D Dirac semimetal phases can also appear in the materials having both TRS and inversion symmetry preserved with extra space group protection, which eventually gives two Weyl points with an opposite chirality at a single Dirac point, as shown in *Figure 17*, and makes it four-fold degenerate.[189] Therefore, these materials with strong SOC can become Weyl semimetals upon breaking either TRS or inversion symmetry. Other derivative topological phases from DSM can be Dirac nodal line and topological superconductor (*Figure 17*). High mobility DSM phases are unstable in stoichiometric single-crystalline systems due to the subtle merging of two low energy Weyl nodes under the protection of multiple symmetries, which made the observation of DSM in real materials challenging.[9] Moreover, DSM phases appear at the critical point of the NI-TI phase conversion boundary; hence, a fine-tuning of either chemical composition or strain is needed to observe DSM in the relevant materials, which often creates disorder in the system.[9] Some prospective materials have a bandgap at Dirac point with an additional Fermi surface.[9] Nevertheless, DSM phases can be obtained in two ways: band inversion mechanism where Dirac nodes appear from the lattice symmetries (mostly rotational symmetry) induced accidental band crossing into an inverted band structure of a topological insulator, and symmetry-enforced Dirac nodes with or without SOC.[189] DSMs have unique quantum properties, which are discussed in the following subsections.

### 8.2.1. Properties of Dirac Semimetals

The key feature of DSM, which was obtained from the theoretical calculations as well as observed in the experiment, is two surface Fermi arcs connecting a pair of Dirac points on the surface, as shown in *Figure 17*. Due to the four-fold degeneracy and zero Chern number of Dirac points, the Fermi arc in DSMs is subtle, not topologically protected, and is considered as a combination of two Fermi arcs of Weyl semimetals.[189] The nature of the Fermi arc of DMSs also depends on the position of the edge states on energy contours.[189] Similar to the Fermi arc, a chiral anomaly at the Dirac node of DSMs is a subtle topological property due to the coexistence of two opposite chiralities at a similar momentum position, which is protected by rotational or reflection symmetries. Besides the unique topological characteristics, DSMs have striking electrical transport features like ultrahigh mobility and large magnetoresistance.[190,191] A very high mobility of 9×10$^6$ cm$^2$/Vs was reported for Cd$_3$As$_2$ at 5K along with 0.25 mm mean-free path at 6K.[190,192] The ultrahigh mobility in DSM is mainly due to the backscattering-less energy states for a carrier at zero magnetic fields, which can provide up to $10^4$ times longer transport lifetime that the quantum lifetime.[190]

DSMs can be the source for driving different topological states like topological insulators, topological nodal lines, and Weyl semimetals (*Figure 17*),[388] as DSM states require multiple symmetry protection, and breaking of each symmetry can result in creating other topological phases. It was also reported that pressure and doping tuning into DSM materials could create a superconducting gap in the bulk band structure and support Majorana fermions; hence, can convert DSMs into topological superconducting insulators.[189]

### 8.2.2. Magnetoresistance of Dirac Semimetals

Magnetoresistance (MR) is defined as the proportional change of the electrical resistance by an external magnetic field. The MR can be further classified as transverse MR and longitudinal MR according to the angle between the B and E fields. DSMs exhibit a large MR in the presence of an external magnetic field. This giant MR in DMSs is caused by the break of protected topological edge or surface states and the change in the Fermi surface under the applied magnetic field.[190] A large linear MR is also observed in low mobility DSMs at even room temperature, which can arise from the magnetic field-induced shifting of the Weyl Fermi surfaces.[192] Similar types of linear MR can be exhibited by gapless semiconductors with linear energy dispersion where all electrons stay at the lowest Landau level.[193] MR of DSMs can also demonstrate Shubnikov-de Hass (SdH) and Zeeman splitting up to a quantum limit at a specific magnetic field, and beyond the limit, MR becomes linear, which indicates the occupation of the lowest Landau level.[192] Many important transport parameters of materials can be extracted by analyzing the SdH oscillations of the MR, such as carrier mobility, carrier density, transport time, and non-zero Berry phases,[192] where the nontrivial Berry phase is an indicator of the existence of Dirac fermions.[192] The Linear variation of MR can also be demonstrated by disordered material systems with inhomogeneous mobility distribution.[194] SdH oscillations in MR measurement can also be used to probe the Fermi surface through systemic electrical magnetotransport measurements.[191]

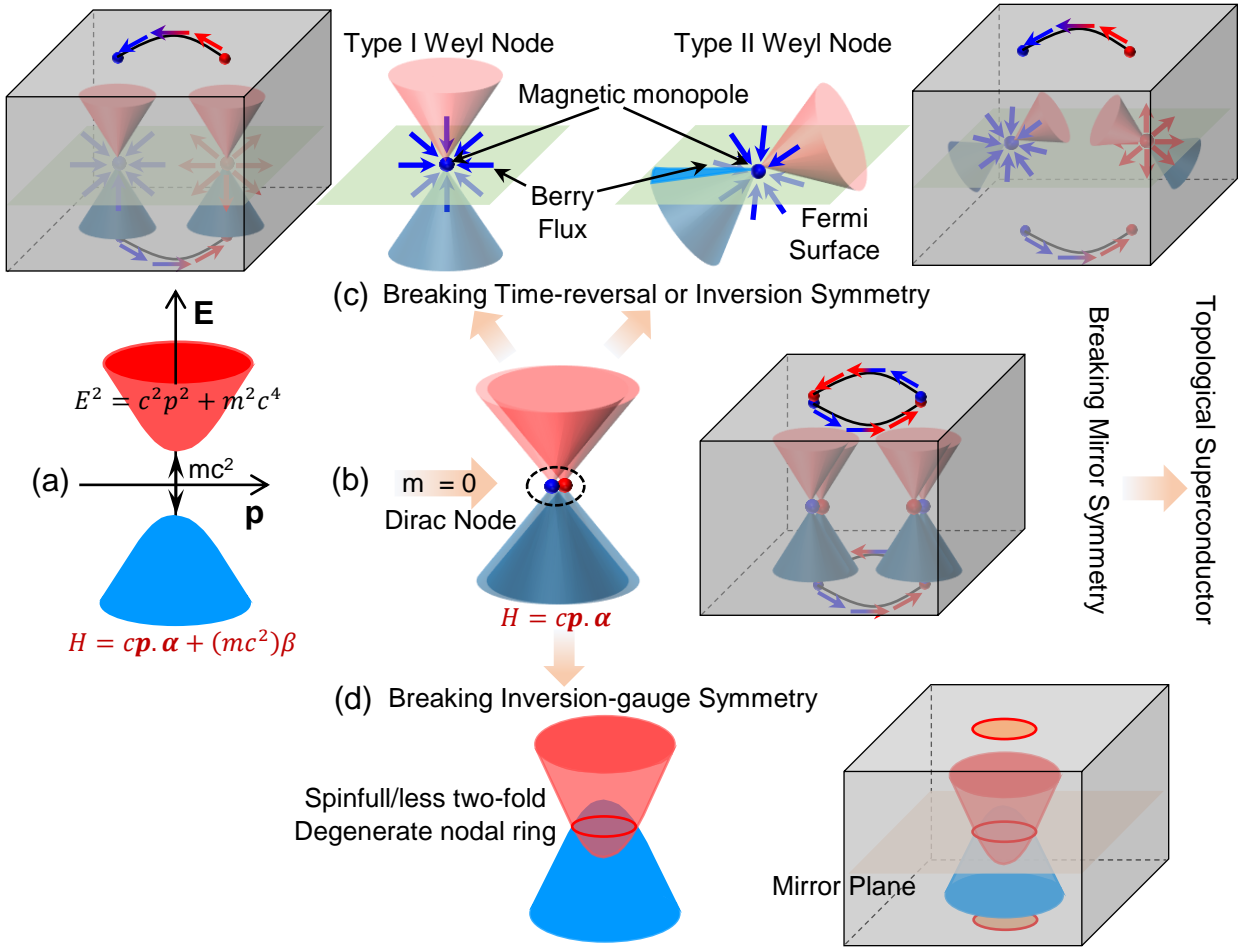

Figure 17: (a) Band Structure of a free Dirac Particle with an energy gap associated with its mass, (b) Dirac particle shows relativistic nature at massless condition where one Dirac node generates two (degenerate) Weyl nodes of opposite chirality, (c) Dirac node can split into two types of Weyl node pairs (type I and type II Weyl nodes) under the breaking of TRS or inversion symmetry, which creates Weyl fermions with a characteristic magnetic monopole and Berry flux, and (d) Formation of Dirac nodal ring due to the inversion-gause symmetry breaking. Bulk and surface states along with their characteristic features are shown for all topological phases. Mirror symmetry breaking leads to a topological superconductor.

### 8.3. Weyl Semimetals

The theoretical prediction of massless Weyl fermions was made in 1929 by Hermann Weyl. They were predicted to appear as pairs with opposite chirality. Weyl fermions in quantum matters appear in Weyl nodes, which show a nontrivial topological effect on Berry curvature and act as magnetic monopoles with Berry flux (*Figure 17*). Weyl semimetals (WSMs) were predicted as the materials to realize Weyl fermions in 2011[76] and experimentally demonstrated in 2015.[19,20] In general, Weyl nodes, the band touching point between conduction and valence band near Fermi level with linear dispersion relation, appear at some region of the Brillouin zone, which is determined and protected by the symmetry. As mentioned, Weyl semimetals can be obtained from DSMs by breaking inversion or TRS symmetry, which can provide the band degeneracy at TRIMs. The number of Weyl points in WSMs is determined by symmetry. In TRS WSMs, a Weyl point and its time reversal partner have the same chirality; hence, two more Weyl points with opposite chirality need to exist to satisfy the Fermion doubling theorem. Therefore, WSMs with TRS have multiple of four Weyl points. On the other hand, inversion symmetry protected WSMs have two Weyl points. WSMs can be three types depending on the tilting of Weyl cone due to the effective velocity parameter, $v$, in Weyl Hamiltonian: (i) small $v$ gives Type I WSMs having conventional Weyl points with point-like Fermi surface, as shown in *Figure 17*, where crystal field anisotropy into band dispersion exists near Weyl points, (ii) large $v$ gives tilting in Weyl cone and hence gives Type II WSMs[195] or structured WSMs having a touching point between electron and hole pockets with open constant energy surfaces (*Figure 17*), and (iii) the quadratic term gives the quadratic tilting in Weyl cone and gives Type III WSM[428] which has two contacted electron or hole pockets (*Figure 1*). All types of WSMs have surface Fermi arcs, while Type II WSMs have unique surface states called track states,[189] and Type III WSMs show long Fermi arcs with multi-fold helicoidal surface states.[428] In the literature, another type of WSM phase has been reported, known as critical-type WSM, which has a nodal line state on the Fermi surface.[196,197] Critical-type WSM has also been reported in some literature as type-III WSM, which should not be confused with the illustrated type-III in Figure 1.[198] Generally, a critical-type is formed by a flat band and a dispersive band; it has the critical band dispersion between Type I and Type II. Sometimes, it becomes challenging to find out Weyl nodes in the Brillouin zone as well as the number of Weyl points, as Weyl points usually do not appear at high-symmetry lines and planes, and this problem can be resolved by obtaining band structure without SOC.[72] Like Dirac fermions in DSMs, Weyl fermions in WSMs exhibit unique topological properties, summarized in the following subsections.

### 8.3.1. Properties of Weyl Semimetals

A magnetic monopole is one of the unique features of Weyl fermions. The topology of the Weyl point in the Fermi surface is related to the Berry curvature, which is considered as an effective magnetic field in the momentum space. Weyl nodes on the Fermi surface act as magnetic monopoles: a source and sink for the Berry curvature. Magnetic charge of the Weyl node can be obtained from the integral over Fermi surface enclosing the Weyl node and are known as chirality.[72] Weyl points on the Fermi surface also form the Fermi arc that connects the surface projections of two Weyl nodes in WSMs, as shown in *Figure 17*. The existence of Fermi arcs in the Fermi surface of surface states is ensured by the monopole feature of the Weyl fermions. Topologically protected Fermi arcs on opposite surfaces are connected through the bulk Weyl points. Due to these unique transport properties, WSMs are receiving more interest in many application fields. The shape and energy dispersion characteristics of Fermi arcs in WSMs are usually defined by the surface properties like the terminating ions on the surface, while the length of Fermi arcs and the separation between Weyl points depend on the strength of SOC.[72] It is reported that the electron topology of Weyl nodes can be reflected in phonon spectra as Kohn anomaly.[199] Unlike DSMs, WSMs exhibit strong chiral magnetotransport, as the linear dispersion and nontrivial Berry phase in the band structure, along with strong quantum oscillations, help to determine large magnetoresistance (MR), high mobility, chiral anomaly, and Fermi surface. Typically, angle-dependent quantum oscillations like SdH and de Hass-van Alphen (dHvA) oscillations are utilized to reconstruct the Fermi surface, as the origin of these magnetic field-dependent quantum oscillations is related to the different types of Fermi surface pockets.[72] Like DSMs, WSMs show high mobility, which is arising from the chiral and near massless Weyl fermions. The carrier mobility in WSMs is sensitive to sample quality, which can modify the Fermi level along with the scattering processes.

A charge current flow is supposed to exist under parallel electric and magnetic fields between different chiral Weyl modes separated in *k*-space. Chiral anomaly is caused by the unbalance in chiral charges in this process, and chiral anomaly leads to the non-conservation of electric charge; hence, an axial current flow. Chiral anomaly in WSMs can give better conductivity and introduce negative longitudinal MR, though negative MR can still be observed in a system with ill-defined chiral quasiparticles at the Fermi level.[200] Therefore, further details are discussed in the following subsection to investigate two possible causes for negative MR as shown in *Figure 18*: (i) inhomogeneous current distribution (current jetting) and (ii) chiral anomaly. Type I WSMs can support chiral anomaly transport in all directions, while chiral anomaly disappears in type II WSMs due to the gapped Landau-levels without chiral zero-mode. Type II WSMs can only support chiral anomaly when the magnetic field is normal to a momentum plane.[72] Type II WSMs also exhibit large transverse MR like type I WSMs. Weyl semimetals demonstrate anomalous Hall and MR features in the ultraquantum regime above the quantum limit (QL), coming from the small Fermi surface of WSMs and the instability of Weyl fermions at a high magnetic field.[201,202] Unlike DSMs, WSMs show additional oscillation structures in MR and Hall above QL at low temperatures. WSMs also demonstrate the chiral magnetic effect (CME) similar to the chiral anomaly, anomalous Hall effect, and Axion electrodynamics.[189] CME effect in a WSM causes a non-vanishing current under an applied magnetic field in the direction of connecting Weyl nodes when Weyl nodes have an effective chemical potential difference.[189] Anomalous Hall effect is only demonstrated by WSMs with TRS breaking, which is also sensitive to crystal symmetry and lattice strain.[189]

### 8.3.2. Magnetoresistance in Weyl Semimetals

The magnetoresistance (MR) of Weyl semimetals has been intensively studied in recent years. Specific characteristics like the chiral anomaly, Fermi arc, and the ultrahigh mobility of the Weyl fermions all influence the magnetoresistance of Weyl semimetals (WSMs). In WSMs, Weyl electrons generally coexist with normal electrons near the Fermi surface, and the number of electrons near Weyl points can be changed in the presence of electric and magnetic fields. The enhancement of the longitudinal conductivity along a weak magnetic field can be represented as:[203]

$$\sigma_l = \frac{e^2 v^3 (eB)^2 \tau}{4\pi^2 \hbar c^2 E_F^2} \qquad (55)$$

Here, *v* is the Fermi velocity near the Weyl node, $\tau$ is the internode scattering time, and $E_F$ represents the Fermi energy from the Weyl node energy. Equivalently, it will bring a reduction to the longitudinal magnetoresistance. Direct observations of this negative longitudinal MR showed that this phenomenon is sensitive to the misalignment between *E* and *B* fields since the perpendicular component of *B* can easily create a large positive transverse MR. Many effects will change the behavior of the longitudinal MR of a Weyl semimetal. One of the most appealing effects is the chiral anomaly, which will induce a negative longitudinal MR in the Weyl semimetal. Negative MR has a sharp dependency on the angle between electric and magnetic fields and has a relation with the Fermi level. Any small changes in the Fermi energy (mostly induced by doping or defects) or even ideal stoichiometric compensated crystals can change the Fermi surface topology significantly.[192] It is essential to have complete information on Fermi energy and Fermi surface topology to relate the negative MR with chiral anomaly transport. Negative MR is also found to be independent of the direction of the field with respect to the crystalline axis.[192] It can be generalized that negative MR in topological materials under parallel electric and magnetic fields is caused by the Berry curvature.[192] WSMs also exhibit large MR with SdH oscillations like DSMs.

As discussed before, the Weyl fermions with Chern number C = 1 at its Weyl points have a linear dispersion relation. If the Fermi surface is close to the Weyl points, these Weyl fermions, with low effective mass and high mobility, will contribute to the conduction. These Weyl fermions are sensitive to the external magnetic field since the $B$ field will effectively reduce the mobility due to the low effective mass. This effect has been observed in all the materials of the TaAs family, which has been proven to be Weyl semimetals. The NbP is reported to have a mobility of μ = 5 × $10^6$ $cm^2V^{-2}s^{-1}$ at 1.85K.[204] The corresponding MR is around 8.6×$10^5$ %. Another example is TaAs; it has high mobility around μ = 5×$10^5$ $cm^2V^{-2}s^{-1}$ at 2K.[205] Usually, SdH oscillation, dHvA oscillation, and Hall coefficient measurement are applied to determine the physical parameters that are related to the carrier mobility. In the following sections, we will discuss several different effects and their impacts on the MR of Weyl semimetals.

### 8.3.3. Weak Localization and Weak Antilocalization

In the Fermi liquid theory, two effects can change the longitudinal MR, for instance, weak localization (WL) and weak anti-localization (WAL).[192] These two effects originate from the phase difference of an electron wave function traveling along a specific closed path in two different directions. Neglecting the spin-orbit coupling (SOC), the phase change of the wave function along these two directions are the same, which will give an increased probability of trapping the electron in the closed-loop, hence a lower conductivity. In the presence of the SOC effect, the phase difference between these two directions differs, which will bring in a destructive interference; hence, a higher conductivity. With perpendicular magnetic field $B$, both WL and WAL are reduced.[192]

A Weyl semimetal is a symplectic system, having either inversion-symmetry breaking or time-reversal symmetry breaking. The symplectic system is predicted to have weak antilocalization.[192] However, the electron-electron scattering effect gives rise to the weak localization. The overall behavior of the longitudinal MR of the Weyl semimetal can be described as follows. When the external $B$ field is around 0, the WAL effect will dominate and gives a $-\sqrt{B}$ law to the magnetoconductivity. When $B$ is big enough, the weak localization will win the competition, thus giving a $B^2$ relationship to the magnetoconductivity. In a simple two Weyl nodes example, the WAL/WL induced longitudinal conductivity change can be represented as:[206]

$$\delta\sigma_l = C_1 \frac{B^{5/2}}{B_c^2 + B^2} + C_1 \frac{B_c^2 B^2}{B_c^2 + B^2} \qquad (56)$$

Here, $C_1$ and $C_2$ are fitting parameters, which are negative for WAL and positive for WL.[206] $B_c$ is a critical field, which is dependent on the temperature and the type of scattering. In the discussion of the temperature dependence of the longitudinal MR in the above model, the following equation is proposed:[206]

$$\delta\sigma(T) = c_{ee}T^{1/2} - c_{qi}T^{p/2} \qquad (57)$$

This equation describes the competence between the quantum interference induced WAL and scattering induced WL. Here, $c_{ee}$ and $c_{qi}$ are positive parameters related to the correction from quantum interference and electron-electron interaction and disorder scattering, respectively. The parameter $p$ depends on the decoherence mechanism and dimensionality. Other different models are also proposed to fit the longitudinal MR.[207] In ref. 207, the chiral anomaly is taken as a correction of the WAL in effect when discussing the longitudinal MR. Also, the team used the two carriers' model to fit the experimental data of the longitudinal MR when considering the temperature dependence. Therefore, perhaps the theory of this effect needs more careful discussion.

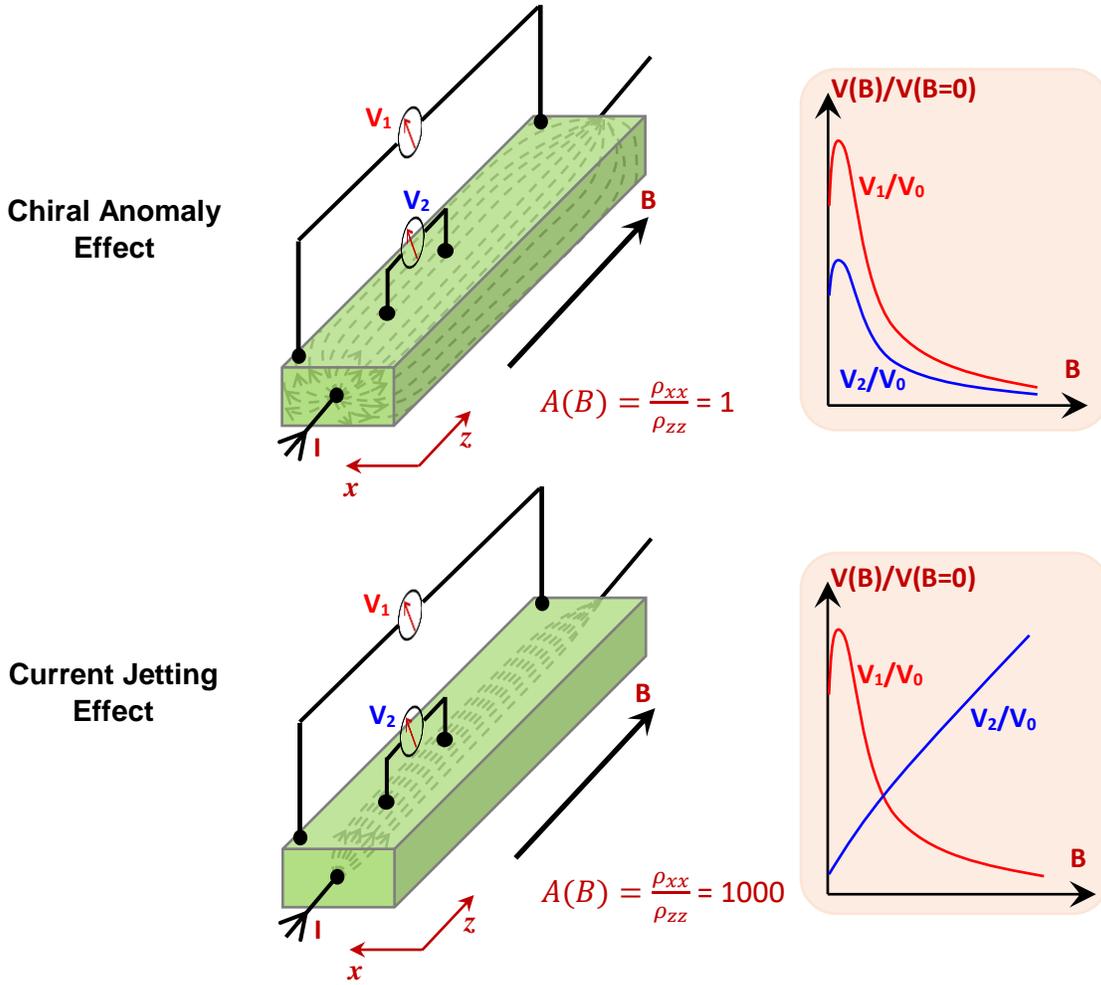

Figure 18: The magnetic field induced inhomogeneous current distributions under chiral anomaly and current jetting effect for different resistivity tensor anisotropies $A = \rho_{xx}/\rho_{zz}$. $\rho_{xx}$ is the resistivity tensor component perpendicular to the magnetic field, and $\rho_{zz}$ is the component along the field.

### 8.3.4. Current Jetting

The transport properties of typical Weyl semimetals are too complicated to be characterized by a single formula due to different scattering mechanisms and different Fermi pocket structures. Another factor that even makes it harder is the *current jetting* effect in the high mobility semimetals.[204] Without a uniform current injection to the sample, the resistivity tensor perpendicular to the direction of the external $B$ field will get dramatically enhanced, whereas the current component in the direction of the $B$ field will stay a relatively low value. This effect is called the current jetting, which will cause an inhomogeneous current distribution (see *Figure 18*). This effect, if not treated carefully, could cause a false measurement of the longitudinal MR.[204] *Figure 18* illustrates the measurement of longitudinal resistances at different voltage probe pairs under chiral anomaly and current jetting effect. As shown, both voltage $V_1$ and $V_2$ generally drop with the increase in *B*-field for chiral anomaly effects, whereas $V_1$ drops and $V_2$ increases with the increase in *B* due to the current jetting effect. From the figure, we can see the point-like current injection will bring a problem if the voltage detecting contacts are not at the same points. For instance, in the second subfigure, if the detecting contacts are placed at the edges, there will be no voltage difference, which will falsely give a 0 longitudinal resistance. Here, the magnetic field induced resistance anisotropy, $A = \rho_{xx}/\rho_{zz}$, plays an important role to create the current jetting, where current flow and *B*-field are in *z* direction. TaAs family materials were studied for the current jetting induced longitudinal MR with different setups of contacts, and the negative longitudinal MR was explained even in the absence of well-defined chiral nodes due to the Fermi pockets covering both Weyl nodes.[204]

### 8.4. Other Topological Semimetals

Topological nodal line semimetals (TNLSMs) are the other type of topological semimetals which have a line of band crossing instead of point-like band crossing in DSMs and WSMs (*Figure 17*). Besides time-reversal and/or inversion symmetries, some crystal symmetries like mirror symmetry and glide symmetry define the types of nodal line semimetal phases (*Figure 1*). Typically, both mirror symmetries protected TNLSMs and screw rotation protected double nodal line semimetals are defined with Z topological invariants while time-reversal, inversion, and spin rotation symmetries protected nodal semimetals are defined with $Z_2$ invariants.[208] Conversion into different topological phases like topological semimetals, topological insulators, and other TNLSMs can be achieved by breaking the protecting symmetry of the TNLSM phase. Therefore, TNLSMs can provide the evolvement into new topological phases by breaking different types of symmetries. As TNLSMs generally do not have topology protected boundary states, experimental observation is one of the major challenges of TNLSMs.[209] ARPES is utilized to determine nodal band structure with limited resolution, and quantum oscillation measurements are performed to define the Fermi surface map and Berry phase.[208] Optical and magneto-optical measurements are also performed to investigate nontrivial bulk response, surface states, and Berry curvature effects.[210] The predicted and physically realized material system for TNLSMs is a tuned superlattice of the normal insulator and a topological insulator with TRS breaking.[208] Despite difficulties in the observation of the nodal ring in Dirac NLSMs, the quantized π Berry phase guarantees the stability of the nodal ring.[211] Dirac NLSMs exhibit parity anomaly, which can cause an electrical field-induced anomalous transverse current due to a small inversion breaking.[211]

### 9. Materials for realizing topological phases

In this section, we summarize the different materials that host different topological phases. Based on the fundamental discussion of the topological phases, materials with strong spin-orbit coupling (SOC) are the most probable candidates for observing topological phases that can be satisfied with materials with heavier elements. Besides SOC, band inversion at Kramers' points is another key ingredient for a topological material, which can be determined by the crystal structures. In *Table 4*, we summarized the state-of-the-art topological materials along with their topological classes, dimensionality, experimental methods to observe topological phases, and the presence of an external field or SOC. General discussions on the main topological material classes are given in the following subsections. Several material databases have been developed in recent years that are informative and useful for researchers in this field. Some material databases are given in the refs. [212-216].

### 9.1. $Z_2$ Topological Insulators

Without breaking the time-reversal symmetry, the spin-orbital coupling (SOC) in some materials is strong enough to induce a band inversion, which has been a typical reason for the nontrivial topological state in the materials known as $Z_2$ topological insulators.[12,13,24] Due to the topology change on the interface, topological insulators have metallic surface states. $Bi_{0.9}Sb_{0.1}$, $Bi_2Se_3$, and $Bi_2Te_3$ are three of the earliest 3-D topological insulators of this kind. For $Bi_{0.9}Sb_{0.1}$, the unit cell is rhombohedral. In the L points of its BZ, the band inversion occurs.[127] According to the experimental results from its surface states, this material is proven to be a TI. $Bi_2Se_3$ family is another intensively studied TIs.[116,217] Experimental results have uncovered the existence of Dirac fermion in its surface BZ,[84] where the backscattering events are forbidden. This mechanism enhances the transport properties. By changing doping strategies, the thermoelectric, magnetic, and transport properties can be enhanced in these materials.[182,218] In 2010, ternary thallium-based semimetal chalcogenides $TlVVI_2$ were theoretically predicted as another family of topological insulators as well.[219] $TlBiTe_2$ has been predicted to also be a superconductor by proper doping. As attested by the follow-up experimental results, $TlBiTe_2$ turned out to be a semimetal with a Dirac cone in its BZ as well as a candidate for a topological superconductor.[220] Pb-based chalcogenide series are also theoretically predicted to be topological insulators. By checking the existence of topological phase transition versus the strength of the SOC coupling and the lattice constants, $PbBi_2Se_4$, $PbSb_2Te_4$, and $Pb_2Sb_2Te_5$ are predicted to be topological insulators.[221] An unoccupied Dirac cone has been observed in $PbSb_2Te_4$. The first predicted TCI is the PbSnTe compound. In 2012, materials from this family such as SnTe, $Pb_{0.77}Sn_{0.23}Se$, and $Pb_{0.6}Sn_{0.4}Te$ were tested by experiments and showed having Dirac cone surface states.[111,112,176] The mirror symmetry in SnTe induced the existence of band inversions in all 4 L points, indicating the Chern number to be $n_M = 2$. By properly choosing the surface orientation, one could see a double Dirac cone as 2 L point projections are on top of each other. PbSe monolayer, as a 2-dimensional TCI, has also been predicted.[222]

Recently, many other candidates have been predicted to be topological insulators. $TaSe_3$ was predicted to have many topological phases, including weak TI, TI, and Dirac semimetal.[223] Also, this material was known as a superconductor for a long period; therefore, it could potentially be a promising platform for the study of the topological phase transitions. If we are not restricting ourselves in $Z_2$ symmetry, higher-order topological insulators (HOTIs) can be found. The topological properties of HOTIs are protected by symmetries that involve spatial transformations, possibly augmented by time-reversal symmetry.[224] They thus generalize TCIs, which have been encompassed in a recent exhaustive classification of TIs in the study of Bradlyn *et al.*,[56] and H. C. Po *et al.*[225]

### 9.1.1. Higher-order topological insulators

As discussed in previous sections, symmetries play essential roles in the classifications of topological phases. Based on the absence or presence of time reversal, particle-hole, or chiral symmetry, insulators and superconductors have been classified under the so-called 10-fold way (Bott periodic table in Sec. 7.5). In addition to these internal symmetries, the topological classification of band structures has also been extended to include crystalline symmetries. Due to the vast array of crystal symmetries encapsulated by the 230 crystalline space groups, a large number of topological crystalline phases have been proposed, such as mirror Chern insulators,[226] quantized electric multipole insulators,[227] nodal chain metals,[228] and hourglass fermions.[229] High-order topological insulators (HOTIs) are also a subset of these topological crystalline phases.[224,230,231] Recent theoretical advances in these HOTIs have greatly transformed the prospect of material discovery. To identify and classify these materials, concepts such as symmetry-based indicator groups and higher-order topological indicators have been developed.[225,232]

For a band insulator of spinful electrons that are symmetric under a space group (SG) and TRS (class AII in Bott periodic table), topological indices corresponding to weak and strong TI phases, protected by TRS, can be defined. As discussed in Sec. 5.3, for centrosymmetric SGs, the Fu-Kane formula allows one to compute these indices using only the parities of the bands. This is one particular instance of the relation between the symmetry representations in $k$-space and the band topology. Recently, this relation has been expanded by exploiting the mismatch between the real and momentum-space descriptions of band structure, leading to novel forms of band topology (*i.e.,* higher-order topological insulators) in the 230 SGs for nonmagnetic[225] (and even for all 1651 magnetic SGs[233]) compounds. By comparing the symmetry representations of the bands against those of trivial insulators, the symmetry-based indicator group $X_{BS}$ (BS is band structure) is computed for all 230 SGs, which takes the form:

$$X_{BS} = Z_{n_1} \times Z_{n_2} \times \cdots \times Z_{n_N} \qquad (*)$$

where $N$ and $\{n_1, n_2, \ldots, n_N\}$ depend on the assumed space group. While $X_{BS}$ has been exhaustively computed for all SGs in reference,[225] the concrete physical interpretation for some of the nontrivial classes was left unclear. By recreating topological indicators into more explicit forms, the physical meanings of these classes are better revealed.[232] A summary of these classes is listed in Table 3. In this table, it is assumed that the mirror Chern number, $C_M^{k_z}$ ($M = \pm i$), is related to the mirror symmetry of the *xy*-plane. For primitive lattice systems, both $k_z = 0$ and $k_z = \pi$ planes are mirror symmetric and can individually support $C_M^{(k_z=0)}$ and $C_M^{(k_z=\pi)}$. For body-centered and face-centered crystals, however, $k_z = 0$ is the only mirror symmetric plane, and there is only one mirror Chern number $C_M^{(k_z=0)}$. Also, the "key space groups" plays the role of the minimal space groups for which a certain nontrivial phase is possible. It should be noted that $X_{BS}$ in equation (*) can be factorized into *weak factors* ($w$) and *strong factor(s)*:

$$X_{BS} = X_{BS}^{(w)} \times X_{BS}^{(s)} \,;$$

$$X_{BS}^{(s)} = Z_{n_N} \,; X_{BS}^{(w)} = Z_{n_1} \times Z_{n_2} \times \cdots \times Z_{n_{N-1}}. \qquad (*)$$

In this equation, every factor in $X_{BS}^{(w)}$ can be completely characterized either by the weak invariants $v_i$ ($i$=1, 2, 3), or the weak mirror Chern number $C_M^{(k_z=0,\pi)}$. According to this table, new topological indices (invariants) are defined as follows:

- *Z₄ index for inversion symmetry*: The strong topological invariant $v_0 \in Z_2$ can be promoted to a new topological invariant $\kappa_1$ mod $4 \in Z_4$ in the presence of inversion symmetry. If $n_K^-$ ($n_K^+$) (due to the Kramers pairing, $n_K^\pm$ is even) is the number of occupied bands with odd (even) parity at each TRIM $K$, the Fu-Kane formula (Eq. (30)) for the strong topological invariant $v_0$ is defined in the following new form:[232]

$$(-1)^{v_0} = \prod_{K \in TRIMs}(+1)^{\frac{1}{2}n_K^+}(+1)^{\frac{1}{2}n_K^-} = (-1)^{\frac{1}{2}\sum_{K \in TRIMs} n_K^-}. \qquad (I)$$

Given the above relation, the new topological invariant $\kappa_1$ can be introduced, which is simply the sum of the inversion parities of occupied bands:

$$\kappa_1 \equiv \frac{1}{4} \sum_{K \in TRIM} (n_K^+ - n_K^1) \in Z. \quad (II)$$

Using the total number of occupied bands $n \equiv n_K^+ + n_K^1$, can be redefine $\kappa_1$ as,

$$\kappa_1 \equiv 2n - \frac{1}{2} \sum_{K \in TRIMs} n_K^-. \quad (III)$$

By comparing equations (I) and (III), we find $(-1)^{\nu_0} = (-1)^{\kappa_1}$. Although this result shows that $\kappa_1 = \nu_0$ mod 2, $\kappa_1$ contains more information than $\nu_0$ as it is *stable* mod 4. This means that any trivial insulator has $\kappa_1 = 4n$ ($n \in Z$), and adding or subtracting trivial bands does not alter $\kappa_1$ mod 4. Besides, weak topological phases may realize $\kappa_1 = 2$ mod 4, but the most interesting case is when $\kappa_1 = 2$ mod 4 while $(\nu_1, \nu_1, \nu_3)$ all vanish. The space group that contains only IS in addition to translations is called $P\bar{1}$. The index $\kappa_1$ can be defined for every centrosymmetric space group, i.e., for every supergroup of $P\bar{1}$ (space groups containing $P\bar{1}$ as a subgroup).

- $Z_2$ *index for four-fold rotoinversion:* The key group spaces $P\bar{4}$ and $I\bar{4}$ possess neither inversion nor mirror symmetry. Therefore, the indicator $X_{BS} = Z_2$ cannot be accounted for by the Fu-Kane parity formula or the mirror Chern number.[226] Hence, a new topological invariant $\kappa_4$ mode 2 $\epsilon$ $Z_2$ in terms of the eigenvalues of four-fold rotoinversion $S_4$, i.e. the four-fold rotation followed by inversion is proposed. This index shows that the $Z_2$ nontrivial phase is a strong TI.

For spinful electrons, the four possible values of $S_4$ eigenvalues are $e^{i(\alpha\frac{\pi}{4})}$ ($\alpha = 1, 3, 5, 7$). Hence to introduce the new invariant, there are four moments in the BZ invariant under $S_4$, which we denote by $K_4$. These points are (0, 0, 0), (π, π, 0), (0, 0, π), and (π, π, π) for primitive lattice systems. The new invariant is the sum of the $S_4$ eigenvalues of occupied bands over the moments in $K_4$:

$$\kappa_4 \equiv \frac{1}{2\sqrt{2}} \sum_{\alpha, K \in K_4} e^{i(\alpha\frac{\pi}{4})} n_{K_4}^\alpha \in Z, \quad (IV)$$

where $n_{K_4}^\alpha$ is the number of occupied bands with the eigenvalue $e^{i(\alpha\frac{\pi}{4})}$ at momentum $K \in K_4$. It is noteworthy that this quantity is always an integer in the presence of TRS, and it is stable modulo 2 against the stacking of trivial bands. Moreover, $\kappa_4$ mod 2 agrees with $\nu_0$.

- $Z_8$ *index with the combination of $\kappa_1$ and $\kappa_4$*: The key space groups $P4/m$ and $I4/m$ contain both inversions $I$, and four-fold rotation ($C_4$), with the inversion center on the rotation axis. So, the product of $I$ and $(C_4)^2$ is a mirror $M_z$ about the $xy$-plane, and one can define the mirror Chern number $C_M^{(k_z=0)}$. The eigenvalues of the four-fold rotation determine $C_M^{(k_z=0)}$ mod 4.[234] Although $X_{BS}$ studied in Ref. 225 contains a $Z_8$ index, it is not detectable by the mirror Chern number. Here, Ref. 232 shows that the combination of $\kappa_1$ and $\kappa_4$ is responsible for this enhanced factor. Since these space groups have both inversion and rotoinversion ($S_4 = IC_4$), $\kappa_1$ and $\kappa_4$ can be defined separately using Eqs. (II) and (IV), respectively. Moreover, the inversion center coincides with the roto-inversion center. So, in this case, we find that the difference,

$$\Delta = \kappa_1 - 2\kappa_4 \qquad (V)$$

is stable modulo 8, not just 4, against the stacking of trivial bands. So, $\Delta$ mod 8 should be understood as a new topological invariant, which can be reconciled with the strong index $Z_8$ (< $X_{BS}$).[232] In fact, for any trivial insulator that is symmetric under either *P4/m* and *I4/m* key space groups, the two invariants $\kappa_1$ and $\kappa_4$ are not independent, and $\kappa_1 = 2\kappa_4$ mod 8 always holds. Furthermore, when all the mirror Chern numbers and the weak invariants $(\nu_1, \nu_1, \nu_3)$ vanish, $\Delta$ can still be $4 \neq 0$ mod 8.

- $Z_{12}$ *and* $Z_3$ *indices with the combination of* $\kappa_1$ *and* $C_M$ *under six-fold rotation, screw, or roto inversion*: The key space group *P6/m* has a six-fold rotation $(C_6)^3$ in addition to inversion $I$. The product of $I$ and $C_6$ (i.e., $IC_6$) is the mirror $M_z$ protecting the mirror Chern number $C_M^{(k_z=0,\pi)}$. Although the eigenvalues of the six-fold rotation can only detect mirror Chern numbers mod 6, $X_{BS}$ contains a $Z_{12}$ index. In this case, the combination of $C_M^{(k_z=0)}$ mod 6 and $\kappa_1$ mod 4 can fully characterize this $Z_{12}$ index. Besides, when the rotation $C_6$ is replaced by the screw $S_{6_3}$, i.e., $C_6$ followed by a half translation in *z*, the resulting key space group is $P6_3/m$. In this case, the product of $I$ and $(S_{6_3})^3$ is also a mirror symmetry. The corresponding $X_{BS}^{(s)}$ includes the same $Z_{12}$ index. Finally, key space group $P\bar{6}$ generated by the six-fold roto-inversion $S_6 = IC_6$ does not have inversion symmetry, and $\kappa_1$ is not defined. The mirror Chern numbers $C_M^{(k_z=0,\pi)}$ protected by $M_z = (S_6)^3$ can be diagnosed by three-fold rotation $C_3 = (S_6)^2$ modulo 3, and this fully explains $X_{BS}^{(s)} = Z_3$.

Table 3: Summary of symmetry-based indicator of band topology. Space groups are grouped into their parental key space group in the first column. The second column lists the key topological indices characterizing nontrivial classes in $X_{BS}$. The $X_{BS}$ can always be factorized into weak factors $X_{BS}^{(w)}$ and a strong factor $X_{BS}^{(s)}$.

| Key space group | key topological indices (invariants) | $X_{BS}^{(w)}$ | $X_{BS}^{(s)}$ | Space groups |
|---|---|---|---|---|
| $P\bar{1}$ (No. 2) | $\nu_{1,2,3}, \kappa_1$ | $\mathbb{Z}_2 \times \mathbb{Z}_2 \times \mathbb{Z}_2$ | $\mathbb{Z}_4$ | **2**, 10, 47 |
| | | $\mathbb{Z}_2 \times \mathbb{Z}_2$ | $\mathbb{Z}_4$ | 11, 12, 13, 49, 51, 65, 67, 69 |
| | | $\mathbb{Z}_2$ | $\mathbb{Z}_4$ | 14, 15, 48, 50, 53, 54, 55, 57, 59, 63, 64, 66, 68, 71, 72, 73, 74, 84, 85, 86, 125, 129, 131, 132, 134, 147, 148, 162, 164, 166, 200, 201, 204, 206, 224 |
| | | 0 | $\mathbb{Z}_4$ | 52, 56, 58, 60, 61, 62, 70, 88, 126, 130, 133, 135, 136, 137, 138 141, 142, 163, 165, 167, 202, 203, 205, 222, 223, 227, 228, 230 |
| $P\bar{4}$ (No. 81) | $\kappa_4$ | 0 | $\mathbb{Z}_2$ | **81**, 111–118, 215, 218 |
| $I\bar{4}$ (No. 82) | $\kappa_4$ | 0 | $\mathbb{Z}_2$ | **82**, 119–122, 216, 217, 219, 220 |
| $P4/m$ (No. 83) | $\nu_{1,2,3}, C_M^{(k_z=0,\pi)}, \Delta$ | $\mathbb{Z}_2 \times \mathbb{Z}_4$ | $\mathbb{Z}_8$ | **83**, 123 |
| | | $\mathbb{Z}_2$ | $\mathbb{Z}_8$ | 124 |
| | | $\mathbb{Z}_4$ | $\mathbb{Z}_8$ | 127, 221 |
| | | 0 | $\mathbb{Z}_8$ | 128 |
| $I4/m$ (No. 87) | $\nu_{1,2,3}, C_M^{(k_z=0)}, \Delta$ | $\mathbb{Z}_2$ | $\mathbb{Z}_8$ | **87**, 139, 140, 229 |
| | | 0 | $\mathbb{Z}_8$ | 225, 226 |
| $P6/m$ (No. 175) | $C_M^{(k_z=0,\pi)}, \kappa_1$ | $\mathbb{Z}_6$ | $\mathbb{Z}_{12}$ | **175**, 191 |
| | | 0 | $\mathbb{Z}_{12}$ | 192 |
| $P6_3/m$ (No. 176) | $C_M^{(k_z=0)}, \kappa_1$ | 0 | $\mathbb{Z}_{12}$ | **176**, 193, 194 |
| $P\bar{6}$ (No. 174) | $C_M^{(k_z=0,\pi)}$ | $\mathbb{Z}_3$ | $\mathbb{Z}_3$ | **174**, 187, 189 |
| | | 0 | $\mathbb{Z}_3$ | 188, 190 |

## 9.2. Topological Semimetal Material Families

Inspired by graphene, prediction and experimental realization of 3-dimensional materials with Dirac or Weyl fermions in the bulk states have been appealing topics. In the BZ of a Weyl semimetal, the valance and conduction bands are touching in some regions. These band touching points are nondegenerate due to breaking either the time-reversal symmetry or the inversion symmetry if we neglect the spin degeneracy. These touching points are often referred to as Weyl nodes and are stable under small perturbations. Weyl nodes show up in pairs with opposite Chern numbers at their closed vicinity. By tuning the Hamiltonian parameters, Weyl nodes can be merged into pairs and result in Chern number 0 at the merged points. These points are not stable, as they are not topologically protected, but by introducing an additional group symmetry, those points can become stable, which are then called Dirac points. An alternative way of understanding these materials is to think of them as properly tuned topological insulators with the bulk gap closed. In 2011, a multilayer topological insulator structure was proposed to be a realization of the Weyl semimetal.[235] By inserting ordinary insulator layers in between thin films of magnetic doped topological insulators, a pair of Weyl nodes can be formed in the BZ.

For the material consideration, typical Type I Weyl semimetals are the TaAs family, including TaAs, TaP, NbAs, and NbP.[72,236,237] $WTe_2$, $MoTe_2$, $WP_2$, and $MoP_2$ are typical Type II Weyl semimetals.[72] As introduced before, the Dirac cone is tilted inside the BZ. High conductivity and large transverse magnetoresistance are also observed. Sputtered $WTe_x$ was also reported to be a Weyl semimetal. For the Weyl semimetals that break time-reversal symmetry, which is also referred to as magnetic Weyl semimetals, typical materials are $Co_3Sn_2S_2$, $YbMnBi_2$, GdPtBi, $Y_2Ir_2O_7$, and $HgCr_2Se_4$.[238,239] The Weyl nodes can be realized by applying an external magnetic field. In principle, the magnetic Weyl semimetals could have the minimal Weyl nodes of two, whereas the inversion symmetry breaking Weyl semimetals have minimal Weyl nodes of four. This makes the magnetic Weyl semimetals a simpler platform for the study of the Fermi arc and chiral anomaly.

### 9.3. Topological Superconductivity

The research for understanding unconventional superconductivity has received new momentum from the study of topological materials. It is predicted that the superconducting state can also be classified based on the topological invariants and properties.[165] Topological superconductivity could host anyons with non-abelian excitations, e.g., Majorana Fermions, which are the center of a groundbreaking proposal for fault-tolerant topological quantum computation.[240] Superconductors are frequently inversion and time-reversal symmetric, and the parity of the superconducting order parameter defines the nontrivial nature of topology in superconductors. The pairing order parameter for topological superconductors is odd, e.g., *p*- or *f*-wave, under inversion symmetry. The odd-parity requires spin-triplet pairing in the absence of spin-orbit interaction. Ferromagnetic fluctuations near a ferromagnetic instability could potentially induce spin-triplet pairing. This mechanism has been observed in the superfluid He.[241] Superconductivity rarely emerges near a ferromagnetic instability. Uranium-based superconductors ($UGe_2$, URhGe, UCoGe, and $UTe_2$) are the unique example of spin-triplet superconductivity emerging near a ferromagnetic instability.[242] A hybrid device approach, however, could proximitize ferromagnetic fluctuations in a superconducting condensate. Heterostructuring with magnetic materials breaks time-reversal symmetry and could potentially induce spin-triplet pairing.[243,244]

Next, we discuss the possibility of topological superconductivity in noncentrosymmetric and "ferroelectric" superconductors with spin-orbit interaction. Noncentrosymmetric superconductors show a wide range of exciting quantum phenomena, including odd-parity superconductivity,[245] helical superconducting state,[246] and above Pauli limit ($\mu_B H_{c2}/K_B T_c$ ~1.84) upper critical field.[247] A combination of large spin-orbit interaction and ferroelectric state, i.e., antisymmetric spin-orbit coupling, could potentially host mixed-parity superconductivity in complex oxides, including $Cd_2Re_2O_7$, $LiOsO_3$, and $SrTiO_3$.[248,249] A nonzero Chern number is predicted in 2D Rashba superconductors with an applied magnetic field.[250,251]

Superconductivity in $SrTiO_3$ emerges near a polar instability. The ferroelectric superconductivity in $SrTiO_3$ under a magnetic field satisfies conditions for topological superconductivity, i.e., broken inversion and time-reversal symmetries.[248,249] It is predicted that the intersection of ferroelectricity and superconductivity harboring exciting quantum phenomena in $SrTiO_3$, including mixed-parity superconductivity,[252] topological Weyl superconducting state.[253] The condition for a nonzero Chern number (($\sqrt{(4t+\mu)^2 +|\Delta|^2} < \mu_B H z < \sqrt{\mu^2+|\Delta|^2}$)), however, could occur beyond critical magnetic field of superconductivity.[254] A more recent theoretical study predicts Weyl topological superconductivity in applied magnetic fields exceeding the Pauli limit.[253] Superconductivity above this limit has been observed experimentally in SrTiO$_3$.[255,256] Ferroelectric superconducting state in SrTiO$_3$ is a promising candidate for topological superconductivity and, more recently, experimental evidence of mixed parity superconductivity[257] and polar nanodomains[258,259,260] have been observed in this material.

Hybrid devices where proximity-induced interface superconductivity in topological insulators could harbor Majorana fermions is another path to engineer topological superconductivity. The seminal work by Fu and Kane revealed a time-reversal symmetric 2D state with a spinless $p_x+ip_y$ superconducting state at the interface of an $s$-wave superconductor and a topological insulator.[261] As stated by Fu and Kane, the $s$-wave pairing generates triplet $p$-wave pair correlations. This occurs due to the spin-rotation symmetry violation by the spin-momentum locking, $\sigma \cdot p$. This state hosts Majorana-bound states at vortices. Revealing the Majorana zero-bias peak requires a proximity superconductor with a hard gap and an abrupt and defect-free interface.

Table 4: A selected list of recent studied topological materials. TI: topological insulator. HO: hidden order. TSM: topological semimetal. QSHI: quantum spin Hall insulator. TKI: topological Kondo insulator. TCI: topological crystalline insulator. TSC: topological superconductor. DSM: Dirac semimetal. WSM-I and WSM-II: type-I and type-II Weyl semimetal. TRS: time-reversal symmetry. IS: inversion symmetry. ARPES: angle-resolved photon emission spectroscopy

| Class | Materials | Symmetry *Breaking Protected* | Material Dimension | Experimental Method | B field/ Magnetic doping | Presence of SOC | Ref. |
|---|---|---|---|---|---|---|---|
| $Z_2$ TI | $Bi_{1-x}Sb_x$, α - Sn HgTe, $Pb_{1-x}Sn_xTe$ β - HgS | *TRS* | 2 and 3 | ARPES | N | Y | 12,112,123, 127,262 |
| $Z_2$ TI | $Bi_2Sb_3$, $Bi_2Te_3$, $Sb_2Te_3$ | *TRS* | 3 | ARPES, Micro-Raman; Quantum oscillation | N | Y | 217,263,264, 265 |
| $Z_2$ TI | $A_mB_{2n}C_{m+3n}$; (A = Pb, Sn, Ge; B = Bi, Sb; C = S, Se, Te) | *TRS* | 3 | ARPES STM | N | Y | 266,267,268 |
| $Z_2$ TI | $A_2Se_3$, $A = Bi_xIn_{1-x}$ | *TRS* | 3 | ARPES | N | Y | 269,270 |
| $Z_2$ TI | $A_2B_2C$ (A = Bi, Sb; B, C = S, Se, Te) | *TRS* | 2 and 3 | ARPES, STM | N | Y | 271,272,273, 274 |
| $Z_2$ TI | BiTeX; (X= I, Br, Cl) | *TRS* | 3 | ARPES | N | Y | 275,276,277 |
| $Z_2$ TI, TSC | $A_5B_{3m}$; (A = PbSe, B = $Bi_2Se_3$) | *TRS* | 3 | ARPES, STM | N | Y | 278,279 |
| $Z_2$ TI, WSM, TCI | $TlBC_2$; (B = Bi, Sb C = S, Se, Te) | *TRS* | 3 | ARPES | N | Y | 280,281 |
| $Z_2$ TI, WSM | $LaBiTe_3$, $GdBiTe_3$ | *TRS* | 3 | ARPES | N | Y | 282,283 |
| $Z_2$ TI | $Bi_{14}Rh_3I_9$ | *TRS* | 3 | ARPES | N | Y | 284,285 |
| $Z_2$ TI | $A_mB_n$; (A = $Bi_2$, B = $Bi_2Se_3$) | *TRS* | 3 | ARPES | N | Y | 286,287 |
| $Z_2$ TI, TSC | $TaSe_3$ | *TRS* | 3 | ARPES, Quantum Oscillation | N | Y | 288 |
| $Z_2$ TI | $Li_2BC$; (B= Cu, Ag, Au, Cd; C= Sb, Bi, Sn) | *TRS* | 3 | ARPES | N | Y | 289 |
| $Z_2$ TI | LiAuSe, KHgSb | *TRS* | 3 | ARPES | N | Y | 290 |
| $Z_2$ TI | $RhBi_2$ | *TRS* *IS* | 3 | ARPES, First Principle Calculation | N | Y | 291 |
| $Z_2$ TI | $ABiO_3$ (A= Y, Ba, Ir) | *TRS* | 3 | ARPES | N | Y | 292,293 |
| $Z_2$ TI | $A_mX$, PuY; (X = N, P, Sb, Bi; Y = Se, Te) | *TRS* | 3 | ARPES | N | Y | 294 |
| $Z_2$ TI | $Sn_{n+1}Ir_nO_{3n+1}$ $Sr_2IrRhO_6$ | *TRS* | 3 | ARPES | N | Y | 295,296 |
| $Z_2$ TI | $β-Ag_2Te$ | *TRS* | 2 and 3 | ARPES, Quantum Oscillation | N | Y | 297,298 |
| 2-D TI | Graphene and silicene family | *TRS* *IS* | 2 | ARPES | N | Y | 299,300,301, 302,303,304, 305,306 |
| 2-D TI | CaAs/Ge | *TRS* | 2 | ARPES | N | Y | 307 |
| 2-D TI | XBi, (X = Ga, Tl, In) | *TRS* | 2 | ARPES | N | Y | 308,309 |
| 2-D TI | $H_2/F_2$- XBi, | *IS* | 2 | | N | Y | 310 |
| | (X = Ga, Tl, In) | | | | | | |
| 2-D TI TCI WSM | Bi$_4$Br$_4$, Bi$_4$I$_4$ | *TRS* *RS* *IS* | 2 | ARPES | N | Y | 311,312 |
| 2-D TI | AB; (A = Ge, Sn, Pb, Bi; B = H, I, Br, Cl, F, OH) | *TRS* | 2 | ARPES | N | Y | 313 |
| 2-D TI | Hg/CdTe Quantum well | *TRS* | 2 | ARPES | N | Y | 314,315 |
| 2-D TI | InAs, GaSb, AlSb | *TRS* | 2 | ARPES | N | Y | 316 |
| 2-D TI | (KTaO$_3$)$_7$/(KPtO$_3$)$_2$ | *TRS* | 2 | ARPES | N | Y | 317 |
| 2-D TI | Na$_2$IrO$_3$; LiIrO$_3$ | *TRS* | 2 | ARPES | N | Y | 318 |
| HO | URu$_2$Si$_2$ | *TRS* | 3 | ARPES | N | Y | 319 |
| TSC | LaAlO3 / SrTiO3 | *TRS* | 2 | ARPES | N | Y | 320 |
| TSM | (SrTiO$_3$)$_7$/(SrIrO$_3$)$_2$ | *TRS* | 2 | ARPES | N | Y | 317 |
| TI | CrO$_3$; TiO$_3$ | *TRS* | 2 and 3 | ARPES | N | Y | 321 |
| TMI WSM | A$_2$Ir$_2$O$_7$; (A = Nd, Sm, Eu, Y) | *TRS;* *TRS* | 2 | ARPES | N | Y | 322,323 |
| TI,DSM | Pb$_2$It$_2$O$_{7-x}$ | *TRS* | 2 | ARPES | N | Y | 324 |
| QSHI, DSM | Zr/HfTe$_5$ | *TRS* | 2 | ARPES | N | Y | 325,326,327 |
| TKI | SmB$_6$, YB$_6$, YB$_{12}$ | *TRS* | 3 | ARPES, Quantum Oscillation | N | Y | 328,329,330, 331 |
| TKI | SmS | *TRS* | 3 | ARPES | N | Y | 332,333 |
| TCI | PbSe, SnSe, PbTe, SnTe | | 3 | ARPES, Quantum Oscillation | N | Y | 334,335,336 |
| MTI | Mn/Fe/Cr/Gd/V/Cu Doping Bi$_2$Se$_3$ | *TRS* | 2 and 3 | ARPES, STM | N | Y | 337,338,339, 340,341,342 |
| ATI | EuB$_2$As$_2$; (B = In, Sn) | *TRS* | 3 | ARPES | N | Y | 343,344 |
| ATI, WSM | MnBi$_{2n}$Te$_{3n+1}$ | *TRS* | 3 | ARPES | N | Y | 344,345,346 |
| TSC | Sr$_2$RuO$_4$ | *TRS* | 3 | ARPES | N | Y | 347 |
| TSC | Cu$_x$Bi$_2$Se$_3$ | *TRS* | 3 | ARPES, Quantum Oscillation, STM | N | Y | 348,349,350, 351 |
| TSC | Sn$_{1-x}$In$_x$Te | *TRS* | 3 | ARPES, Muon Spin Spectroscopy | N | Y | 352,353,354 |
| TSC | CePt$_3$Si | *TRS* | 3 | ARPES | N | Y | 355 |
| TSC | Li$_2$(Pd$_{1-x}$Pt$_x$)$_3$B | *TRS* | 2 and 3 | ARPES | N | Y | 356,357 |
| TSC | Cu$_x$A$_5$B$_6$; (A = PbSe B = Bi$_2$Se$_3$) | *TRS* | 3 | ARPES | N | Y | 358 |
| TSC | UPt$_3$ | *TRS* | 3 | ARPES | N | Y | 359 |
| TSC | YPtBi | *TRS* | 3 | ARPES, Quantum Oscillation | N | Y | 360 |
| DSM | Bi$_2$Se$_3$ | *TRS* *IS* | 3 | | N | Y | 361 |
| DSM | A3Bi; A =Na, K, Rb | *TRS* *IS* | 3 | ARPES, Quantum Oscillation, STM | N | Y | 362,363 |
| DSM | BiO$_2$ | *TRS* *IS* | 3 | ARPES, Quantum Oscillation, STM | N | Y | 15 |
| DSM | Cu$_3$PdN | *TRS* *IS* | 3 | ARPES, Quantum Oscillation, STM | N | Y | 364 |
| DSM | CaAgX (X = P, As) | *TRS* *IS* | 3 | ARPES, Quantum Oscillation, STM | N | Y | 365 |
| DSM | GeTe-Sb$_2$Te$_3$ | *TRS* | 3 | ARPES, Quantum | | | 366 |
| | | *IS* | | Oscillation, STM | | | |
| DSM | PtSe$_2$ | *TRS IS* | 3 | ARPES, Quantum Oscillation, STM | N | Y | 367,368 |
| DSM | Cd$_3$As$_2$ | *TRS IS* | 3 | ARPES, Quantum Oscillation, STM | N | Y | 369,370 |
| DSM | Sr/Ca/Ba/EuMnBi$_2$ | *TRS IS* | 3 | ARPES, Quantum Oscillation, STM | N | Y | 371,372,373, 374 |
| DSM | LaAgSb$_2$ | *TRS IS* | 3 | ARPES, Quantum Oscillation, STM | N | Y | 375,376 |
| DSM | CoSb$_3$ | *TRS IS* | 3 | ARPES, Quantum Oscillation, STM | N | Y | 377,378 |
| DSM | NiTe$_2$ | *TRS IS* | 3 | ARPES, Quantum Oscillation, STM | N | Y | 379 |
| DSM | TiSb$_2$ | *TRS IS* | 3 | Quantum Oscillation | N | Y | 380 |
| DSM | KV$_3$Sb$_5$ | *TRS IS* | 2 | ARPES, Quantum Oscillation | N | Y | 381 |
| DSM | SrAuSb | *TRS IS* | 3 | ARPES, Quantum Oscillation | N | Y | 382 |
| DSM | EuMg$_2$Bi$_2$ | *TRS IS* | 3 | ARPES, Quantum Oscillation | N | Y | 383 |
| DSM | XInPd$_2$ (X = Ti, Zr and Hf) | *TRS IS* | 3 | ARPES, Quantum Oscillation | N | Y | 384 |
| DSM | Ca$_3$NBi; Ca$_3$PbO | *TRS IS* | 3 | ARPES, Quantum Oscillation, STM | N | Y | 385 |
| DSM | ZrXY (X=Si, Ge; Y=S, Se, Te) | *TRS IS* | 3 | ARPES, Quantum Oscillation, STM | N | Y | 386 |
| DSM | Ca$_3$Ru$_2$O$_7$ | *TRS IS* | 3 | ARPES, Quantum Oscillation, STM | N | Y | 387 |
| DSM | MgTa$_2$N$_3$ | *TRS IS/DCS* | 3 | First Principle Calculation | N | Y | 388,389 |
| WSM-II | Ta$_3$S$_2$ | *TRS IS* | 3 | ARPES, Quantum Oscillation | N | Y | 390 |
| WSM-II | Mo$_x$W$_{1-x}$Te$_2$ | *TRS IS* | 3 | First Principle Calculation | | | 391,392 |
| WSM-II | TaIrTe$_4$ | *TRS IS* | 3 | ARPES, Quantum Oscillation, Raman | N | Y | 393,394 |
| WSM-II | AB$_2$; A = W, Mo; B = Te, P | *TRS IS* | 3 | ARPES, Quantum Oscillation, Raman | N | Y | 395,396,397, 398 |
| WSM-II | LaAlGe | *TRS IS* | 3 | ARPES | N | Y | 399,400 |
| WSM-II | InNbX$_2$(X = S, Se) | *TRS IS* | 3 | ARPES, Quantum Oscillation, STM | N | Y | 401 |
| WSM-II | NbIrTe$_4$ | *TRS IS* | 3 | ARPES, Quantum Oscillation, STM, Raman | N | Y | 402,403 |
| WSM-I | Bi$_2$Se$_3$ | *TRS IS* | 3 | First Principle Calculation | Y | Y | 404 |
| WSM-I | AB; A=Ta,Nb; B= As, P | *TRS IS* | 2 and 3 | ARPES, Quantum Oscillation, STM | N | Y | 405,406, 407,408, 409 |
| WSM-I | Ag$_2$S | *TRS IS* | 3 | First Principle Calculation | N | Y | 410 |
| WSM | $Hg_{1-x-y}Cd_xMn_yTe$ | *TRS IS* | 3 | Magnetic Field angle dependent Hall Conductivity | Y | Y | 411 |
| WSM | CaPd | *TRS IS* | 3 | First Principle Calculation | N | Y | 196 |
| WSM | $Hf_xZr_{1-x}Te_2$ | *Rotational* | 3 | First Principle Calculation | N | Y | 198 |
| WSM-II | $YbMnBi_2$ | *TRS IS* | 3 | ARPES, Quantum Oscillation, STM | Y | Y | 412,413 |
| Magnetic WSM-I | $Co_3Sn_2S_2$ | *TRS IS* | 3 | ARPES, Quantum Oscillation, STM | Y | Y | 414 |
| WSM-I | GdPtBi, NdPtBi | *TRS IS* | 3 | ARPES, Quantum Oscillation, STM | Y | Y | 415 |
| WSM | $Y_2Ir_2O_7$ | *TRS IS* | 3 | ARPES, Quantum Oscillation, STM | Y | Y | 416 |
| WSM-I | $EuCd_2As_2$ | *TRS IS* | 3 | Photoemission Spectroscopy | Y | Y | 417 |
| WSM-II | $Mn_3Ge$, $Mn_3Sn$ | *TRS IS* | 3 | ARPES, Quantum Oscillation | Y | Y | 418,419,420 |
| Magnetic WSM-I | $Co_2MnGa$ | *TRS IS* | 3 | ARPES, Quantum transport | Y | Y | 421,422 |
| Magnetic WSM | $Sr_{1-x}Mn_{1-y}Sb_2$ | *TRS IS* | 3 | ARPES, Quantum transport | Y | Y | 417,423 |
| WSM-I | RhSi | *TRS IS* | 3 | First Principle Calculation | N | Y | 424 |
| WSM-I | CoSi, CoGe | *TRS IS* | 3 | ARPES, Quantum transport | N | Y | 425,426 |
| WSM | $(Eu_{1-x}Sm_x)TiO_3$ | *TRS IS* | 3 | Quantum transport | Y | Y | 134,427 |
| WSM-III | $(TaSe_4)_2I$ | *TRS IS* | Quasi 1D | ARPES, First Principle Calculation | N | Y | 428 |
| WSM-III | $X_2RhF_6$ (X = K, Rb, Cs) | *TRS IS* | Quasi 1D | First Principle Calculation | N | Y | 429 |

## 10. Topological Phase Conversion

If topological materials are going to make transformational technologies, it is critical to be able to control their conversion. Topological phase conversion has twofold importance in quantum matter research (i) at the fundamental physical level, developing quantum/topological phases into a trivial matter with different ground and symmetry states,[9,88] and (ii) various device applications with reliable control. The conversion process requires an understanding of the formation of the electronic and spin ground states along with the topological surface states and their protection mechanisms versus symmetry. The relation between trivial and nontrivial topological phases is theoretically well-known, which helps to determine the conversion techniques. The major challenge to achieve topological phase conversion is to realize it in a physical system with a robust and reliable control mechanism. Therefore, the objective of this section is to highlight the recent advancements in topological phase conversion techniques both theoretically and physically, along with the associated roadblocks and possible solutions. Finally, we will discuss the prospect of topological phase conversions from the context of device applications.

### 10.1. State-of-the-art Topological Phase Conversion Techniques

Symmetry-driven topological phases of matter demonstrate their exotic nontrivial topological characteristics arising from the corresponding topological edge/surface states, which are also subjected to symmetry protection or breaking. This implies that the key factor in the topological phases of matter is the techniques to create or break a certain symmetry of matter. For conventional or trivial phases of matter, the temperature is the most common external force or stimulus to achieve the conversions among phases. To control the local ordering mediated conventional phases, one can use both temperature and pressure as an acting thermodynamical force. But global ordering dependent topological phase conversion is quite different than the conventional phase conversions. As discussed in the earlier sections and illustrated in *Figure 1*, all topological phases have particular types of symmetry protection/breaking, which determine the types of conversion processes. For instance, time-reversal symmetry breaking is related to the magnetic field, or crystal symmetry breaking can be related to the pressure/strain or an electric field. In this section, we mainly confine our discussion to 3D topological phase conversions, which can be easily attained at room temperature. On the other hand, 2D topological phases appear in 2D material systems or at the interface of the heterostructures. The quantum Hall insulator phases (IQH and FQH) are subjected to the presence of an external magnetic field, while conversion between the trivial insulator and quantum spin Hall insulator phase into a quantum well structure with certain materials can be achieved by the tuning of the well-width. In the 2D Dirac semimetals, the breaking of the high-symmetry points can be done by applying strain or an electric field, but this conversion is limited to the 2D material systems like graphene with bidirectional conversion between the trivial and nontrivial 2D Dirac semimetal phases. In contrast, multiple phase conversions can be realized in bulk materials with different external stimuli, which will be discussed in the following subsections. The first subsection summarizes the theoretical studies on topological phase conversion, providing more insight into understanding the relationship between different trivial and nontrivial phases. The following subsections summarize the theoretically and experimentally realized techniques and discuss the state-of-the-art of topological phase conversion research.

### 10.1.1. Theoretical Models

Modification of the band structure is the key factor in achieving the topological phase transition into a material. Due to the recent advances in calculating complex band structures under the external stimuli, theoretical computations offer effective means to explore the band structure tuning mechanisms and their corresponding limiting ranges. In the context of the topological phase studies, many works in the literature have demonstrated the possibilities of the tuning of the band structure theoretically by varying chemical composition,[186] doping,[182] strain,[430] pressure[431], or an external electric or magnetic field.[432] Most of the theoretical works demonstrated the predicted phase diagram based on different tuning parameters. Different works of literature reported their theoretical predictions based on different approaches like fitting the experimental data with various models,[433] calculating the phase boundaries from Hamiltonian with external tuning parameter,[16,67,432] and calculating the band structures with different tuning conditions.[431,434] There are several software packages such as Z2Pack,[435] WannierTools,[436] Irvsp,[437] qeirreps,[438] WloopPHI,[439] Pythtb,[440] NodeFinder,[441] etc. widely used to determine the topological band-structure properties along with topological invariants and electronic surface and bulk band structures. Some other open-source software tools based on first-principle calculations like ABINIT and Quantum-espresso can also be used for theoretical modeling of the topological phases of matter. To demonstrate the prediction of topological phase transitions via different models, different computer-aided models like DFT-based Vienna ab initio package (VASP)[442] or Drude-Lorentz model using RefFIT[433] program have been widely used. Murakami et al. demonstrated the topological phase diagram from the Hamiltonian with an external tuning parameter in momentum space where the external parameter was defined separately under different symmetry conditions.[432] Wang et al. calculated the Berry curvature defined topological charge or the Chern number enclosed by the Fermi surface ($C_{FS}$) from the Hamiltonian to predict the topological phase diagram.[16] Chern number enclosed by Fermi surface has the following relation with the Berry curvature[16]

$$C_{FS} = \frac{1}{2\pi} \int_{FS} [\nabla_k \times A(k)] . dS \qquad (58)$$

Where $A(k)$ is the Berry connection calculated from the wavefunctions at the Fermi level. A nonzero $C_{FS}$ indicates the existence of nontrivial topological phases. In general, all the theoretical predictions of topological phase transitions are associated with the first principal calculation that begins with defining the Hamiltonian with the correct external tuning parameters. However, the determination of the phase diagrams showing the transition between trivial to nontrivial phases can be accomplished with different approaches. Besides demonstrating the phase diagram among the known topological phases, theoretical models can predict new topological phases that can appear during the adiabatic deformation of one phase into another one. In general, theoretical modeling requires rigorous efforts with huge simulation capacity, while the practical realization of the topological phase transition into materials is also challenging in terms of both deformations of one phase into another and reliable, controllable tuning mechanisms.

### 10.1.2. Experimental Mechanisms for Phase Conversion

Since the observation of the topological phases into a physical system, several efforts have been made to obtain the phase conversion into a real system experimentally.[31,270,430,431] Researchers investigated different external stimuli to obtain the conversion and showed some success in attaining the goal.[270,430,431] Topological phase transition (TPT) can be happened between non-trivial and trivial phases or between two different topological phases. As mentioned earlier, topological phase conversion is the tuning of the electronic band structure to change the topological ground state of the associated phase. Fundamentally, all the parameters related to the band structure characteristic of the system can be utilized for a phase conversion. But, the realization of those theoretical phase conversion techniques in a real system is rather more challenging; as such, not all theoretically predicted band structure tuning parameters have to lead to a phase conversion between certain phases. The typical external stimuli for topological phase conversions are electric-field (E-field), magnetic field (B-field), doping, pressure, and modification of the layer thickness. *Figure 19* summarizes the different experimental mechanisms to attain the topological phase conversion.

According to *Figure 19*, the trivial phase of the matter (normal insulator (NI)) can be transitioned into TIs, DSMs, and WSMs via E-field or pressure. Here, one important note is that a high E-field is usually required for obtaining topological phase conversion via E-field tuning, which may cause a material breakdown. Moreover, pressure-induced phase conversion needs a good interface between the topological layer and pressure inducing layer, as surface properties can alter the topological performance of the system. One of the typical topological phase conversion mechanisms is through varying the doping composition and concentration. For example, the bandgap in TlBi($S_{1-\delta}Se_\delta$)$_2$ was tuned by varying $\delta=0.2$, 0.6, and 0.8; consequently, the material system transitioned from trivial to non-trivial phases.[280] Doping induced topological phase conversion mechanism has no tunability or robustness. Apart from the illustrated experimental mechanisms, topological phase conversions can also be attained by laser or microwave pumping to produce a nonequilibrium topological state or a Floquet TI.[443,444] All topological phase conversion mechanisms have their unique challenges, which can limit the success of attaining the expected phase conversion. In the Table 5, we summarize several experimentally realized topological phase conversions along with the material system, physical mechanism, and the device structure.

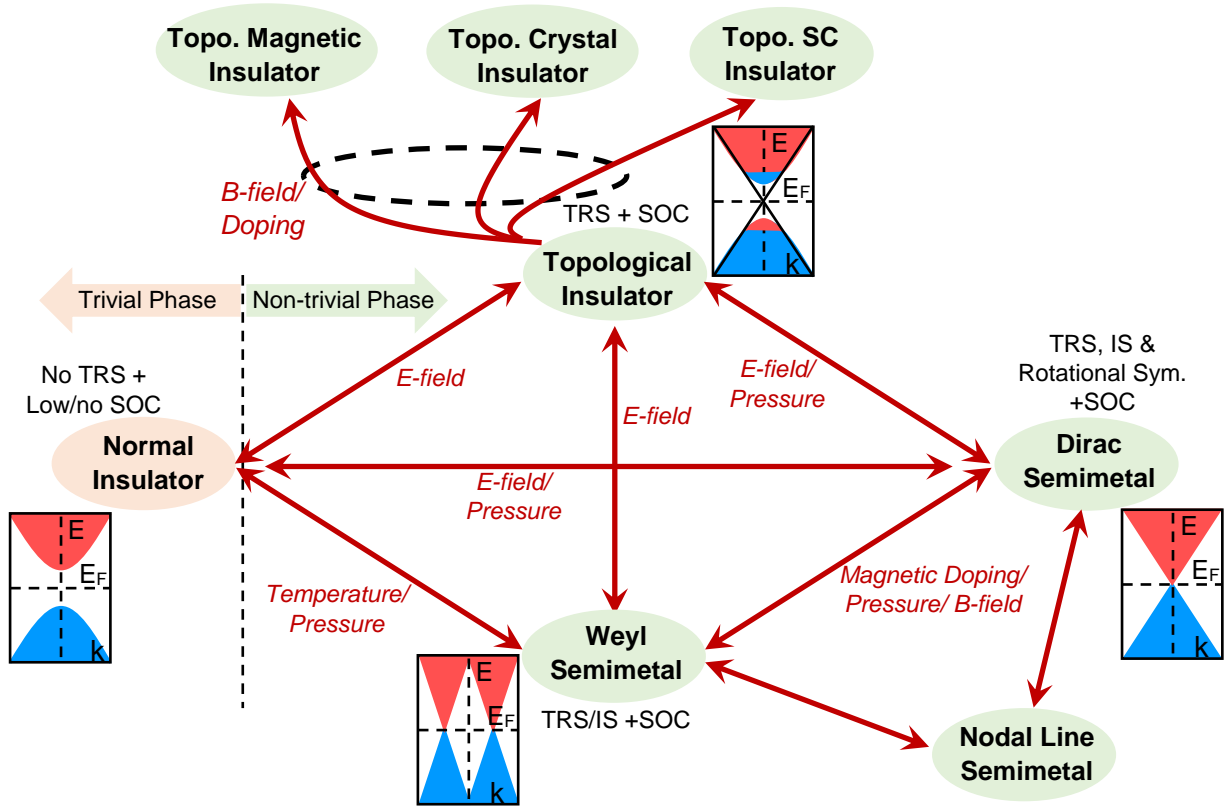

Figure 19: Phase conversion diagram for different topological phases and their associated physical mechanisms to obtain the conversion.

Table 5: Different topological phase conversion mechanisms along with the material system, stimuli, and device structure

| Phase transition | Materials | Means | Device structure | Ref |
|---|---|---|---|---|
| TI-DSM-NI | Na$_3$Bi | E-field | BL-Na3Bi(001)/Si(111) | 31 |
| NI-TI | (Bi$_{1-x}$In$_x$)$_2$Se$_3$ | Thickness and Composition | QLs (Bi$_{1-x}$In$_x$)$_2$Se$_3$ | 270 |
| NI-TI | InN/GaN | Polarization from symmetry and lattice-mismatched, & E-field induced Rashba SOC | GaN/InN(3-4ML)/GaN QW | 434 |
| DSM-WSM-NI | Pyrochlore oxide | Strain | Pyrochole lattice along {111} | 445 |
| DSM-WSM | Na$_3$Bi | B-field | Na$_3$Bi | |
| WSM-WSM | TaAs | Pressure-induced | TaAs | 431 |
| DSM-TI | Na$_3$Bi | Strain | Na$_3$Bi | 430 |
| NI-TI | Phosphorene | E-field | 3-4L Phosphorene | 442 |
| TI-TCI | Theoretical | B-field | Nanodisk | 446 |
| QHSI-NI | Silicine | E-field | Monolayer | 447 |
| TI-TI | Silicine | Polarized light | Monolayer | 304 |
| WSM-TI | MoTe$_2$ | Temperature | -- | 448 |
| WSM-NI | WTe$_2$ | Terahertz light pulse | WTe$_2$/SiO$_2$/Si | 397 |

The success in the realization of the topological phases into practical devices clearly depends on the ability to control the phase conversion of the material experimentally. As an example, a universal phase conversion process is demonstrated for the Bi$_2$Se$_3$ system in *Figure 20*. A sample physical structure to obtain the conceptual conversion process is also shown and discussed in the following subsection.

### 10.2. Physical Structure for Topological Phase Conversions

The appearance of the topological phases into a material system is quite different than the conversion of topological phases via various mechanisms. Certain material systems (shown in *Table 4*) possess topological phases inherently, while phase conversion is attained by external stimuli acting on the topologically trivial and non-trivial phases. The application of the external stimuli may require a complex physical structure to realize the topological phase conversion. There are several key features to determine the performance of the phase conversions, such as robustness, sustainability, tunability, and time-response.[397] Often, the external stimuli create an extreme condition that may cause the physical breakdown of the system. For example, Na$_3$Bi thin layers (monolayer or bilayers) need around 2.2 V/nm (i.e., 2.2×10$^7$ V/cm) E-field to achieve the phase conversion. Therefore, a physical structure, along with the topological layer, is needed to maintain the sustainability of the device under extreme high field conditions. On the other hand, B-field, laser pumping, microwave pumping, and temperature offer non-destructive ways to obtain phase conversions without any external physical layers. However, these techniques can be bulky, expensive, or not compatible with the existing semiconductor fabrication process, which can be the major roadblocks to use the topological phase conversions into practical device applications. Here, it is worth mentioning that laser and microwave pumping can offer ultrafast techniques for attaining phase conversions.[397] Nonetheless, due to the practical considerations, the E-field and strain mediated topological phase conversions are often preferred despite their approaching their extreme limits. Moreover, E-field tuned topological phase conversion is more suitable for making topological transistors. Such transistors are expected to outperform the traditional semiconductor transistors due to the exotic features associated with the topological phases. Such physical structures are also sometimes called a *topological switch*.[397] Generally, physical layers around the topological layer have manifold roles in the topological phase conversions, i.e., they protect the topological layers from physical breakdown, provide controllable conversion process on the topological layers, provide isolation between external stimuli and the physical topological layers, or work as the gate layer for controlling the switching function.

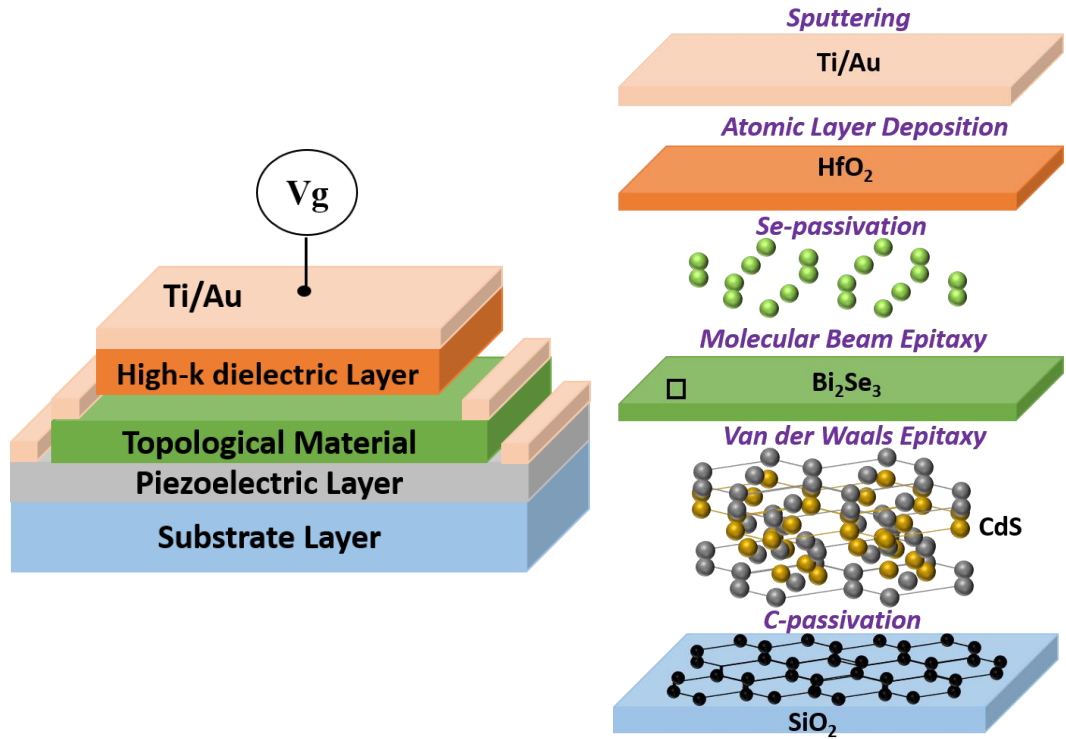

Figure 20: Conceptual multi-layers structure for topological phase conversion via E-field and strain along with the suggested material systems and their growth processes.

The typical challenges in multi-layers topological device structures are related to the interface quality at heterostructure interface, lattice-matched materials, and the thickness of the layers. *Figure 20* demonstrates a conceptual device structure based on the topological $Bi_2Se_3$ layer. This conceptual structure is for tuning the topological phase of $Bi_2Se_3$ via E-field and strain. To achieve the phase conversion structure, we are considering $Bi_2Se_3$ as the channel layer, CdS as a strain-induced layer, $HfO_2$ as a high-k dielectric gate oxide layers, Ti/Au layers as metal contacts, and the $SiO_2$ layer as the substrate. To maintain the interface quality of the heterostructures, one may use a passivation layer followed by the van der Waals epitaxy (VDWE). Here, Se-passivation can be performed on the $Bi_2Se_3$ layer.[451] To apply a high electric field for tuning the band gap, one has to add a high-k dielectric layer to protect the device from physical breakdown.[451] Typically, 1 V/nm field is required for 100 meV band closing.[31,452] The thickness of the dielectric layer can be optimized to stay away from voltage breakdown. Lattice matched CdS with $Bi_2Se_3$[453] can induce the Weyl nodes into the system by breaking the inversion symmetry. Finally, the C passivation on $SiO_2$[456] can help to avoid the accumulation of the strain in the topological layer from the substrate due to a lattice mismatch.

### 10.3. Topological Phase Conversion in Device Applications

As mentioned, using the topological phases into practical applications requires controllable and reliable switching of the topological phase in a device structure. Therefore, topological phase conversion techniques play critical technological roles. The exotic topological characteristics are considered excellent prospects for many existing and next-generation device applications. Despite their attractive features, topological materials are still lagging in device applications due to some critical challenges.

Before they can be used for technological applications, one has to consider parameters such as the compatibility of the topological material with the existing devices, their fabrication and batch processing, compatibility of the topological or quantum behavior under external stimuli, working temperature range, cost-efficiency, practical structure and ease of use.[457]

Much effort has been made in condensed matter research to overcome the roadblocks associated with the topological materials for device applications. In this section, we summarize several state-of-the-art application fields for which topological characteristics show prospects. *Figure 21* demonstrates the physical device structures for different applications.

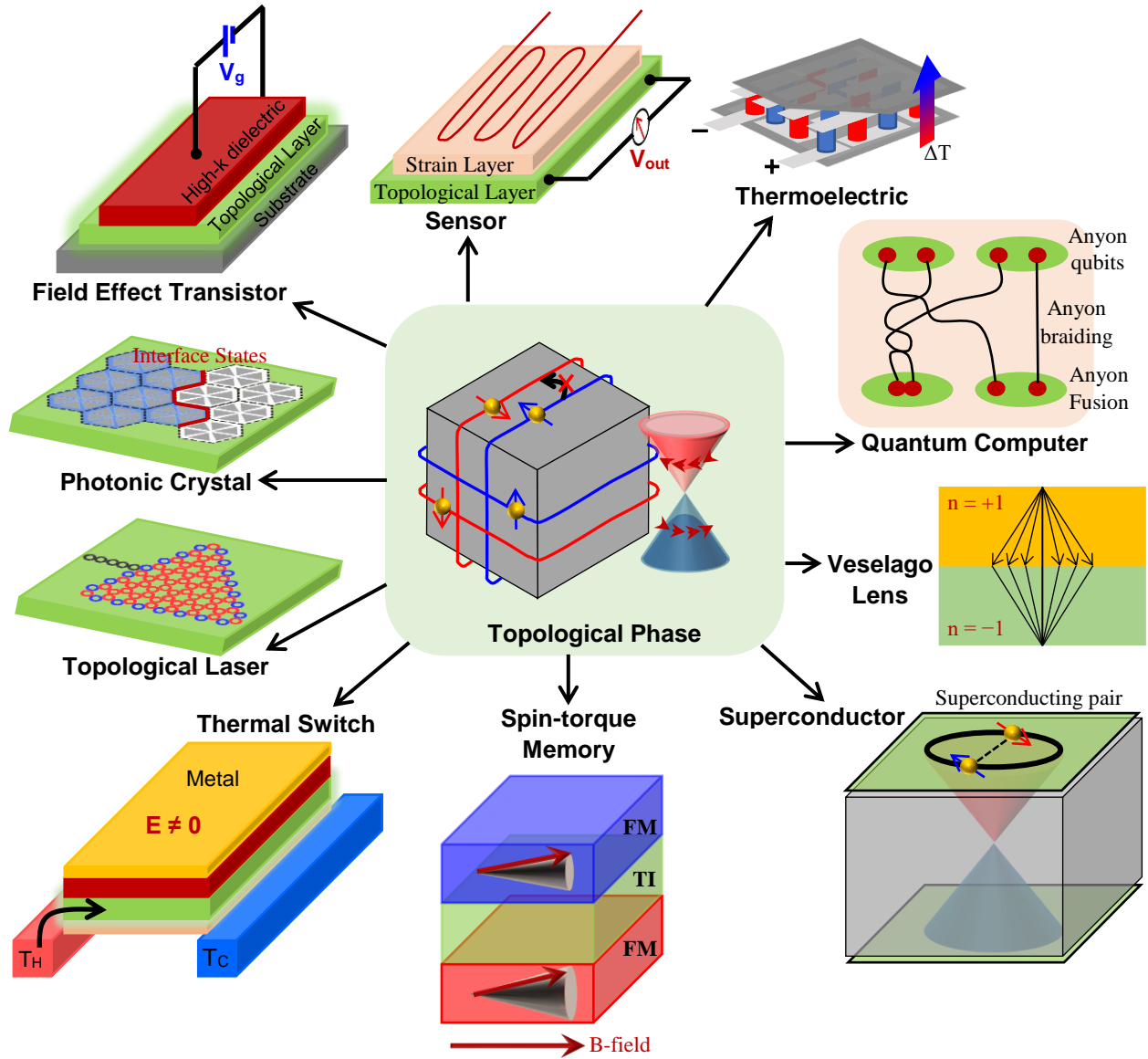

Figure 21: Various device application fields for topological phases of matter along with the physical device structures.

One of the common application fields for topological phase conversion is in electronics with switches as a logic device. As such, many studies are focused on building field-effect transistors (FET) functioning based on the topological phase transition. The topological transistors can have an ultrahigh-speed operation due to the high carrier mobility and dissipationless operation due to the absence of backscattering. A topological transistor generally consists of one or two high-k dielectric gate layers and a substrate along with the topological channel. Typically, a topological insulator to normal insulator phase conversion via E-field tuning is used to realize the switching behavior.[31] Topolocal transition with dual TI layers protected by more than one symmetry is also reported where the breaking of one protected symmetry can manipulate the spin texture of TSS preserved by the other symmetry.[458] Besides TIs, topological crystal insulators (TCIs) can be another phase to realize transistor operation, as TCIs have crystal symmetry-protected topological states rather than time-reversal symmetry protection.[334] Unlike in the bulk semiconductor transistors, carrier transport occurs at the surfaces of the topological layer via topological surface states. Therefore, a high-quality interface at the heterostructure of the topological transistor is vital for the device performance.

Another exotic application of topological materials is for quantum bits (qubits) in *Quantum computers*. The efforts on building quantum computers introduce several ways to make the quantum version of a classical computer conceptually. Topological states provide unique features of suppressing the noise and decoherence of the qubits, two major roadblocks in the performance of quantum computers.[73] The non-abelian statistical state of matter can act as a physical system to realize qubits.[459] This state of the matter is often known as non-abelian anyons; a collective composite particle resides in a quantum system where a bandgap exists between the ground and excited states. Chiral Majorana fermions also hold a prospect in performing as qubits.[460] Still, the challenge in quantum computers is to realize the qubit system physically. Their complex principles and operations are still not fully absorbed by the researchers. Therefore, this is one of the active topological research fields with a great prospect in the near future.

*Thermoelectricity* is another active area of research in the quest for finding high-performance materials for thermal to electrical or vice versa solid-state energy conversion.[461,462,463] The major challenge in current thermoelectric research is to overcome the fermionic limitation on the intercoupled thermoelectric material properties.[464] To achieve a non-fermionic improvement in thermoelectric performance, people have further considered the contributions from magnons and paramagnons as new ways for a high thermoelectric power factor.[465,466,467] Many thermoelectric materials and topological systems consist of heavy atoms; as such, several topological materials can also make good candidates for thermoelectric applications.[468,469,470,471] However, further materials engineering and optimization are required for topological thermoelectric materials to achieve their highest performance due to the large bipolar contributions in thermopower and thermal conductivity. Bipolar effects can significantly reduce thermoelectric performance.[464] Apart from thermoelectric devices, *thermal switches* are another thermal device application for topological materials to control the entropy flow through the surface and the bulk states under the external stimuli.

In addition to electronic and spin properties, topological materials also demonstrate exotic optical properties, namely, Kerr and Faraday rotation, ultrahigh bulk refractive index, unusual electromagnetic scattering, near-infrared frequency transparency, and ultra-broadband surface plasmon resonances.[472] These exceptional optical behaviors show prospects of topological materials in unique *optoelectronic* applications like plasmonic solar cells, ultrathin holograms, plasmonic and Fresnel lens, Veselago lens, tips for scanning tunneling microscope, broadband photodetectors, and nanoscale waveguides.[43,472] On the other hand, backscattering-less topological states can potentially make the next-generation *photonics* applications like light source devices, light-transmission devices, and optical signal processing devices.[473,474] Due to the importance of laser devices in various application fields,[475,476,] utilizing the topological materials in laser applications receive great attention among both optoelectronic and condensed matter societies.[477,478] Nevertheless, like the other applications, topological optoelectronic or photonic device structures also have roadblocks associated with the structural interface quality, efficient and reliable control, and complete understanding of the topological optical properties.

The emergence of *Spintronics* in device applications provides unique advantages of near-zero Joule heating and ultrafast operation. Several spin properties of topologically protected robust surface states like strong spin-momentum locking, high charge current to spin current conversion efficiency, long spin diffusion length, strong magnetoresistance, and high spin filtering efficiency make the topological materials suitable for spintronics applications.[479] Both topological insulators and semimetals may enable future energy-efficient spintronic devices. Typical topological spintronic device applications can be the memory device, spin-filter transistor, and spin-torque transistors, which can be attained via topological insulators, Dirac, and Weyl semimetals.[299]

*The superconductor* is another exciting research field. The trivial superconducting materials have the Bose-Einstein condensed Cooper pairs.[480] However, the topological phase into superconductors has a distinct nature from the Cooper pairs. Typically, topological superconductors have a fully-open gap between the ground and excited states along with non-zero topological invariants.[480] Topological superconducting states hold parity-hole symmetry and contain non-Abelian Majorana zero mode coming from the spinless pairing of the time-reversal breaking *s*-, *p*- or *d*-waves with spin-orbit coupling.[481] A Majorana fermion in topological superconductors has unique properties. It is a quasiparticle that is its own antiparticle and can emerge from the superposition of an electron and a hole having the same direction or spin.[482,483] Generally, topological insulators in 2D and 3D demonstrate the superconducting states in 1D or 2D topological surface states, respectively. But topological crystalline insulators, Dirac and Weyl semimetals, can also exhibit superconducting states.[480] Topological superconductors exhibit different unique properties like zero-bias conductance peak, quantized thermal Hall conductivity, nematicity, anomalous Josephson effects, and odd-frequency Cooper pairs. Majorana superconducting mode can also be utilized as physical qubits for quantum computers. Despite the progress in topological superconductors, more understanding of the underlying physics of other superconducting states is still required. As the physical environment is different from theoretical estimation, more experimental studies are needed for topological superconducting materials.

Besides the abovementioned applications, there are other applications, such as different types of sensors, lab black-hole environment, and on-chip high-performance micro-ultrasonic devices.[43,484,485] For sensing applications, topological materials are found ideal for gas sensors,[485] magnetic sensors,[446] wireless sensors,[486] and strain sensors.[487]

Recently, several studies are performed on the nanostructuring of the topological phases to evaluate their performances in nano-devices.[488] However, nano-structured topological phases have their unique challenges from fabrication and characterizations.[488] Extensive ongoing research on topological materials, phases of matter, and the phase conversion mechanisms can pave the way for more exciting device applications. To realize the topological/quantum phases of matter into device applications successfully, more theoretical and experimental investigations are needed to understand the underlying physics, fabrication processes, control mechanisms, and suitable material systems.

## 11. Conclusion

Recent discoveries of topological materials in which macroscopic properties originate explicitly from quantum many-body effects can open a new landscape for developing transformational technologies not reachable by classical or even nanostructured materials. Topological materials are the prime manifestations of massless relativistic fermions in condensed matter physics, first predicted by P. Dirac and H. Weyl. Since the emergence of the topological materials, there have been growing interests in their technological applications, such for future smart materials, due to their salient features like global symmetry protected metallic surface-states with spin-polarized relativistic fermions with zero-mass and high mobility, defect tolerance, and near dissipationless transport properties.

Protecting or breaking the global symmetry of the material can lead to different topological phases such as the topological insulator, Dirac semimetal, Weyl semimetal, and nodal line semimetal. Numerous materials have already been reported experimentally and theoretically to have topological phases. $Z_2$ index-defined topological phases of matter are strongly related to the Berry connection and Berry curvature of the effective Brillouin zone, defined by the time-reversal invariant momenta (TRIM) points based on time-reversal symmetry (TRS) and Kramers' theorem. Besides TRS, discrete crystal symmetry (DCS) can also introduce relativistic fermions in the different crystal systems. The spin-orbit coupling (SOC) provides mixing of *s* and *p* orbitals that can lead to band inversion, another important behavior of a topological phase.

If topological materials are going to make transformational technologies, it is critical to be able to control their properties. As such, a great deal of studies is focused on finding methods for the topological phase conversion, which not only increases the number of topological physical systems but also opens the pathway for the development of unique devices like near dissipationless transistors, spintronic devices, topological thermoelectric devices, fault tolerance quantum computers, etc. For successful device operation, the external stimuli for the topological phase conversion should be capable of controlling the TRS and SOC of the system reliably and efficiently. Most convenient external stimuli are in the extreme range of the host system, like a very high electric or magnetic field or an excessive strain beyond the limit of the material. Realizing such stimuli in practice may require unconventional or complex device structures to develop a viable topological device that can be controlled without failure. Moreover, the ease of operation, scalability, and compatibility with conventional devices are other essential requirements for topological devices. Besides the appropriate materials and the control mechanisms, it is also critical to develop the fabrication and integration methods for realizing a complete topological device.

## Acknowledgments

The authors acknowledge the funding support by the Air Force Office of Scientific Research (AFOSR) under Contract No. FA9550-19-1-0363 and the National Science Foundation (NSF) under Grants No. ECCS-1351533, No. ECCS-1515005, and No. ECCS-1711253.